\documentclass[twocolumn]{aastex63}

\usepackage{verbatim}
\usepackage{morefloats}
\usepackage{apjfonts}
\usepackage{amsmath}
\usepackage{xcolor}
\usepackage{ragged2e}
\usepackage[normalem]{ulem}

\newcommand{\kms}{\rm\ km\ s^{-1}}
\newcommand{\msun}{M_\odot}
\usepackage{float}
\usepackage{enumitem}
\usepackage{natbib}
\DeclareUnicodeCharacter{2212}{-}

\shortauthors{Omoruyi et al.}
\shorttitle{`Beads on a String' Star Formation Tied to a Powerful AGN Outburst}

\begin{document}

\title{`Beads on a String' Star Formation Tied to one of the most\\  Powerful AGN Outbursts Observed in a Cool Core Galaxy Cluster}

\correspondingauthor{Osase Omoruyi}
\email{osase.omoruyi@gmail.com}

\author[0000-0002-3649-5362]{Osase Omoruyi}
\affiliation{Center for Astrophysics $|$ Harvard \& Smithsonian, 60 Garden St.,
Cambridge, MA 02138, USA}
\author[0000-0002-5445-5401]{Grant R.~Tremblay}
\affiliation{Center for Astrophysics $|$ Harvard \& Smithsonian, 60 Garden St.,
Cambridge, MA 02138, USA}
\author[0000-0003-2658-7893]{Francoise Combes}
\affiliation{Observatoire de Paris, LERMA, Collège de France, CNRS, PSL University, Sorbonne University, 75014 Paris, France}
\author[0000-0003-4932-9379]{Timothy A. Davis}
\affiliation{Cardiff Hub for Astrophysics Research \&\ Technology, School of Physics \&\ Astronomy, Cardiff University, Cardiff, CF24 3AA, UK}
\author[0000-0003-1370-5010]{Michael D. Gladders}
\affiliation{Department of Astronomy and Astrophysics, University of Chicago, 5640 South Ellis Avenue, Chicago, IL 60637, USA}
\author[0000-0001-8121-0234]{Alexey Vikhlinin}
\affiliation{Center for Astrophysics $|$ Harvard \& Smithsonian, 60 Garden St.,
Cambridge, MA 02138, USA}
\author[0000-0003-0297-4493]{Paul Nulsen}
\affiliation{Center for Astrophysics $|$ Harvard \& Smithsonian, 60 Garden St.,
Cambridge, MA 02138, USA}
\affiliation{ICRAR, University of Western Australia, 35 Stirling Hwy, Crawley, WA 6009, Australia}
\author[0000-0003-3203-1613]{Preeti Kharb}
\affiliation{National Centre for Radio Astrophysics, Tata Institute for Fundamental Research,
Savitribai Phule Pune University, Ganeshkhind, Pune 411007, India}
\author[0000-0002-4735-8224]{Stefi A. Baum}
\affiliation{University of Manitoba, Dept. of Physics and Astronomy, Winnipeg, MB R3T 2N2, Canada}
\author[0000-0001-6421-054X]{Christopher P.~O'Dea}
\affiliation{University of Manitoba, Dept. of Physics and Astronomy, Winnipeg, MB R3T 2N2, Canada}
\author[	
0000-0002-7559-0864]{Keren Sharon}
\affiliation{Department of Astronomy, University of Michigan, 1085 S. University Ave, Ann Arbor, MI 48109, USA}
\author[0000-0001-5529-7305]{Bryan A.~Terrazas}
\affiliation{Columbia Astrophysics Laboratory, Columbia University, 550 West 120th Street, New York, NY 10027, USA}
\author[0000-0003-1056-8401]{Rebecca Nevin}
\affiliation{Fermi National Accelerator Laboratory, P.O. Box 500, Batavia, IL 60510, USA}
\author[0000-0001-7120-2433]{Aimee L.~Schechter}
\affiliation{Department of Astrophysical and Planetary Sciences, University of Colorado, Boulder, CO 80309, USA}
\author[0000-0003-3175-2347]{John A.~Zuhone}
\affiliation{Center for Astrophysics $|$ Harvard \& Smithsonian, 60 Garden St., Cambridge, MA 02138, USA}
\author[0000-0001-5226-8349]{Michael McDonald}
\affiliation{Kavli Institute for Astrophysics and Space Research, Massachusetts Institute of Technology Cambridge, MA 02139, USA}
\author[0000-0003-2200-5606]{H{\aa}kon Dahle}
\affiliation{Institute of Theoretical Astrophysics, University of Oslo, P.O. Box 1029, Blindern, NO-0315 Oslo, Norway}
\author[0000-0003-1074-4807]{Matthew B.~Bayliss}
\affiliation{Department of Physics, University of Cincinnati, Cincinnati, OH 45221, USA}
\author[0000-0002-7898-7664]{Thomas Connor}
\affiliation{Center for Astrophysics $|$ Harvard \& Smithsonian, 60 Garden St., Cambridge, MA 02138, USA}
\author[0000-0001-5097-6755]{Michael Florian}
\affiliation{Steward Observatory, University of Arizona, 933 North Cherry Ave., Tucson, AZ 85721, USA}
\author[0000-0002-7627-6551]{Jane R.~Rigby}
\affiliation{NASA Goddard Space Flight Center, 8800 Greenbelt Road, Greenbelt, MD 20771, USA}
\author[0000-0003-3295-6595]{Sravani Vaddi}
\affiliation{Arecibo Observatory, NAIC, HC3 Box 53995, Arecibo, Puerto Rico, PR 00612, USA}

\begin{abstract}
\noindent With two central galaxies engaged in a major merger and a remarkable chain of 19 young stellar superclusters wound around them in projection, the galaxy cluster SDSS J1531+3414 ($z=0.335$) offers an excellent laboratory to study the interplay between mergers, AGN feedback, and star formation. New \textit{Chandra} X-ray imaging reveals rapidly cooling hot ($T\sim 10^6$ K) intracluster gas, with two ``wings'' forming a concave density discontinuity near the edge of the cool core. LOFAR $144$ MHz observations uncover diffuse radio emission strikingly aligned with the ``wings,'' suggesting that the ``wings'' are actually the opening to a giant X-ray supercavity. The steep radio emission is likely an ancient relic of one of the most energetic AGN outbursts observed, with $4pV  > 10^{61}$ erg. To the north of the supercavity, GMOS detects warm ($T\sim 10^4$ K) ionized gas that enshrouds the stellar superclusters but is redshifted up to $+ 800$ km s$^{-1}$ with respect to the southern central galaxy. ALMA detects a similarly redshifted $\sim 10^{10}$ M$_\odot$ reservoir of cold ($T\sim 10^2$ K) molecular gas, but it is offset from the young stars by $\sim 1{-}3$ kpc. We propose that the multiphase gas originated from low-entropy gas entrained by the X-ray supercavity, attribute the offset between the young stars and the molecular gas to turbulent intracluster gas motions, and suggest that tidal interactions stimulated the ``beads on a string" star formation morphology.

\end{abstract}
\vspace{-5mm}

\section{Introduction} 
\label{sec:intro}

Near the centers of galaxy clusters lie the most luminous and massive elliptical galaxies in the Universe -- Brightest Cluster Galaxies (BCGs).  These galaxies are broadly characterized as "red and dead" due to their smooth ellipsoidal shapes and significant fractions of old stars \citep{edwards_stellar_2016, kormendy_elliptical_2016}. However, under certain conditions, the hot ($ > 10^6$ K), diffuse plasma found between galaxies in clusters, known as the intracluster medium (ICM), cools rapidly, fueling new star formation and black hole activity near the BCG. 

Early X-ray observations revealed that in approximately half of galaxy clusters, the intracluster gas harbored dense central regions with temperatures cooler than the cluster virial temperature and cooling times shorter than the age of the Universe \citep[e.g.,][]{fukazawa_metal_1994, kaastra_spatially_2004, sanderson_statistically_2006}. These observations led to the development of the simple ``cooling flow" model \citep{cowie_radiative_1977, fabian_subsonic_1977}, which postulated that in the absence of a heat source to compensate for the rapidly cooling gas, several $100\ M_\odot\rm\ yr^{-1}$ of plasma should cool to form large cold gas reservoirs of $5-50 \times 10^{11}$ $M_\odot$ near the cluster center \citep[e.g.,][]{fabian_cooling_1994}. 

Follow-up ultraviolet, optical, and infrared observations, however, found less than 1\% of the predicted amount of cooled gas and highly suppressed star formation rates of $\sim 1-100$ $M_\odot$ yr$^{-1}$ near BCGs \citep[e.g.,][]{johnstone_optical_1987, romanishin_imaging_1987, mcnamara_star_1989, crawford_nature_1993, allen_starbursts_1995, crawford_rosat_1999, mittaz_uv_2001, rafferty_feedback-regulated_2006, odea_infrared_2008, mcnamara_1010_2014, donahue_ultraviolet_2015, mittal_constraining_2015, mcdonald_revisiting_2018}. Moreover, high spectral resolution imaging and spectroscopy from the 1999 launches of the \textit{Chandra} and XMM-Newton X-ray space observatories found that the supposed "cooling-flow" clusters exhibited minimal-to-no evidence for cooling down to temperatures of $\sim 0.1$ keV and below \citep[e.g.,][]{peterson_high-resolution_2003, pinto_discovery_2014}. These discrepancies suggested that either the remaining cooling is hidden from view and/or that some steady form of heating compensates for the bulk of the radiative cooling (for reviews, see \citealt{mcnamara_heating_2007, mcnamara_mechanical_2012, fabian_observational_2012, fabian_hidden_2022}). 

Various heating sources have been suggested for quenching cooling flows, such as thermal conduction, gas motions generated by mergers, and cosmic rays. However, feedback from active galactic nuclei (AGN) has since emerged as the most important source. The effects of "radio-mode" AGN feedback are nearly ubiquitous in cool cores, with 19 of the 20 brightest "cooling flow" clusters harboring distinct X-ray cavities \citep{fabian_observational_2012}. These cavities, also known as ``bubbles", are often filled with radio emission from the lobes of the BCG's central AGN, suggesting that the lobes displaced the surrounding ICM to create the observed cavities \citep{birzan_systematic_2004}. As the bubbles buoyantly rise, the work done by the expanding radio lobes heats the surrounding X-ray plasma, offsetting radiative losses from the ICM \citep{churazov_cooling_2002}. 

Although AGN mechanical feedback is routinely invoked as a mechanism for quenching star formation in simulations of galaxy formation \citep[e.g.,][]{somerville_physical_2015}, it does not completely offset ICM cooling, resulting in residual cooling at either low ($< 1\%$) rates \citep[e.g.,][]{tremblay_residual_2012} or in elevated episodes as the AGN varies in power \citep[e.g.,][]{odea_hubble_2010, tremblay_feedback_2011, mcdonald_deeplessigreaterchandralessigreaterlessigreaterhstlessigreater-cos_2015}.  Observations have revealed massive flows of atomic and molecular gas that appear entrained around the rims of jet-blown cavities \citep[e.g.,][]{russell_close_2017, tremblay_galaxy-scale_2018}, or closely trailing behind them \citep[e.g.,][]{vantyghem_molecular_2016, russell_alma_2016}, suggesting an incredibly efficient coupling between the radio jets, the ICM, and the cooled multiphase gas. Given that the cooling hot plasma, warm ionized gas, cold molecular gas, and radio emission from the AGN should each retain imprints of their shared journey within the ICM, multi-wavelength observational studies of cool core clusters have become standard practice for studying the interplay between nebular emission, star formation, and AGN activity, commonly referred to as the ``AGN feedback cycle" \citep[e.g.,][]{odea_infrared_2008, mcnamara_1010solar_2014, mcdonald_deep_2015, russell_close_2017, tremblay_galaxy-scale_2018, pasini_bcg_2019, ciocan_vlt-muse_2021, calzadilla_testing_2022, tamhane_radio_2023, masterson_evidence_2023}.

The AGN feedback cycle, however, is not without potential disruption. Under the hierarchical model of structure formation, both BCGs and galaxy clusters assemble via a sequence of major and minor mergers driven by gravity \citep[e.g.,][]{peebles_primeval_1970, rosati_evolution_2002, voit_observationally_2005, kravtsov_formation_2012}. Galaxy-galaxy interactions have long been predicted by theoretical \citep[e.g.,][]{martinet_evidences_1995} and numerical analysis \citep[e.g.,][]{toomre_galactic_1972, barnes_dynamics_1992, mihos_gasdynamics_1996} to generate tidal torques that drive gas inflows. These gas inflows may then be accreted by the central supermassive black hole, fueling nuclear activity \citep{sanders_ultraluminous_1988}.  Multiple observational
studies show significant AGN enhancement in merging galaxies 
\citep[e.g.,][]{alonso_active_2007, woods_minor_2007, weston_incidence_2017}. Contrasting findings exist, however, with some studies not observing a higher merger rate in AGNs \citep{grogin_agn_2005, kocevski_candels_2011}. On larger scales, cluster-cluster mergers can disperse and reheat cooling gas in the cluster core. These clusters are often home to diffuse radio emission, indicating the presence of relativistic particles (i.e., cosmic rays) and cluster-wide magnetic fields amplified by the shocks and turbulence injected into the ICM by the merger. Simulations predict that such turbulence has the potential to stimulate condensation, which then fuels AGN activity \citep[e.g.][]{gaspari_raining_2017, gaspari_shaken_2018}.

\begin{figure*}
\centering
\includegraphics[width=\linewidth]{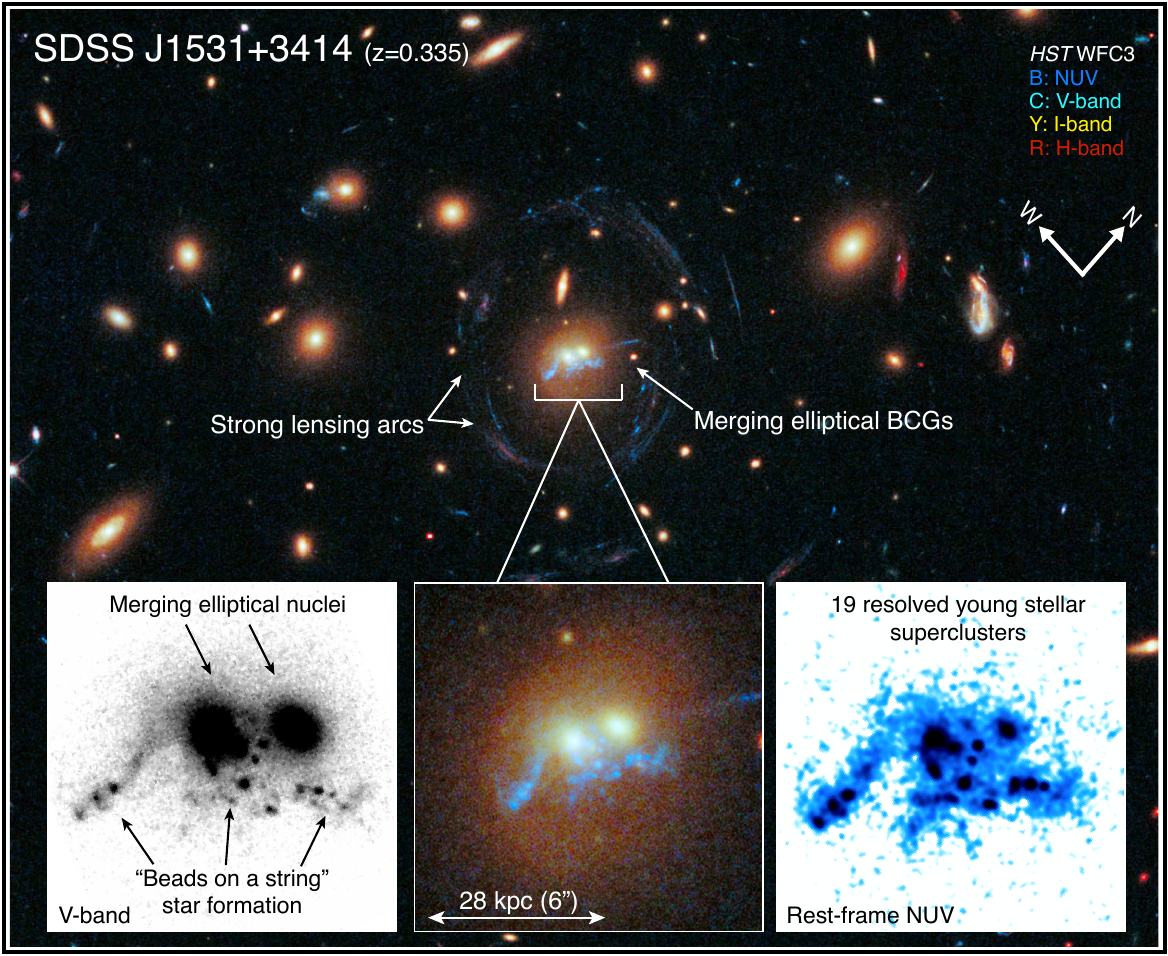}
\caption{Hubble's Wide Field Camera 3 (WFC3) view of the SDSS J1531+3414 galaxy cluster (hereafter SDSS 1531), the focus of this paper. Near-UV (NUV), V-band, H-band, and I-band emission are shown in blue, cyan, red, and yellow, respectively. The cluster features remarkable strong-lensing arcs, numerous elliptical and spiral galaxies, and the focus of this paper: merging elliptical brightest cluster galaxies (BCGs). From left to right, the three inset panels show a closer view of the merging elliptical nuclei and ``beads on a string" star formation in the V-band, the BCGs in all bands, and the 19 resolved young stellar superclusters in the rest-frame NUV.}
\label{fig:hst}
\end{figure*}

Observational studies of cool core clusters have traditionally biased ``relaxed" systems due to their primary identification via X-ray surveys. Although the detection of mass-selected samples of clusters via the Sunyaev-Zel'dovich effect \citep{andrade-santos_fraction_2017} has facilitated the study of a larger sample of dynamically disturbed systems \citep[e.g.,][]{olivares_x-ray_2023}, these systems remain comparatively understudied. To contribute to the evolving understanding of AGN feedback in disturbed systems, we present a multi-wavelength study of a recently discovered cool core cluster that features remarkable signatures of cooling-powered star formation amid a major merger between the two central galaxies: the SDSS J1531+3414 (hereafter SDSS 1531) galaxy cluster.

\begin{deluxetable*}{cccccccccc}
\tabletypesize{\footnotesize}
\tablecaption{\textsc{Summary of Observations}\label{tab:observations}}
  \tablehead{
    \colhead{Waveband} &
    \colhead{Observatory} &
    \colhead{Instrument} &
    \colhead{Filter / Config.} &
    \colhead{Central $\lambda$ / Line} &
    \colhead{Integration Time} &
    \colhead{Obs. Date} &
    \colhead{Comment} \\
    \colhead{(1)} &
    \colhead{(2)} &
    \colhead{(3)} &
    \colhead{(4)} &
    \colhead{(5)} &
    \colhead{(6)} &
    \colhead{(7)} &
    \colhead{(8)} 
}
  \startdata
X-ray & \textit{Chandra}    &  ACIS-S &  VFAINT & 0.5-7 keV &  62.73 ks & 20-21 Oct 2015 & Ambient intracluster gas\\
\nodata & \nodata   & \nodata  &  \nodata & \nodata  &  59.46 ks & 28-29 Oct 2015 & \nodata \\
\hline
Optical & \textit{HST}  & WFC3 / UVIS &  F390W  & 3923 \AA &  2256 s & 6 May 2013 & Young stellar component   \\
\nodata & \nodata    &  \nodata &  F606W  & 5887 \AA &  1440 s & \nodata & Includes [O~\textsc{ii}], H$\beta$   \\
\nodata & \nodata    &  \nodata &  F814W  & 8026 \AA &  1964 s & \nodata & Includes H$\alpha$+[N~\textsc{ii}]   \\
\nodata & \nodata    &  WFC3 / IR &  F160W  & 1.537 $\mu$m &  912 s & \nodata & Old stellar component   \\
\nodata & Nordic Optical & ALFOSC  &  Grism \#7 / 1\arcsec   &  5260 \AA\  &  2400 s  & 29 Apr 2014                       & Redshift confirmation \\
\nodata & \nodata & \nodata  &  Grism \#5 / 2\farcs5   &  7000 \AA\    & \nodata   & \nodata                       & SFR estimate \\
\nodata & GMOS-N    &  EEV DD &  R400  &  700 nm  &  1800 s & 27 June 2014 & Stellar \& emission line kinematics \\
\hline
Radio & ALMA    &  Band 6 &  Compact, Extended & 259.023213 GHz &  105 min & 22 Apr 2016 & CO$(3-2)$; Cold filaments\\
\nodata & \nodata    &  \nodata &  \nodata & 260 GHz &  \nodata & \nodata & Continuum; Non-detection\\
\nodata & \nodata    &  \nodata &  \nodata & 245 GHz &  \nodata & \nodata & \nodata\\
\nodata & \nodata    &  \nodata &  \nodata & 243 GHz &  \nodata & \nodata & \nodata\\
\nodata & IRAM 30-m    &  EMIR &  $-$ & 86 GHz &  2880 s & 22 Dec 2013 & CO$(1-0)$; Non-detection\\
\nodata & EVLA    &  L-Band &  C-array & 1.5 GHz &  37 min & 22 Mar 2014 & Non-detection\\
\nodata & VLA    &  \nodata &  B-array & 1.4 GHz &  165s & 18 Jun 1994 & few diffuse radio sources\\
\nodata & \nodata    &  \nodata &  D-array & 1.4 GHz &  30s & 04 May 1995 & Non-detection\\
\nodata & LOFAR    &  HBA &  Dual Inner & 143.65 MHz &  8 hr & 14 September 2018 & Several diffuse radio sources\\
\enddata
 \tablecomments{Summary of all new and archival observations presented in this paper. The observations are presented in descending order of wavelength, from X-ray to radio.
   (1) Waveband;
   (2) Facility name;
   (3) instrument (and aperture / detector/ band) used for observation;
   (4) imaging filter or spectroscopic configuration;
   (5) waveband, filter/grism central wavelength, or emission lines covered by observation;
   (6) on-source exposure time;
   (7) date of observation;
   (8) comment specific to observation.
}
\vspace*{-9mm}
\end{deluxetable*}

SDSS 1531 is a strong-lensing cluster of galaxies at $z = 0.335$  \citep{hennawi_new_2008,oguri_subaru_2009,gralla_sunyaev-zeldovich_2011,bayliss_geminigmos_2011,postman_cluster_2012}. The cluster core was imaged as part of the {\it Hubble Space Telescope}
({\it HST}) strong-lensing imaging program \citep{sharon_strong_2020}, revealing a remarkable $\sim 28$ kpc-scale network of young stellar superclusters (or tidal dwarf galaxies) wound around and in between two merging giant elliptical galaxies of roughly equal stellar mass in projection (see Figure \ref{fig:hst}). The superclusters are reminiscent of the “beads on a string” star formation frequently observed in the arms of spiral galaxies, resonance rings, and the tidal arms that bridge interacting galaxies \citep[e.g.,][]{elmegreen_extension_1996}. Our initial Letter on this system 1) resolved 19 young stellar superclusters in the \textit{HST} NUV filter (Figure \ref{fig:hst}; rest-frame NUV inset plot, bottom right); 2) determined that the stellar superclusters and central elliptical galaxies comprising the BCGs are pinned to nearly the same redshift, indicating that the observed features are coplanar and not the result of a projection effect and 3) estimated an extinction-corrected $5-10$ M$_\odot$ yr $^{-1}$ star formation rate (SFR) \citep{tremblay_30_2014}. In a complementary work that analyzed the cluster's strong-lensing properties, \cite{sharon_mass_2014} concluded that the overwhelming majority of the observed NUV emission is too bright to stem from counter images of the lensed galaxies or a faint central image of a background source.

In this work, we present new data from the Atacama Large Millimeter and submillimeter Array (ALMA), the \textit{Chandra} X-ray Observatory, the Gemini North telescope's Gemini Multi-Object Spectrograph (GMOS), the Low Frequency Array (LOFAR) and the Very Large Array (VLA) to facilitate a multi-wavelength view of SDSS 1531's dynamic environment and uncover the origin and evolution of the "beads on a string" star formation complex. This paper is organized as follows: Section \ref{sec:observations} describes our procedure for reducing and analyzing the new observations from \textit{Chandra}, LOFAR, VLA, GMOS, and ALMA. Section \ref{sec:results} presents spatial and spectral results for each gas phase. We use 1) \textit{Chandra} X-ray observations to create thermodynamic profiles and spectral maps that reveal the general structure and properties of the hot ICM; 2) LOFAR and VLA radio surveys to identify radio emission associated with AGN activity or larger-scale cluster-activity, 3) GMOS integral field unit (IFU) observations to create emission line maps that display the spatial orientation of the warm ionized gas, unveil its kinematics and allow us to explore potential sources of ionization; and 4) ALMA observations to create maps revealing the morphology, kinematics, and mass distribution of the cold molecular gas. Section \ref{sec:beads_origin} synthesizes the results and proposes scenarios for the origin and fate of the star formation complex. 

Throughout this study we assume $H_0$ = 70 km s$^{−1}$ Mpc$^{−1}$, $\Omega_M$ = 0.27, and $\Omega_\Lambda$ = 0.73. In this cosmology, 1\arcsec corresponds to 4.773 kpc at the redshift of the southern BCG ($z = 0.335$), where the associated angular size and luminosity distances are 984.4 and 1756.1 Mpc, respectively, and the age of the Universe is 9.728 Gyr. The spectral index, $\alpha$, is defined such that the flux density at frequency $\nu$ is $S_\nu \propto \nu^{\alpha}$.

\section{Observations and Data Reduction} \label{sec:observations}
In this section, we describe the new and archival observations from  \textit{Chandra}, VLA, LOFAR,  GMOS, and ALMA. All new and archival observations of SDSS 1531 are summarized in Table \ref{tab:observations}.  All Python codes/Jupyter Notebooks used and created to analyze the data are publicly available in an online repository.\footnote{\url{https://github.com/osaseo/Beads2023_Code}}

\subsection{Chandra X-ray Observations}
\label{subsec:chandraobs}

As part of Cycle 16 program 16800783 (PI: Baum), the \textit{Chandra} X-ray Observatory observed SDSS 1531 on 2015 October 20-21 (\dataset[ObsID 17218]{https://doi.org/10.25574/17218}) and 2015 October 28-29 (\dataset[ObsID 18689]{https://doi.org/10.25574/18689}) for a total of 122.2 ks (62.73 and 59.46 ks, respectively). All exposures centered the cluster core on the nominal aimpoint of the back-illuminated Advanced CCD Imaging Spectrometer (ACIS)-S3 chip.

\begin{figure}
\centering
\includegraphics[width=\linewidth]{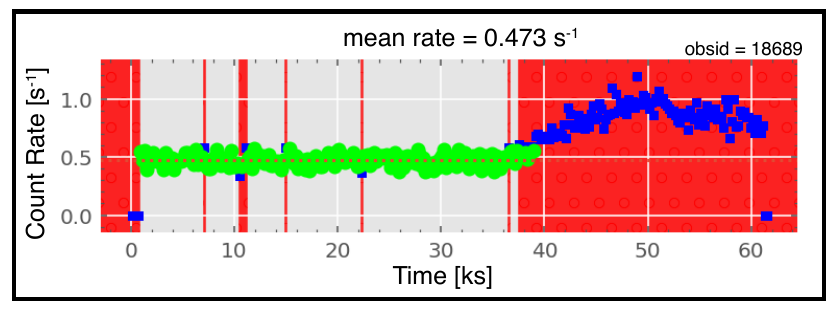}
\caption{Count rate vs time in \textit{Chandra} ObsID 18689. $\geq 3\sigma$ flares are persistently detected from $\sim 45-60$ ks, requiring the complete removal of the last 15 ks, one-quarter of the total exposure time.}
\label{fig:chandra_flare}
\end{figure}

To reduce and analyze the data from both ObsIDs, we used \textsc{ciao} v4.13 (\textit{Chandra} Interactive Analysis of Observations; \citealt{fruscione_ciao_2006}) and \textsc{CALDB} v4.9.3. We reprocessed the data with \textit{chandra\_repro}, cleaning the ACIS background in \textsc{VFAINT} mode. Flares were identified and filtered using the ChIPS routine \textsc{lc\_clean}. To completely remove the $\geq 3\sigma$ flares detected during the last 15 ks of ObsID 18689 (see Figure \ref{fig:chandra_flare}), we excluded this period from analysis. Point sources were identified through a wavelet decomposition technique \citep{vikhlinin_catalog_1998} and visually inspected before masking. After cleaning, we retained a total of 65 ks (44.1 ks and 20.9 ks for ObsIDs 17218 and 18689, respectively), yielding a total $\sim 8200$ net counts in a 60\arcsec radius centered on the cluster's X-ray emission peak. The counts present are sufficient to constrain the surface brightness profile within the central region, classify SDSS 1531 as a cool core cluster, search for tentative X-ray cavities within the cluster center, and perform cursory measurements of key spectral properties, such as the central projected temperature and cooling time \citep{mcdonald_detailed_2019}.

\begin{figure}
\centering
\includegraphics[width=\linewidth]{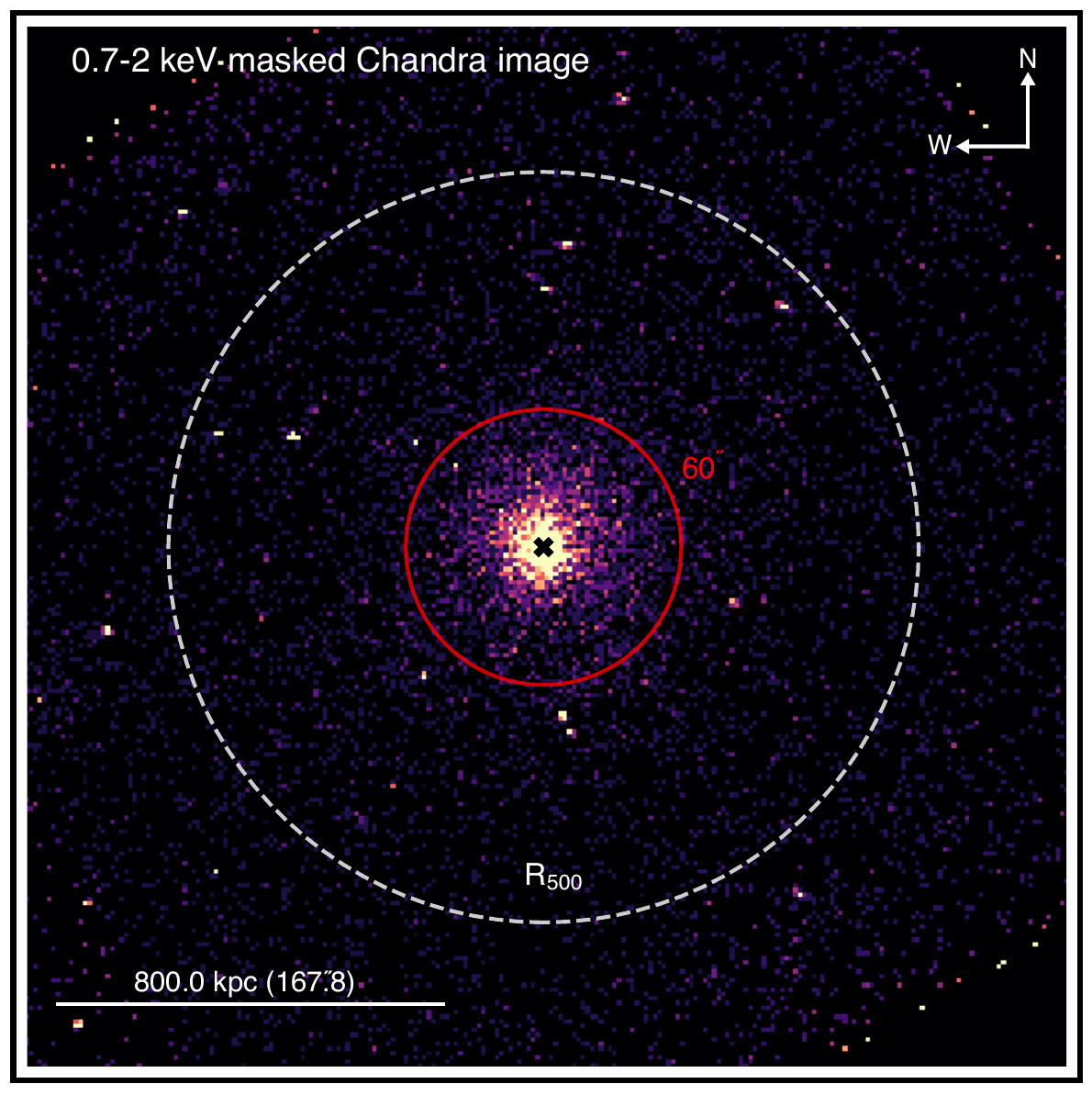}
\caption{\textit{Chandra} X-ray image of SDSS 1531 in the 0.7–2 keV band. We primarily analyze the spectra extracted from 20 linearly spaced annuli within $60\arcsec$ (red solid circle) of the cluster's central region, marked by the peak of the X-ray emission (black cross). The derived value for $R_{500}$ is marked by the white dashed circle.}
\label{fig:chandra_annuli}
\end{figure}

To analyze the morphology of the intracluster gas, we merged the ObsIDs with the \textsc{merge\_obs} \texttt{CIAO} script. To highlight surface brightness edges, we used the wavelet decomposition technique \citet{vikhlinin_catalog_1998} 
to create a reconstructed $0.5-7$ keV image. We also created an unsharp-masked image by smoothing the data with a $0\farcs98$ Gaussian and subtracting it from the same image smoothed with a $9\farcs8$ Gaussian \citep[e.g.,][]{hlavacek-larrondo_extreme_2012}.

To derive the spectral properties of the intracluster gas, we first extracted spectra from a series of 20 concentric annuli centered on the X-ray emission peak from $1 - 60\arcsec$ in the energy range $0.5-7$ keV (see Figure \ref{fig:chandra_annuli}) for each ObsID. Next, we defined a $46\farcs5$ circle within the same chip but further from the cluster to encompass the appropriate local background, then subtracted it from the annuli.  Each annulus contained $\sim 200 - 800$ net counts across both ObsIDs. Next, we fit and modeled the total $0.5-7.0$ keV spectrum extracted from each annulus to a \textsc{phabs*apec} model in \textsc{XSPEC}. The hydrogen column density and abundance were fixed at $N_H = 1.79 \times 10^{20}$ cm$^{-2}$  (estimated using NASA's HEASARC $N_H$ Column Density tool, \citealt{bekhti_hi4pi_2016}) and $0.3 Z_\odot$ \citep[e.g.,][]{panagoulia_volume-limited_2014}, respectively, to better constrain the fit. The temperature \textit{kT} and normalization parameter were allowed to vary. 

The normalization factor $N(r)$ of the \textsc{apec} model provides an estimate for the 3D density profile through the relation:
\begin{align}
N(r) = \frac{10^{-14}}{4\pi [D_A(1+z)]^2} \int n_e n_p dV,
\label{eqn:norm}
\end{align}

\noindent where $D_A$ is the angular distance of the source, $n_e$ is the electron density, $n_p$ is the proton density, and $V$ is the volume integral performed on the projected annulus along the line of sight. Assuming spherical symmetry, we can estimate the projected $n_e$ as:
\begin{align}
n_e = \sqrt{10^{14} \Big(\frac{4\pi \cdot N(r) \cdot [D_A \cdot (1+z)^2]}{0.82\cdot V}\Big)} .
\label{eqn:electron_density}
\end{align}

From the electron density and temperature of the ICM, we obtained the pressure $P\equiv 1.83n_ekT$, the entropy $K \equiv kTn_e^{-2/3}$, and the cooling time $t_{\text{cool}}$ of the ICM, defined as:

\begin{align}
t_{\text{cool}} (r)\equiv \frac{3}{2} \frac{(n_e + n_p) kT(r)}{n_en_H \Lambda(kT(r), Z)},
\label{eqn:tcool}
\end{align}

\noindent where $\Lambda$ is the cooling function \citep{sutherland_cooling_1993}, $n_e = \sqrt{(n_e n_p)/0.92}$ , $n_p = 0.92n_e$, $n_H = 0.83 n_e$ and we assume $Z=Z_\odot/3$. 

As a consistency check on our values from the above analysis and to obtain a careful estimate of the cluster mass, we follow a procedure similar to that outlined in \cite{vikhlinin_chandra_2006}. First, we generated surface brightness profiles for each ObsID by extracting spectra from concentric annuli with finely spaced radii. The annuli were centered on the peak of the X-ray emission and extended from $1\arcsec$ to $1000\arcsec$.

The resulting surface brightness profile for each ObsID was corrected for spatial variations in temperature, metallicity, and effective area, and expressed as a projected emission measure integral. We then modeled each calibrated surface brightness profile with the following  modified $\beta$-model \citep{cavaliere_distribution_1978, vikhlinin_chandra_2006}:
\begin{align}
 n_p n_e = n_0^2\frac{(r/r_c)^{-\alpha}}{(1 + r^2/r_c^2)^{3\beta - \alpha/2}} \frac{1}{(1 + r^\gamma/r_s^\gamma)^{\epsilon/\gamma}} + \frac{n_{0,2}^2}{(1 + r^2/r_c2^2)^{3\beta_2}}
 \label{eqn:gas_density}
\end{align}
where $\alpha$ and $\beta$ are fit indices, $n_0$ is the core density, $r_c$ is the scaling radius of the core, and $r_s$ is the scaling radius of the extended components.  We fixed  $\gamma =3$ and $\epsilon < 5$ to exclude unphysically sharp density breaks and set all other parameters free. Projecting this three-dimensional model along the line of sight yielded an emission measure profile to fit to the data.

An analytic expression for the three-dimensional gas density profile $\rho_g(r)$ was obtained by fitting the emission measure profile. The gas density was estimated assuming $\rho_g = m_p n_e A/Z$, where $A$=1.397 and $Z=1.199$ represent the mean atomic mass and mean charge for a plasma with 0.3 Z$_\odot$, respectively \citep{mcdonald_growth_2013}. From the gas density profile, we computed the total gas mass $M_g$ within a spherical volume $V(r)$. From this, we obtain an estimate for the classical cooling rate $\dot{M}_{\text{cool}}$, where $\dot{M}_{\text{cool}} \equiv \frac{M_g(<r_{\text{cool}})}{t_{\text{cool}}} $.

The temperature profile was fit using 9 annuli of equal logarithmic width, $r_{\text{out}}/r_{\text{in}} = 1.4$, from the peak of the X-ray emission out to $60\arcsec$. After subtracting the background, each annulus contained $\sim 600 - 1300$ net counts across both ObsIDs. The temperature was modeled using a three-dimensional analytic model described by \citep{vikhlinin_chandra_2006}:
\begin{align}
T_{3D}(r) = T_0 \frac{x + T_{\text{min}}/T_0}{x + 1} \frac{(r/r_t)^{-a}}{\Big[ 1 + (r/r_t)^b \Big]^{c/b}} 
 \label{eqn:tprofile}
\end{align}
where $x = (r/r_{\text{cool}})^{\alpha_{\text{cool}}}$. The temperature profile was then obtained by projecting the model along the line of sight to fit the observed temperature values to $\sim 170$ kpc, where the temperature is well-constrained. 

To calculate the total cluster mass within radius $r$, we utilize a three-dimensional model for the gas density and temperature profiles and employ the hydrostatic equilibrium equation \citep[e.g.,][]{sarazin_x-ray_1988},
\begin{align}
M_{r} =  -3.68 \times 10^{13} M_\odot T(r) r \Big(\frac{d \log \rho_g}{d \log r} + \frac{d \log T}{d \log r} \Big),
 \label{eqn:mass_r}
\end{align}

\noindent where $T$ is in units of keV and $r$ is in units of Mpc. Given $M(r)$, we then compute the total matter density profile $\rho(r) = (4\pi r^2)^{-1} dM/dr$. Since $M(r)$ is most reliably determined in the central region of the cluster where the temperature profile is well-constrained, we obtain the total cluster mass $M_{500}$ from the $Y_X-M$ scaling relation, where $Y_X$ approximates the total thermal energy within $R_{500}$, the cluster radius corresponding to a density contrast of $\delta = 500$ \citep{vikhlinin_chandra_2009}. We list masses derived from the $Y_X-M$ scaling relation as $M_{500} - Y_X$ and masses derived from the hydrostatic equilibrium assumption as $M_{HE}$.

To generate high-resolution temperature, pressure, and entropy maps, we used the automated Python pipeline \textsc{clusterpyxt}\footnote{\url{https://github.com/bcalden/ClusterPyXT}} \citep{alden_galaxy_2019}. We provided the pipeline with the previously described cleaned, merged and exposure-corrected observations and fixed the galactic hydrogen density and cluster metal abundance to the same values described above. Point sources, identified by visual inspection, were carefully selected and fed to the pipeline to be excluded from the analysis. From here, \textsc{clusterpyxt} utilized the \textsc{CIAO} adaptive circular binning (\textsc{ACB}) algorithm to conduct high-resolution spectral fitting and subsequently generate temperature, pressure, and entropy maps. For a more detailed description of \textsc{clusterpyxt}'s spectral fitting process, we refer the reader to \citet{alden_galaxy_2019}.

We present and analyze the resulting X-ray deep image, spectral fits, profiles, and maps in Section \ref{subsec:hot_gas}.

\subsection{Radio Surveys}
\label{subsec:radio_obs}

To identify any diffuse radio emission associated with AGN or larger-scale cluster activity, as well as provide rough estimates of their spectral indices, we have thoroughly examined all publicly available images of the SDSS 1531 cluster field from the VLA and LOFAR radio facilities.

The VLA has observed SDSS 1531 over multiple epochs. The Expanded VLA conducted the most recent observation of the system on 22 March 2014 for 37 minutes in the 1.5 GHz band in the C configuration as part of project 14A-527 (PI: C. O'Dea). The source J1331+3030 (3C286) was used as a flux and phase calibrator for the observation. The data was successfully reduced using CASA (Common Astronomy Software Applications) version 5.1.0 \citep{mcmullin_casa_2007}, and the observation yielded a non-detection with a 3$\sigma$ point source upper limit of 135 $\mu$Jy. The VLA also observed the SDSS 1531 field in the 1.4 GHz band on 18 June 1994 in the B-array configuration (Project ID: AB628) and 04 May 1995 in the D-array configuration (Project ID: AC308).
Both observations were calibrated and imaged after several additional rounds of phase-only and phase+amplitude self-calibration using standard procedures in the Astronomical Image Processing System (AIPS) version 31DEC23. The final B-array image achieved an 
rms sensitivity of $\sim 0.17$ mJy/beam and has a restoring beam of size $2.88\arcsec \times 2.47\arcsec$ at p.a. $=−91.1^\circ$. The final D-array image achieved 
an rms sensitivity of $\sim 0.12$ mJy beam$^{-1}$ and has a restoring beam of size $37.7\arcsec \times 26.2\arcsec$ at p.a. $=78.3^\circ$.

On 14 September 2018, the LOFAR high-band array observed the cluster field at 144 MHz for a total of 8 hours as part of the LOFAR Two-meter Sky Survey (LoTSS) \citep{shimwell_lofar_2022} under project code LC10\_010. The second LoTSS data release (LoTSS-DR2) provided a public high-resolution 6\arcsec mosaic of the pointing P233+35, which contains the full cluster field near the lower center. The LoTSS-DR2 data reduction and image mosaicing pipelines are described in detail in \citet{shimwell_lofar_2022} and references within.

\subsection{GMOS-N Optical IFU Spectroscopy}
\label{subsec:gmosobs}

\begin{figure*}
\centering \includegraphics[width=\linewidth]{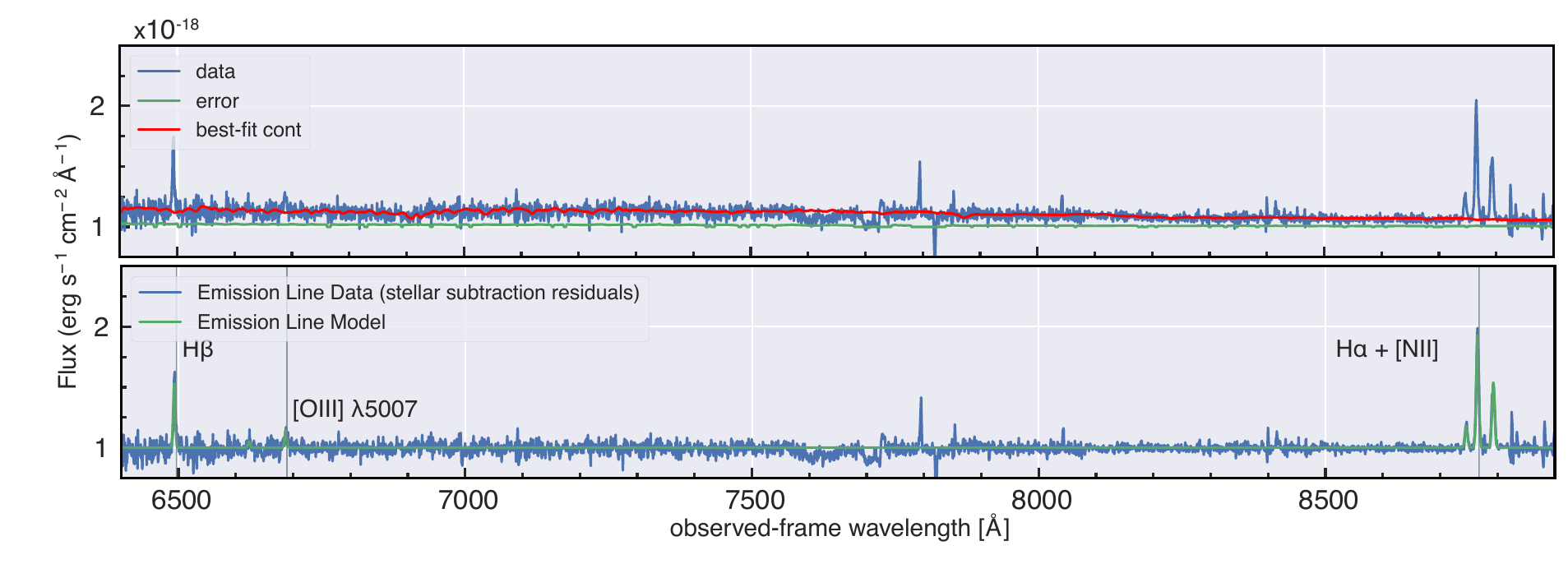}
\caption{An illustration of the best-fit stellar and gas-phase components obtained using \textsc{pyparadise} for one of the brightest spaxels (x=8,y=13) in the GMOS-N IFU datacube of the BCGs. The top panel shows the original spectrum with both stellar and gas emission (dark blue) and the best-fit stellar continuum by \textsc{PYPARADISE} (red). The error spectrum is shown in green. The spectrum of the stellar continuum subtracted gas-only component (dark blue) is depicted in the bottom panel, while the fit performed by the \textsc{pPXF} routine is shown in green.}
\label{fig:gmos_pyp}
\end{figure*}

\begin{figure}[H]
\centering \includegraphics[width=\linewidth]{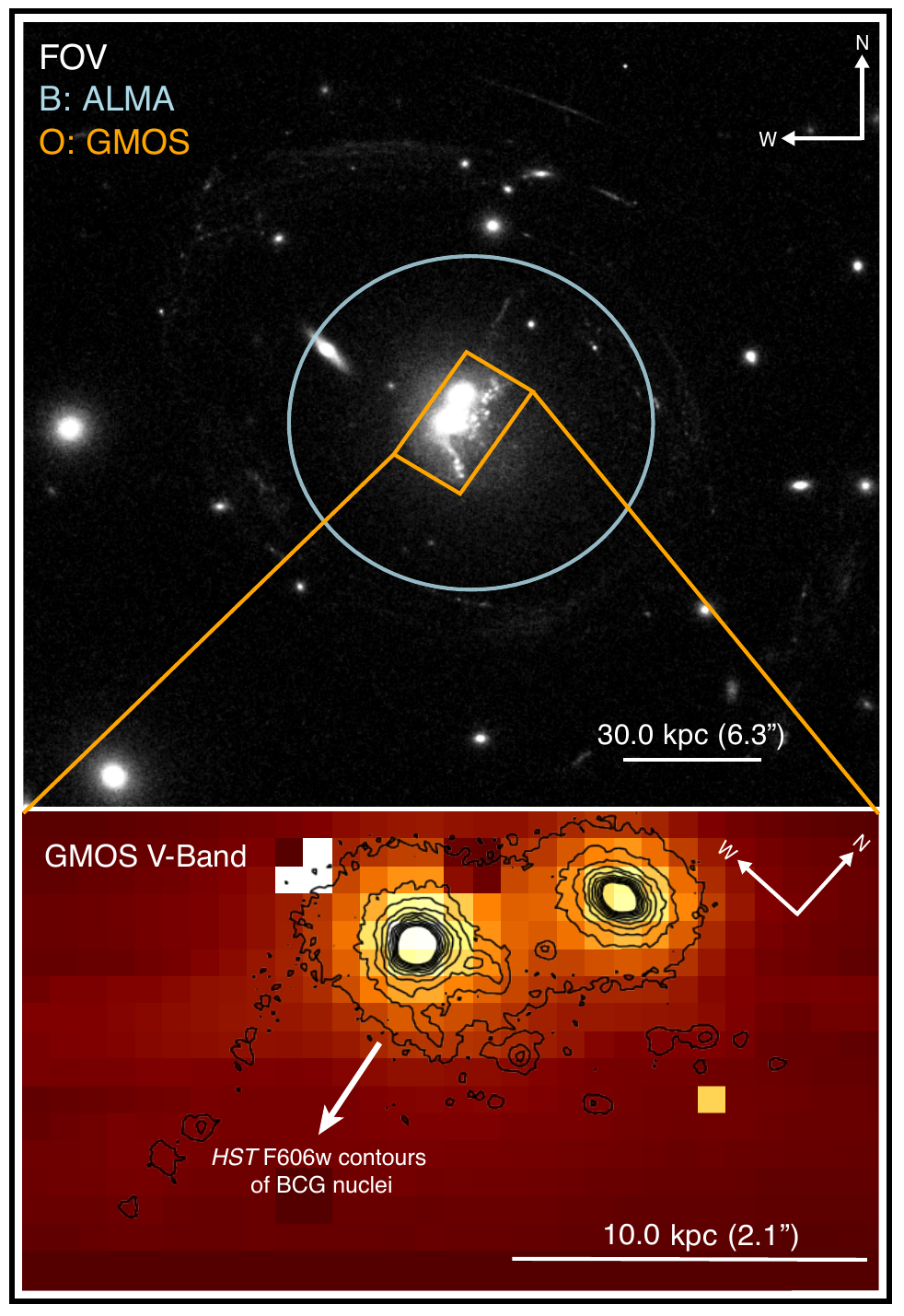}
\caption{\textit{Top:} FOV covered by ALMA (blue) and GMOS (orange), overlaid on the \textit{HST} F606w image of SDSS 1531. The \textit{Chandra}, VLA, and LOFAR observations (FOV not shown) cover the entire cluster field. \textit{Bottom}: GMOS V-Band view of the BCGs, with \textit{HST} contours (black) overlaid. The photocentroids of the optical nuclei of the galaxies in the GMOS cube match those of the \textit{HST} data.} 
\label{fig:hst_alma_fov}
\vspace{-4mm}
\end{figure}

In 2014, the GMOS-N instrument on the ground-based Gemini North telescope observed the SDSS 1531 BCGs in IFU-R mode (Gemini program GN-2014A-DD-3, PI: Tremblay). The 1800s exposure was taken with spatial dithering at a position angle of 330$^{\circ}$. The R400 grating was used and centered at 7000$\textup{~\AA}$, providing spectral coverage from 5800$\textup{~\AA}$ $ < \lambda <$ 9000$\textup{~\AA}$. The resolving power, $R$, of the R400 instrument is $R = \lambda/\Delta \lambda = 1918$ at a blaze wavelength of 764 nm, yielding a velocity resolution of $\sim 160$ km/s (FWHM) at 764 nm. The original field of view (FOV) of the instrument is $3\farcs5 \times 5 \arcsec$; after dithering, the final image area covered $3\farcs8 \times 6\farcs0$, corresponding to 18.1 kpc $\times$ 28.6 kpc at the target redshift (see Figure \ref{fig:hst_alma_fov}).  

The data was reduced using the Py3D data reduction package for fiber-fed IFU spectrographs \citep{husemann_large-scale_2016}. To obtain the most accurate emission line measurements, we decoupled and modeled the stellar and gas components of the galaxy using the package \textsc{PYPARADISE}\footnote{\url{https://github.com/brandherd/PyParadise}}. \textsc{PYPARADISE} is a Python version of the stellar population synthesis fitting code \textsc{PARADISE} \citep{walcher_abundance_2015}. The code models the stellar continuum by iteratively performing non-negative linear least-squares fitting of the stellar spectrum of each spectral pixel (``spaxel") to a wide library of stellar population templates and finds the best-fit line-of-sight velocity distribution with a Markov Chain Monte Carlo method. The best-fit stellar continuum spectrum is subtracted from each 

\noindent spaxel to obtain a gas-only data cube.

Using the publicly available penalized pixel-fitting (pPXF) code \citep{cappellari_improving_2017}, we fit the emission line model of the gas-only datacube produced by \textsc{PYPARADISE}. We fit the emission lines with single Gaussians to determine their flux, velocity, and velocity dispersion, and also obtained formal uncertainties from the covariance matrix of the fitted parameters. Although the observed gas emission lines were bright enough to fit at the native spatial resolution, the stellar continuum necessitated spatial binning using Voronoi tessellation \citep{cappellari_adaptive_2003} due to the low signal-to-noise ratio (S/N). The GMOS cube was tessellated to achieve a minimum S/N of 20 (per bin) in the line-free stellar continuum. Figure \ref{fig:gmos_pyp} shows the modeled stellar and gas-phase components. 

 The products from \textsc{PYPARADISE} and \textsc{pPXF} enabled the creation of spatially resolved flux, velocity, and velocity dispersion maps for the following five emission lines:  H$\alpha$, H$\beta$, [\ion{O}{3}] $\lambda 5007$, [\ion{N}{2}] $\lambda 6548$ and [\ion{N}{2}] $\lambda 6583$. To analyze the kinematics of the lines, we adopt a threshold of S/N $\geq 3$. Section \ref{subsec:warm_gas} discusses the maps created. 

\subsection{ALMA}
\label{subsec:almaobs}

ALMA observed the center of SDSS 1531 in Band 6 as part of ALMA Cycle 3 (ALMA Program ID: 2015.1.01426.S; PI: G. Tremblay) across two scheduling blocks between 22-27 April 2016 and 8-13 September 2016. One spectral window was centered at 259.02322 GHz (rest frame 345.8 GHz at $z = 0.335$) to target the $^{12}$CO ($J = 3-2$) molecular line transition, an excellent tracer of cold molecular hydrogen (H$_2$). Three 1875 MHz spectral windows, centered at rest frequencies of 347.1, 327.075, and 324.405 GHz, were also used to detect line-free continuum. The total integration time on source was 105 minutes. 

The BCGs were completely mapped within ALMA's $\sim 28 \arcsec$ ($\sim 130$ kpc) primary beam, but this array configuration is only sensitive to emission on scales up to $\sim 17$\arcsec ($\sim 80$ kpc) (see Figure \ref{fig:hst_alma_fov}). Notably, no line or continuum emission was detected in the extended configuration observations when imaged alone. Although line emission was detected in the compact configuration, no corresponding continuum emission was found.

\begin{figure*}
\includegraphics[width=\linewidth]{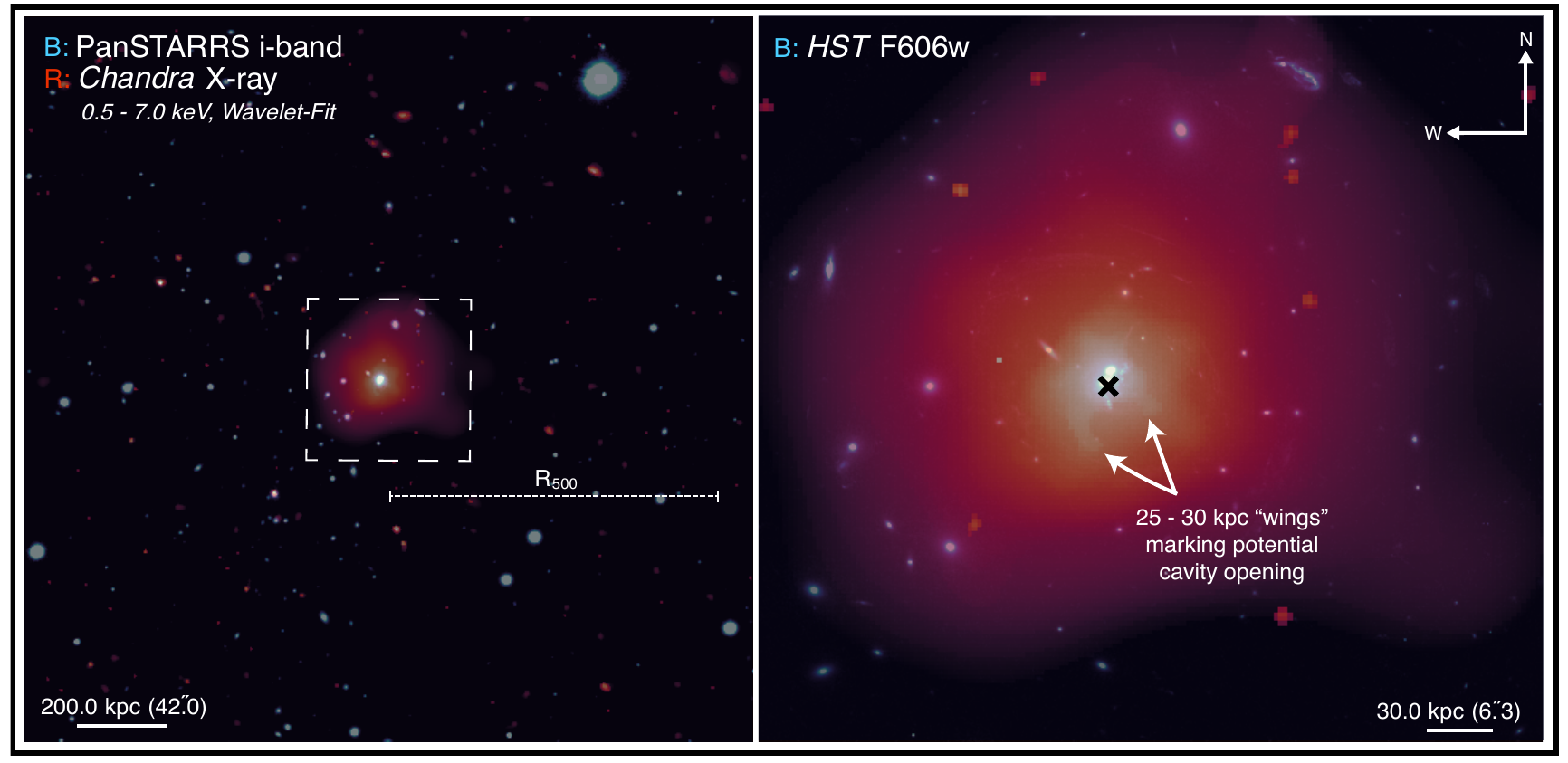}
\caption{\textit{Left:} Pan-STARRs \textit{i}-band image (blue) of SDSS 1531 and its surrounding $\sim$1720 kpc $\times$ 1720 kpc environment. Overlaid in red is a wavelet-fit broadband \textit{Chandra} X-ray surface brightness map in the $0.7-3.0$ keV range, revealing the extended emission from the intracluster gas. The white box outlines the central 50 kpc region shown to the right. \textit{Right:} A closer 100 kpc $\times$ 100 kpc view of the SDSS 1531 cluster in \textit{HST} F606w band (blue), with the same X-ray surface brightness map overlaid. The peak of the X-ray gas (black cross) coincides with the location of the southern BCG. Two 25 kpc (right) and 30 kpc (left) ``wings" below the peak mark a potential cavity opening.}
\label{fig:chandra_flux}
\end{figure*}

The raw ALMA visibilities were processed into calibrated measurement sets using the ALMA automated pipeline reduction script in CASA version 6.2.1.7. The calibrated visibilities were then imaged and deconvolved with the \textsc{CLEAN} algorithm. After testing multiple configurations of weightings (natural and Briggs), binning (10, 20, 40, and 80 km s$^{-1}$ channels), and uv-tapering ($0\farcs6$, $0\farcs8$, $1\farcs0$, and $1\farcs2$), we determined that the natural-weighted 20 km s$^{-1}$ cube without uv-tapering provided the optimal setup, recovering the most emission and maximizing sensitivity. The final data cube presented in this paper achieves an rms sensitivity and angular resolution of 0.22 mJy beam$^{-1}$ per 20 km s$^{-1}$ channel, and a $0\farcs53 \times 0\farcs33$ (2.51 kpc $\times$ 1.59 kpc) synthesized beam at position angle = $-1.15^{\circ}$.  All CO(3-2) fluxes and line widths reported are corrected for the response of the primary beam (\textsc{pbcor=True}). 

We present CO(3-2) integrated intensity, line-of-sight (LOS) velocity, and velocity dispersion maps in Section \ref{subsec:cold_gas}. The moment maps were created from the continuum-subtracted, calibrated spectral cube and constructed using the Python package \textsc{BETTERMOMENTS}\footnote{\url{https://github.com/richteague/bettermoments}}, which applies a quadratic fit to the spectral data \citep{teague_robust_2018}. To recover as much flux as possible while suppressing noise, we spectrally smoothed the data with a top-hat kernel 2 channels wide, applied a Savitzky-Golay filter using a polynomial of order 0, spatially smoothed the data by 2.6 pixels, and applied a sigma clip to all pixels with S/N $< 3.5\sigma$. 

Our observations, presented in Section \ref{subsec:cold_gas}, affirm the presence of a cold star-forming reservoir of gas located just beneath the central elliptical galaxies. 
\subsection{Adopting a Systemic Velocity}
\label{subsec:sys_vel}

The SDSS 1531 BCGs hosts various moving components: the two central galaxies, as well as their stellar and nebular components. \citet{tremblay_30_2014} used archival SDSS spectroscopy centered on the southern BCG's nucleus to pin the stellar redshift to $z = 0.3350 \pm 0.0002$, in agreement with prior redshift measurements for the BCG \citep{hennawi_new_2008} and the cluster \citep{bayliss_geminigmos_2011}. Follow-up spectroscopy from the Nordic Optical Telescope indicated that the velocity of the northern elliptical is blueshifted by 280-300 km s$^{-1}$ with respect to the southern BCG. Though this offset is negligible within the context of the Hubble flow (i.e., the galaxies are certainly interacting), it becomes significant when placed in context with the motions of the cold molecular and warm ionized gas.

To interpret the motions of the stellar components of the BCGs with respect to the motions of the gas, we must select a systemic velocity to be used as a `zero point,' marking the transition from blue- to redshift. In this paper, we adopt a systemic velocity of $cz = 100,430 \text{ km s}^{-1}$, where $c$ is the speed of light and $z = 0.3350 \pm 0.0002$ is the redshift of the southern nucleus. We select the redshift of the southern nucleus instead of that of the northern nucleus because its value is aligned with prior redshift measurements, detailed above. All velocity maps presented in this paper are projected at this value.

Instead of using the stellar redshift, other studies of BCGs in cool cores have set the zero point relative to where the observed molecular CO emission peaks \citep[e.g.,][]{tremblay_cold_2016, vantyghem_enormous_2019}. However, we do not adopt this approach because the bulk of the molecular gas detected lies to the east of the BCGs, rather than concentrated within their central few kiloparsecs \citep[e.g.,][]{vantyghem_massive_2021}. Regardless of the reference point chosen, the data reveals a significant velocity offset between the young stellar superclusters and the gas.

\section{Results}
\label{sec:results}

In this section, we present \textit{Chandra} X-ray observations of the cooling hot ($\sim 10^6$ K) intracluster medium, followed by LOFAR and VLA radio observations tracing emission from AGN and dynamic cluster-scale activity. Next, we present GMOS-IFU optical and ALMA millimeter observations of the warm ($\sim 10^4$ K) ionized and cold ($\sim 10^2$ K) molecular gas near the BCGs. For a detailed discussion of the results, please refer to Section \ref{sec:beads_origin}.

\subsection{Unveiling the Cool Cluster Core}
\label{subsec:hot_gas}

\subsubsection{Large-scale Structure}
\label{subsubsec:chandra_sb} 

In Figure \ref{fig:chandra_flux}, we present a Pan-STARRS \textit{i}-band wide-field view of SDSS 1531 (left) and a \textit{HST} view of the cluster's central region (right) with the background-subtracted, exposure-corrected, wavelet-fit \textit{Chandra} X-ray surface brightness map of the ICM overlaid. The X-ray emission is centered on the BCGs, and has an angular inclination similar to that of the BCGs in the optical. The emission has a predominantly smooth and circularly symmetric surface brightness distribution, except for two thin, $20-30$ kpc "wings" to the southeast and southwest, which create a concave feature. 

\begin{figure}
\centering \includegraphics[width=\linewidth]{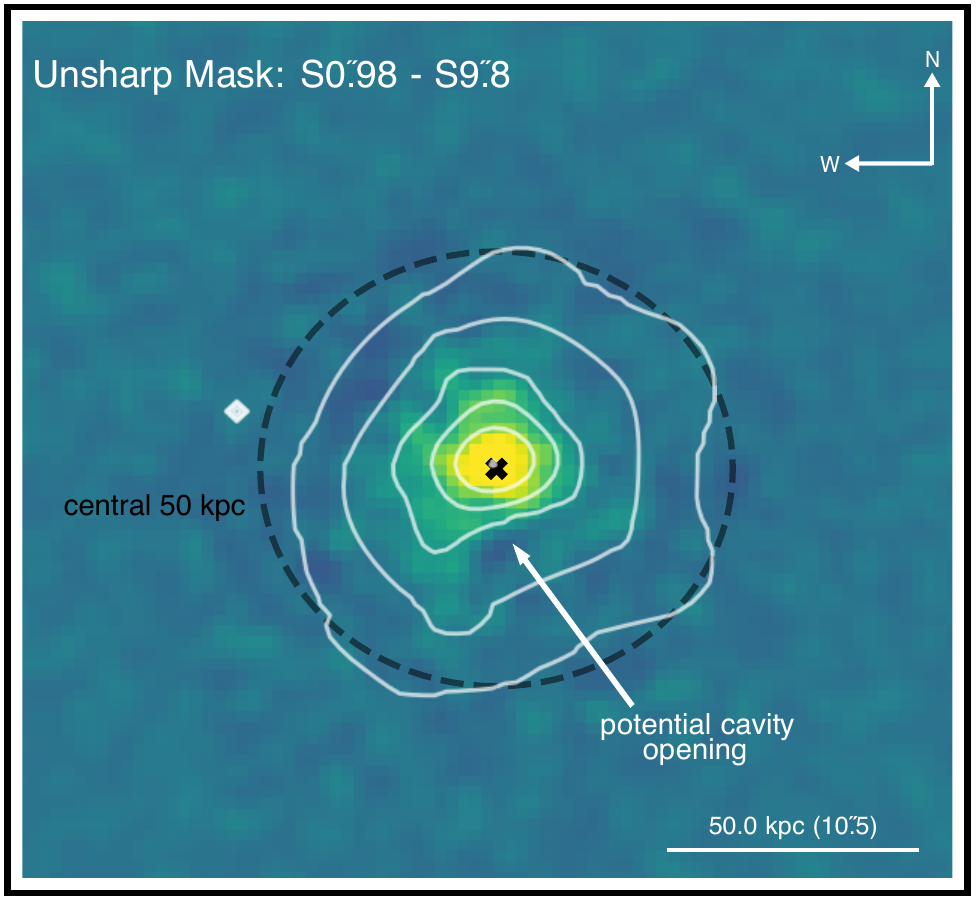}
\caption{Unsharp-masked 0.5–7.0 keV image of the central region of SDSS 1531 with white contours of the surface brightness map from Figure \ref{fig:chandra_flux} overlaid in white contours. The opening to a potential massive cavity is located between the X-ray ``wings."}
\label{fig:chandra_unsharp}
\end{figure}

We searched for potential cavities in the unsharp-masked image shown in Figure \ref{fig:chandra_unsharp} (see Section \ref{subsec:chandraobs} for full details on the image creation). While helpful for identifying hidden structures, unsharp-masked images can also give the illusion of false cavities. Therefore, we only consider unsharp-masked image fluctuations consistent with those in the surface brightness map. The most plausible cavity-candidate lies in the $\sim 16$ kpc radius, concave surface brightness edge created by the two X-ray "wings." However, the radio data presented in Section \ref{subsec:radio} suggests that this feature may instead represent an opening to a large ($\sim 50$ kpc radius) avocado-shaped cavity that the current X-ray data is too shallow to resolve.

\subsubsection{Discontinuity Near the Putative Cavity Edge}
\label{subsubsec:cold_front}

\begin{figure*}
\centering \includegraphics[width=\linewidth]{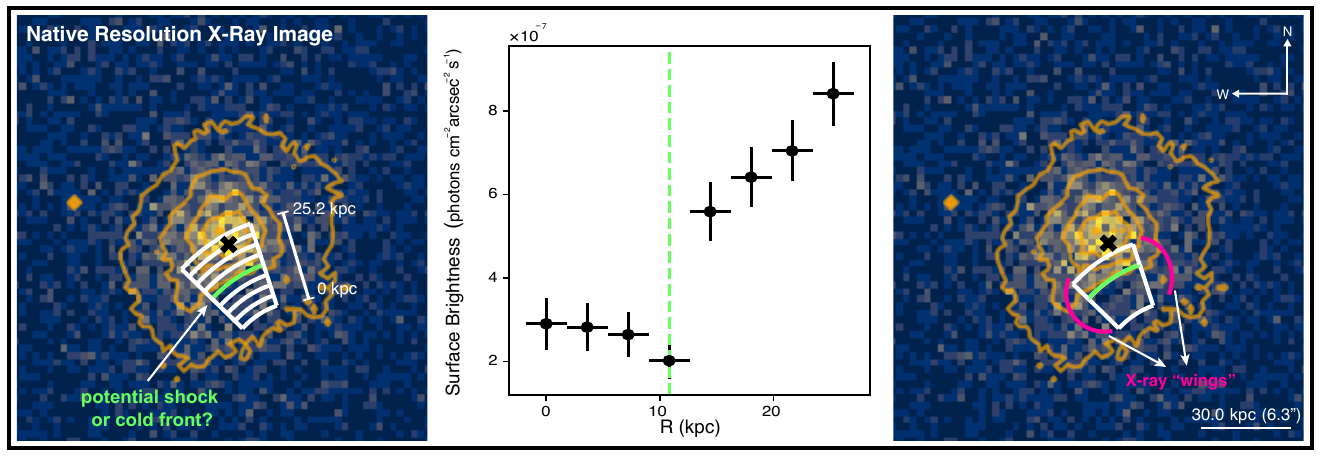}
\caption{Unfiltered 0.5–7 keV \textit{Chandra} X-ray image of SDSS 1531 shown in Figure \ref{fig:chandra_unsharp} with the regions used to extract the surface brightness profile (\textit{left}) and spectra (\textit{right}) overlaid. The dashed green line marks the location of the tentative surface brightness edge. The center plot shows the resulting surface brightness profile and depicts a clear discontinuity in surface brightness at the site of the potential shock/cold front. Due to a limited number of counts, we can only conduct spectral fitting within the two large pie regions shown.}
\label{fig:chandra_cf}
\end{figure*}

To investigate the region surrounding the ``wings" further, we extracted a surface brightness profile from the $0.5 - 7.0$ keV \textit{Chandra} image in the pie regions shown in Figure \ref{fig:chandra_cf}, left. Figure \ref{fig:chandra_cf}, center shows the resulting surface brightness profile. The clear density jump (green dashed line) at $\sim 12.6$ kpc confirms the presence of a surface brightness edge. 

To test whether the density jump represents a shock or a cold front, we extracted spectra of the ICM inside and outside the discontinuity in the regions shown in Figure \ref{fig:chandra_cf}, right.  To obtain projected thermodynamic properties, we fit the 0.5−7 keV band spectra with a \textsc{phabs*apec} model. We measured a clear density jump of $n_{e,\text{in}}/n_{e,\text{out}} = 1.40 \pm 0.05$. We tentatively find continuous pressure at the surface brightness edge ($p_{\text{out}}/p_{\text{in}} = 0.94 \pm 1.22$), and that the ICM temperature close to the cluster center is lower than it is outside the surface brightness edge ($kT_{\text{in}}/kT_{\text{out}} = 0.67 \pm 0.87$). These characteristics are consistent with the properties expected of a cold front. However, it is important to note that these results are not conclusive since there are only $\sim200$ net counts across the selected regions, resulting in significant errors. Therefore, it is still possible that the observed discontinuity could be better explained by a shock rather than a cold front.

\setlength{\tabcolsep}{30pt}
\begin{deluxetable}{lc}
\tabletypesize{\footnotesize}
\tablewidth{\linewidth}
\tablecaption{\textsc{X-ray Properties}\label{tab:chandra_profiles}} 
\tablehead{
\colhead{Property (units)} &
\colhead{Value}
}
\startdata
$M_{500} - Y_X$ ($10^{14} M_\odot$) & $2.4 \pm 0.13$ \\
$R_{500}$ (kpc) & $840 \pm 15$ \\
$kT_0$ (keV) & $2.78 \pm 0.04$ \\
$K_0$ (keV cm$^{2}$) & $18.1 \pm 0.26$ \\
$P_0$ ($10^{-10}$ dyn cm$^{-2}$) & $5.36 \pm 0.076$ \\
$t_{\text{cool, 0}}$ (Gyr) & $0.5 \pm 0.015$ 
\enddata
\tablecomments{Summary of the key X-ray properties of SDSS 1531. $M_{500}$ and $R_{500}$ are calculated assuming hydrostatic equilibrium and the $Y_X-M$ relation from \cite{vikhlinin_chandra_2009}. Central quantities ($kT_0$, $P_0$, $K_0$, $t_{\text{cool, 0}}$) are measured at a radius of 11.9 kpc and extracted from the linearly spaced annuli. }
\vspace*{-13mm}
\end{deluxetable}

\subsubsection{Spectral Maps \& Profiles of the ICM}
\label{subsubsec:chandra_spectral_maps_profiles} 

\begin{figure*}
\centering \includegraphics[width=\linewidth]{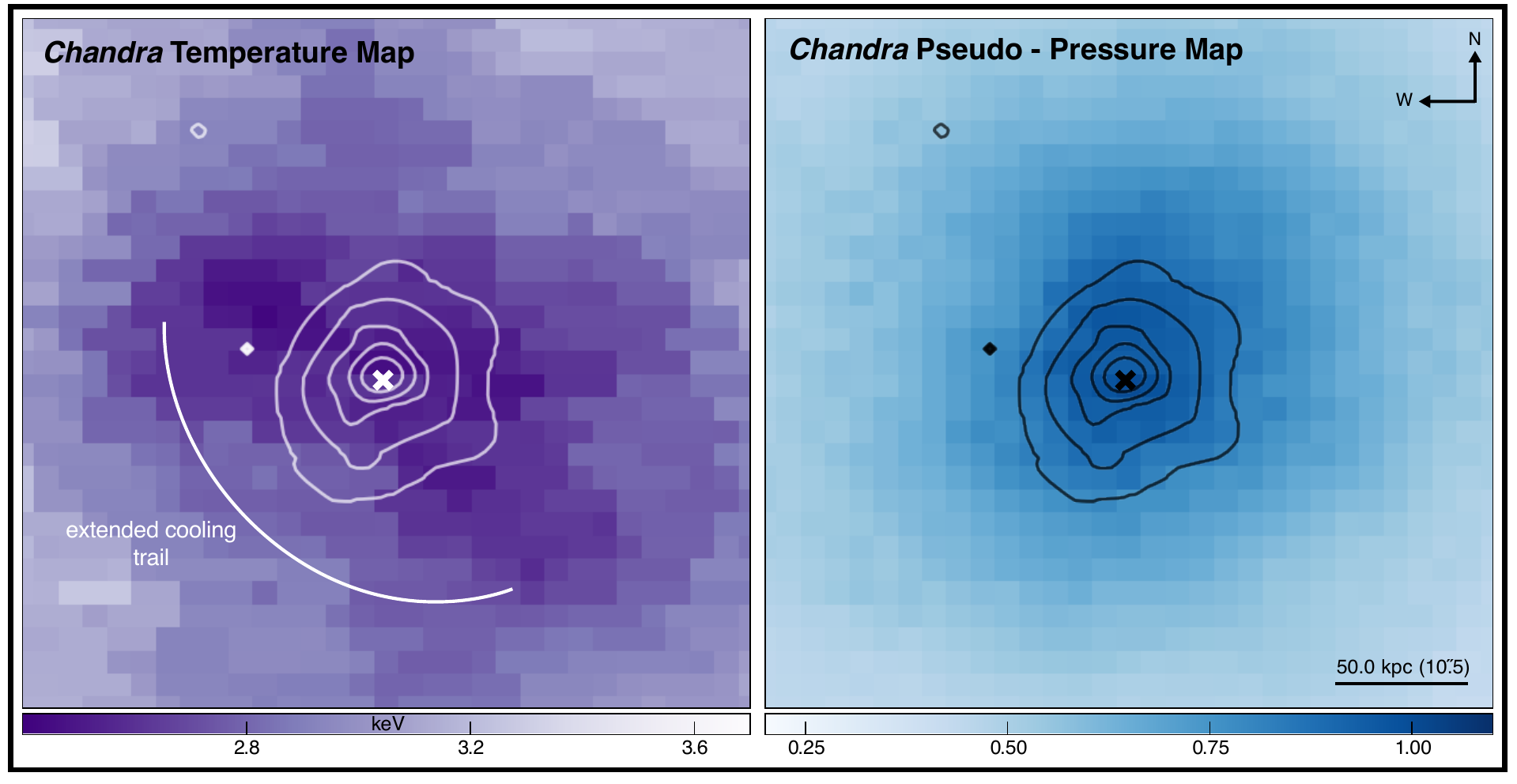}
\caption{ Wide-field temperature (\textit{left}) and pressure (\textit{right}) map of SDSS 1531. Both maps are overlaid with contours of the X-ray surface brightness map shown in Figure \ref{fig:chandra_flux}. Uncertainties range from $7 - 10$\%. The temperature map reveals an extended cooling trail. The distribution of the pressure and entropy (not shown) are roughly circularly-symmetric, with the highest pressure and lowest entropy near the X-ray peak.}
\label{fig:chandra_spectral_maps}
\end{figure*}

We created high-resolution thermodynamic spectral maps of the ICM using \textsc{clusterpyxt} (see Section \ref{subsec:chandraobs} for details). The temperature map (Figure \ref{fig:chandra_spectral_maps}, left) reveals the presence of a central, extended ($\sim 200$ kpc), cool ($\sim 2-3$ keV) butterfly-shaped structure, with its major axis perpendicular to that of the central surface brightness map. The coolest gas is found in a curved trail extending from the Northeast to the Southwest. Outside the cool region, temperatures gradually increase to $\sim 4$ keV, with $\sim 8\%$ uncertainties. The pseudo-pressure (see Figure \ref{fig:chandra_spectral_maps}, right) and entropy maps (not shown) exhibit a uniform circular distribution within the central $\sim 100$ kpc, with the lowest entropy and highest pressure near the center of the cluster.

\begin{figure*}
\centering \includegraphics[width=\linewidth]{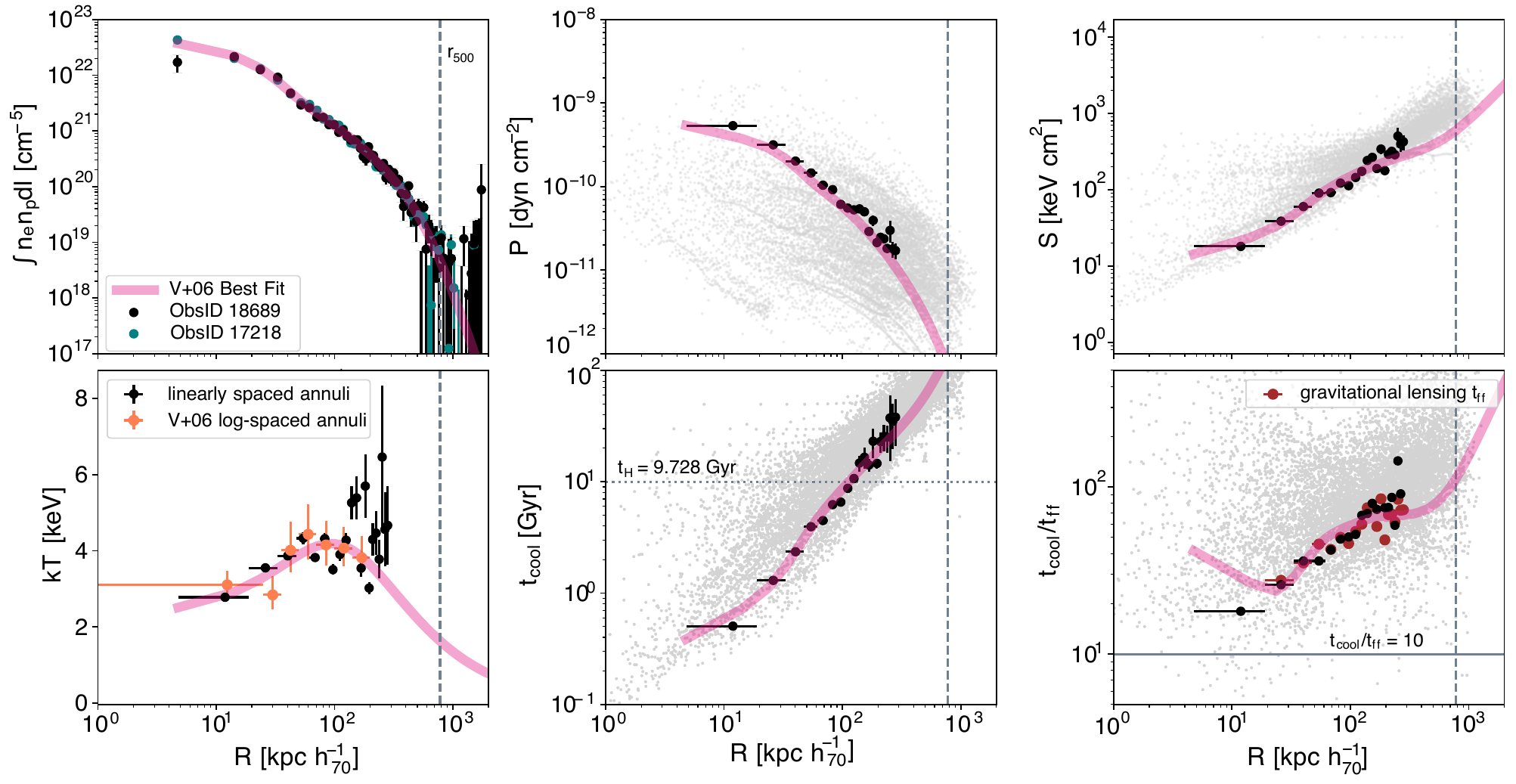}
\caption{\textit{Top left}: Projected X-ray emissivity per unit area for both \textit{Chandra} ObsIDs (black and green points), fit with a modified beta model (thick pink line). \textit{Bottom left}: Projected X-ray temperature profile using linearly-spaced annuli (black points) vs log-spaced annuli (orange points), which are used to fit the temperature model described in (\cite{vikhlinin_chandra_2006}; V+06). \textit{ Middle column}: Pressure (\textit{top}) and entropy (\textit{bottom}) profiles. \textit{Right column}: Cooling time (\textit{top}) and $t_{\text{cool}}/t_{ff}$ (\textit{bottom}) profiles. Brown points represent the free-fall time estimate using the strong-lensing mass profile from \cite{sharon_mass_2014}. The upturn of the V+06 $t_{\text{cool}}/t_{ff}$ profile at low radii is a numerical artifact resulting from preventing the derivative from approaching zero. In the middle and right columns, gray points represent the thermodynamic profiles of nearby clusters from the ACCEPT sample \citep{cavagnolo_intracluster_2009}. All profiles consistently indicate the presence of a strong cool core in SDSS 1531.}
\label{fig:chandra_tepe_prof}
\end{figure*}

We also created thermodynamic profiles of the larger cluster atmosphere, shown in Figure \ref{fig:chandra_tepe_prof}. In Table \ref{tab:chandra_profiles}, we list the derived X-ray properties for SDSS 1531, namely the total cluster mass $M_\Delta$, radius $R_\Delta$, central temperature $kT_0$, central entropy $K_0$, central pressure $P_0$, and central cooling time $t_{cool,0}$. Each property is consistent with that expected of a strong cool core.

In Figure \ref{fig:chandra_tepe_prof} (top left), we plot the projected emission measure profile for ObsID 17218 (green) and ObsID 18689 (black). The emission measure profile is fit with the modified $\beta$-model described by Equation \ref{eqn:gas_density}. Both profiles show excellent agreement between the independent observations and exhibit a central overdensity, suggesting the presence of a cool core. At low redshift, cool core clusters have characteristic central cusps in the X-ray surface brightness distribution. \citet{vikhlinin_lack_2007} characterizes the central cusp as the power-law index of the gas density profile $\alpha = -2 d \log \rho_g / d \log r$ at $r = 0.04 R_{500}$, with clusters known to host strong cool cores having $\alpha > 0.7$. From the fit to the emission measure, we obtain $\alpha = 1.4$, satisfying the cuspiness criteria for the presence of a strong cool core. 

In Figure \ref{fig:chandra_tepe_prof} (bottom left), we plot the projected temperature profile, fit with the analytical model given by Equation \ref{eqn:tprofile}. The density peak in the emission measure profile coincides with a significant temperature decline inside the central 100 kpc, from $T = 4.6 \pm 0.22$ keV at $r \sim 110$ kpc to  $T = 2.4 \pm 0.03 $ keV at $r \sim 10$ kpc.  We also find a three-dimensional temperature drop $T_{\text{min}}/T_0 \sim 0.25$. For comparison, strong cooling flow clusters like Abell 478 and Abell 133 have $T_{\text{min}}/T_0$ ranging from $0.1-0.4$ \citep{vikhlinin_lack_2007}. 

We compared the derived pressure and entropy profiles (see Figure \ref{fig:chandra_tepe_prof}, top center and top right), with those from the Archive of Chandra Cluster Entropy Profile Tables (ACCEPT; \citealt{cavagnolo_intracluster_2009}, C09) for 239 galaxy clusters. The profiles are consistent with those of similar cool core clusters. 

\vspace{-3mm}

\subsubsection{Cooling and Free-fall Timescales }
\label{subsubsec:chandra_timescales}

According to the cooling time profile (Figure \ref{fig:chandra_tepe_prof}, bottom center), the ICM cools within a Hubble time ($t_H$) out to a radius of $\sim 100$ kpc, with the innermost $\sim 30$ kpc cooling within $< 1$ Gyr. Another relevant timescale for cool core clusters is the free-fall time $t_{\mathrm{ff}}$, which describes the characteristic time for a system to collapse. When $t_{\mathrm{cool}} > t_{\mathrm{ff}}$, the system has sufficient time to re-establish hydrostatic equilibrium. However, $t_{\mathrm{cool}} < t_{\mathrm{ff}}$ signals catastrophic cooling, for the gas is unable to rebound fast enough to account for the rapid loss in pressure \citep{nulsen_thermal_1986, mccourt_thermal_2012}. Numerical simulations of thermal instabilities in cluster atmospheres have found the $t_{\mathrm{cool}}/t_{\mathrm{ff}}$ ratio is less stringent, with instability occurring when $t_{\mathrm{cool}}/t_{\mathrm{ff}} \lesssim 10$ \citep{gaspari_cause_2012, voit_cooling_2015}. 

The free-fall time is estimated as:

\begin{align}
t_{\text{ff}} (r)= \sqrt{\frac{2 r^3}{GM(r)}} 
\label{eqn:tff}
\end{align}

\begin{figure*}
\includegraphics[width=\linewidth]{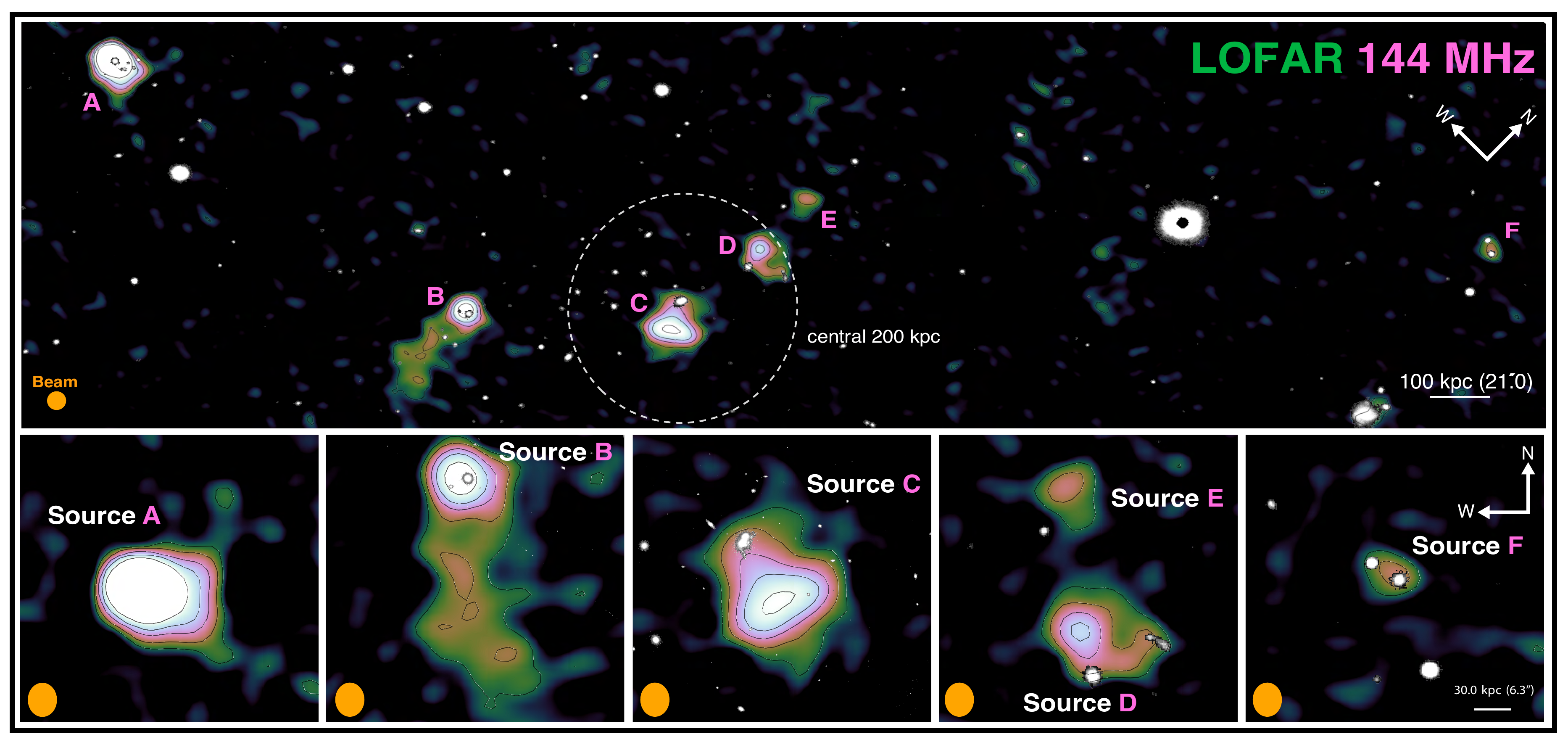}
\caption{\textit{Top:} LOFAR 144 MHz image of SDSS 1531 with $4\arcsec \times 6\arcsec$ beam (orange ellipse) and noise $\sigma_{\rm rms} = 0.15$ mJy/beam. The image unveils 6 prominent sources of diffuse, low-frequency radio emission within the cluster, each labeled with a pink letter. A white dashed circle outlines the central 200 kpc of the cluster. \textit{Bottom:} A closer view of each identified source. The redshift of each potentially associated galaxy is labeled. Source C is most closely aligned with the BCGs. In both top and bottom panels, the contour levels start at $3\sigma_{\rm rms}$ and increase sequentially with steps of 6$\sigma$, 12$\sigma$, 20$\sigma$, and 37$\sigma$.}
\label{fig:lofar_mosaic}
\end{figure*}

\noindent where $r$ is the distance from the center of the cluster, and $M(r)$ the radial cluster mass profile. Since the free-fall time is most informative near the center of the cluster, we consider two mass profiles: one derived from strong-lensing analysis \citep{sharon_mass_2014} and the other from hydrostatic X-ray measurements (see Section \ref{subsec:chandraobs}). The strong-lensing analysis yields a cylindrical mass of $M(<R_E) = 2.32 \pm 0.01 \times 10^{13} M_\odot$ within the Einstein radius $R_E \sim 52.1 \pm 0.5$ kpc. On the other hand, the hydrostatic X-ray mass measurement, obtained at $R=52 \pm 7.1$ kpc, gives $M_{HE}(\lessapprox R_E) = 9.3 \times 10^{12} M_\odot$. To compare the X-ray mass with the strong-lensing cylindrical mass, we recalculate the hydrostatic X-ray spherical mass within a cylindrical volume of height 10 Mpc as done in \cite{sharon_multi-wavelength_2015}. This yields a cylindrical mass of $\sim 1.3 \times 10^{13} M_\odot$. Though small in magnitude on the scale of the total mass, the factor of 2 difference between the X-ray and strong-lensing mass estimates, coupled with the disturbed morphology of the surface brightness map, indicates that the hydrostatic equilibrium assumption does not accurately describe the cluster's core, as expected given the ongoing major merger. 

Figure \ref{fig:chandra_tepe_prof} (bottom right) illustrates the enclosed $t_{\mathrm{cool}}/t_{\mathrm{ff}}$ profiles estimated from the hydrostatic mass (thick pink line and black points) and the strong-lensing mass re-projected to a spherical mass (brown points). Both estimates lead to a minimum $t_{cool}/t_{ff} \sim 20$ for $r\leq 10$ kpc. Although the theoretically expected $t_{cool}/t_{ff}$ ratio often falls near or below unity \citep{gaspari_cause_2012, mccourt_thermal_2012, sharma_structure_2012}, the observed behavior in cool core clusters typically deviates from this threshold, with most cool core clusters harboring cold filaments when $t_{cool}/t_{ff} \leq 20$ \citep{olivares_ubiquitous_2019}. Other works have suggested that thermally unstable cooling can occur when this ratio lies above unity, in the range of $10-40$ \citep{gaspari_cause_2012, mccourt_thermal_2012, sharma_structure_2012}.

\subsection{A Radio View of the Cluster}
\label{subsec:radio}

\subsubsection{Unclassified Extended Radio Emission}
The LOFAR 144 MHz image, shown in Figure \ref{fig:lofar_mosaic}, reveals several sources of diffuse radio emission throughout the cluster, few of which are marginally visible in the VLA images. From this image, we have identified and labeled six distinct radio sources affiliated with the cluster.

Of the six radio sources, Source C is most closely aligned with the BCGs. Its total angular size is 28\arcsec, corresponding to a physical size of 133 kpc. The emission takes on an avocado-esque shape and increases in intensity as it extends southeastward from the BCGs. The emission peaks $\sim 60$ kpc away from the southern BCG, forming a distinct ``hotspot." Notably, the source is undetected in all available VLA images.

Source D is a hybrid blob/extended source with an angular size of 22\arcsec, corresponding to a physical length of $\sim 105$ kpc. The $\geq 6\sigma$ emission resembles the avocado-like shape of Source C, with a similar hotspot near the edge of the source. The $>3\sigma$ emission also allows for an interpretation where Source C resembles a miniature head-tail source whose tail curves around the ``head," where the emission peaks. The source has no confirmed counterpart but is close to a galaxy at $z=0.3357$. Directly north of Source D is Source E, the faintest source detected. Source E also does not have a direct optical counterpart, but its proximity to Source C and the affiliated galaxy suggests it could belong to the cluster. The source has a similar ``avocado"-shaped morphology and a very small hotspot detected at $6\sigma$.

\begin{figure*}
\centering \includegraphics[width=\linewidth]{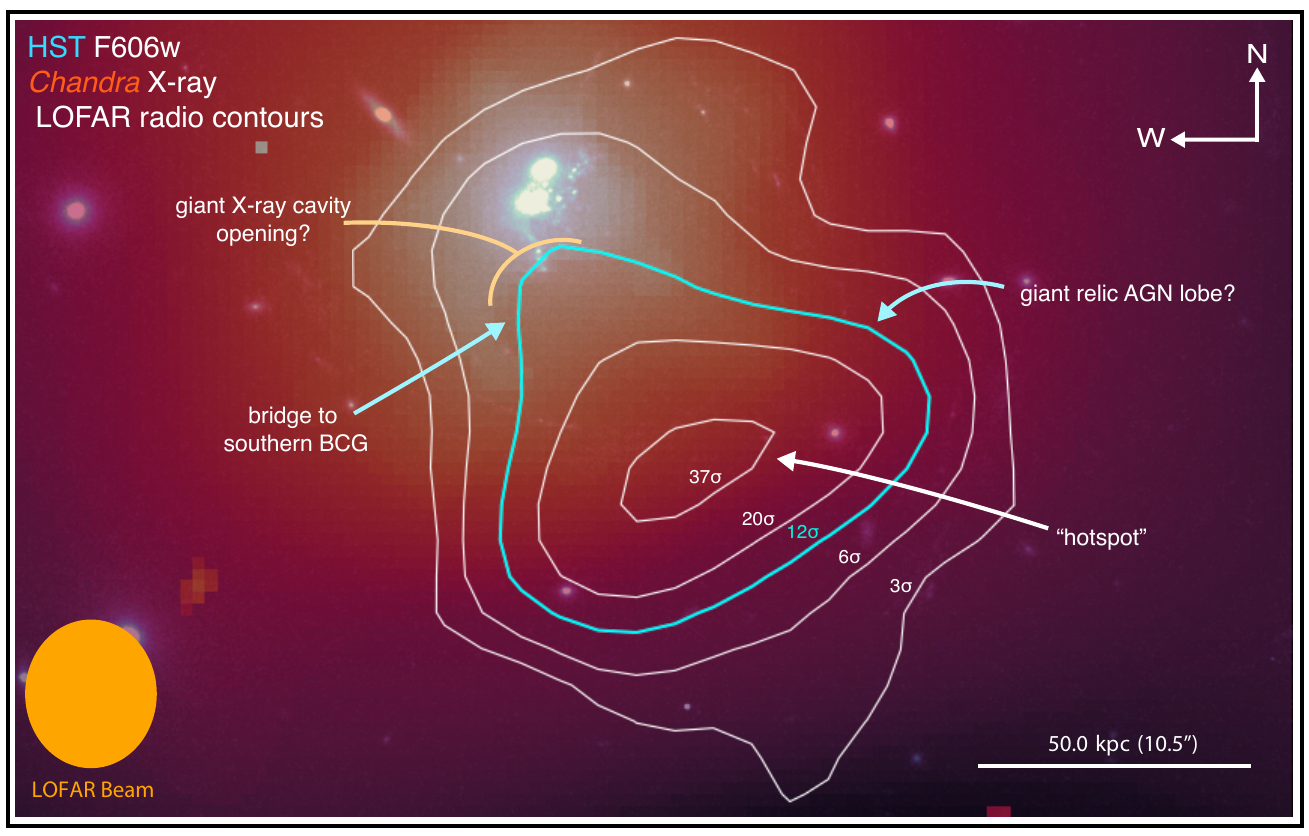}
\caption{LOFAR contours of Source C overlaid on the \textit{Chandra} surface brightness map shown in Figure \ref{fig:chandra_flux}. The $12\sigma$ contour (cyan) aligns remarkably well with the X-ray ``wings," suggesting that the wings represent an opening to a giant X-ray cavity, filled by the relic AGN lobe. }
\label{fig:chandra_lofar}
\end{figure*}

Source B, the largest source within the cluster, has a distinct head-tail morphology. The ``head" of the source comprises a concentrated emission region, with two objects in the center, one of which has $z=0.341$. The tail extends outward up to 45\arcsec, corresponding to a physical length of $\sim 215$ kpc. Only the head of the source is also detected in the VLA B-array image. 

Source A is the most luminous radio source detected and located very close to the cluster's virial radius ($r_{200}$). The emission has an oval morphology and peaks in the center. The source is brightly detected in all VLA images. Several objects appear associated with the emission, one of which has a redshift of $z=0.3376$. Conversely, Source F is the faintest source associated with the cluster and is located opposite Source A near the virial radius.  The round, compact source is juxtaposed with two foreground objects, one of which is located at $z=0.3298$.

\subsubsection{Similarities between the X-ray and Radio Morphology in the Cluster Core}
\label{subsubsec:xray_lofar}

Figure \ref{fig:chandra_lofar} displays $\geq 3\sigma$ LOFAR radio contours of Source C superimposed on the \textit{Chandra} X-ray surface brightness image of the cluster's central field. The most concentrated radio emission, detected at $\geq 12\sigma$ and highlighted in blue, aligns remarkably well with the region delineated by the X-ray ``wings" and is connected by a bridge of radio emission to the southern BCG. The spatial coincidence is reminiscent of radio AGN lobes that push aside the X-ray gas to create cavities, as observed in several other cool core clusters with active AGN, such as Perseus \citep{fabian_chandra_2000} and Abell 2597 \citep{mcnamara_discovery_2001}. If the concave region is indeed a cavity, the \textit{Chandra} observations are likely too shallow to resolve the full extent uncovered by the high-resolution LOFAR image. However, the fact that the radio emission is undetected by the VLA suggests that we may be observing an aged AGN lobe, which we discuss further in Section \ref{subsubsec:old_agn}.

None of the remaining sources identified in the cluster have readily identifiable X-ray counterparts, as they are located in the less dense portions of the ICM.

\subsubsection{Preliminary Source Classification from Integrated Radio Spectra}

\begin{figure*}
\begin{center}
\includegraphics[width=1.0\linewidth]{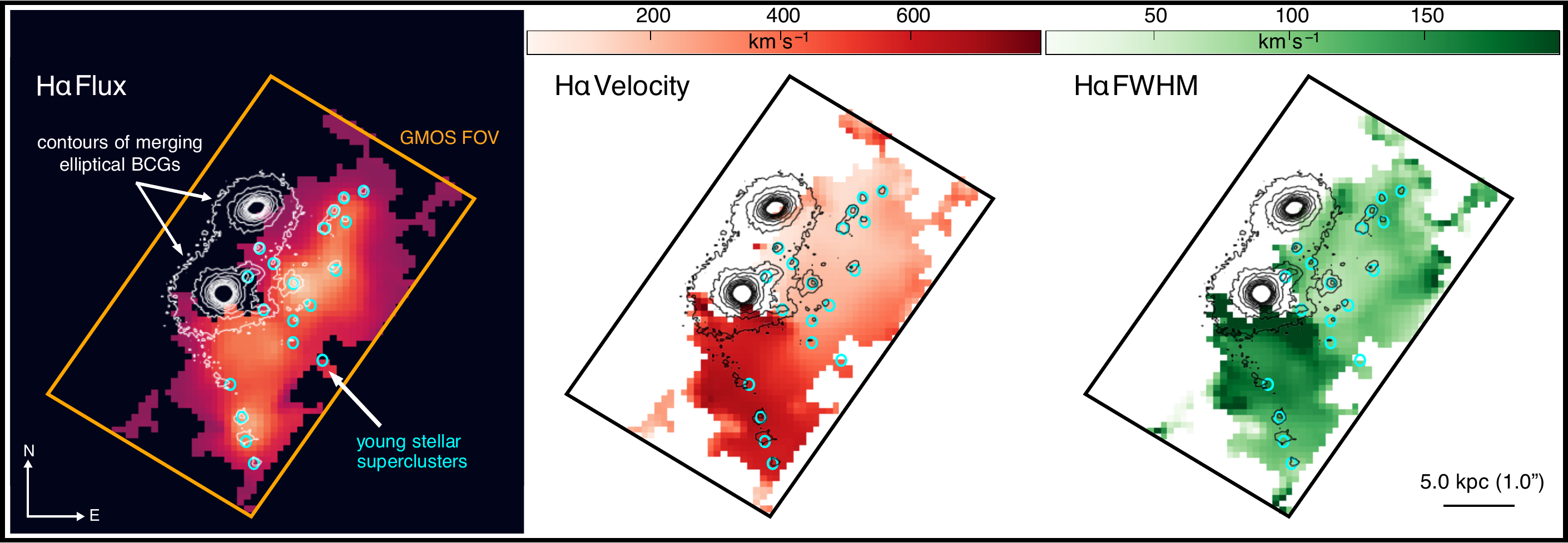}
\caption{\textit{Left to Right:} GMOS maps of H$\alpha$ flux, line-of-sight velocity, and velocity dispersion, extracted after subtracting the stellar continuum as described in Section \ref{subsec:gmosobs}. White (left) and grey (center, right) contours depict the \textit{HST} F606w view of the BCGs. Blue circles indicate the spatial location of the 19 young stellar superclusters (YSCs). The H$\alpha$-emitting gas fully engulfs the YSCs and may extend beyond the eastern edge of the GMOS FOV. The velocities shown are projected around a zero point at $z=0.335$ (e.g., $cz = 100, 430$ km s$^{-1}$), consistent with the velocity calibration used for the ALMA maps in Section \ref{subsubsec:cold_vel}. The gas is redshifted up to $+800$ km s$^{-1}$ with respect to the southern BCG and is most disturbed near the southern nucleus. We directly compare these kinematic maps with the ALMA data in Section \ref{subsubsec:gmos_alma}.}
\label{fig:gmos_ha}
\end{center}
\end{figure*}

Various models have been used to describe the origin of diffuse radio emission in clusters, a list of which is discussed in \citet{kempner_conference_2004}. To make preliminary classifications of the radio sources in the cluster, we estimate the spectral index, $\alpha$, from 144 MHz - 1.4 GHz for each source using the equation:
\begin{align}
    \alpha_{\rm 144 MHz}^{\rm 1.4 GHz} = \frac{\log(S_1) - \log(S_2)}{\log(\nu_1) - \log(\nu_2)}
\end{align}
where $S_1$ is the flux density of the source at $\nu_1 = 1.4$ GHz, and $S_2$ is the flux density of the source at $\nu_2 = $144 MHz. 

\setlength{\tabcolsep}{3pt}
\begin{deluxetable}{lccc}
\tabletypesize{\footnotesize}
\tablewidth{\linewidth}
\tablecaption{\textsc{Integrated Spectral Indices}\label{tab:radio_alpha}} 
\tablehead{
\colhead{Source} &
\colhead{$\nu$} &
\colhead{Flux density} &
\colhead{$\alpha_{144 \rm{MHz}}^{1.4 \rm{GHz}}$ }\\
\colhead{} &
\colhead{(MHz)} &
\colhead{mJy} &
\colhead{} }
\startdata
A - Bright round source & 144 &  73.1 & $-0.15 \pm 0.005$ \\
& 1400 & 52.5 & \\
\hline
B - Head-tail source & 144 & 13.1 &$-0.19 \pm 0.03$  \\
& 1400 & 8.6 & \\
\hline
C- Central avocado-shaped source & 144 & 26.0 & $-1.7 \pm 0.4$ \\
& 1400 & $< 0.5$ & \\
\hline
D - Off-center amorphous source & 144 & 9.9 & $-1.3 \pm 0.4$\\
& 1400 & $< 0.5$ & \\
\hline
E - Faint amorphous source & 144 & 2.4 &$ -0.7 \pm 0.4$ \\
& 1400 & $< 0.5$ & \\
\hline
F - Faint round source & 144 & 1.9 & $-0.6 \pm 0.4$ \\
& 1400 & $< 0.5$ & \\
\enddata
\tablecomments{Integrated spectra for each radio source identified in Figure \ref{fig:lofar_mosaic}. The integrated spectral index is calculated from 144 MHz to 1.4 GHz. For sources undetected with the VLA, the $3\sigma = 0.5$ mJy upper limit is used to estimate the spectral index.}
\vspace{-13mm}
\end{deluxetable}

To obtain integrated flux densities in the LOFAR and VLA B-array images, rectangular regions were defined around each source. For sources undetected in the VLA image, we assume that the flux densities of the undetected sources were no more than $3\sigma_{\rm rms}$ at 1.4 GHz ($\sigma_{\rm rms} =0.135$ mJy beam$^{-1}$), setting an upper limit of $0.052$ mJy beam$^{-1}$. The rms noise at 144 MHz and 1.4 GHz were used to estimate the spectral index error.  Table \ref{tab:radio_alpha} lists all integrated flux densities and derived spectral indices for each source.

Sources A and B have relatively flat spectra $\alpha < -0.20$. Given Source B's tailed morphology, optical counterpart at $z=0.336$,  and flat spectral index, the source is likely a radio jellyfish \citep[e.g.,][]{roberts_lotss_2021}. Source A may be an early-stage radio jellyfish.

The integrated spectral index estimated for Source C  is $-1.7 \pm 0.4$. Given the diffuse emission's projected location in the cluster's central region, steep spectral index, striking overlap with the concave X-ray surface brightness edge, and emission bridge connecting the southern BCG to the emission peak, the bulk of the emission in Source C likely stems from an AGN relic. Source D also has a steep spectrum, with $\alpha \sim -1.3 \pm 0.4$. If Source D is not a separate radio source, it could be Source C's missing relic lobe counterpart, a scenario we explore further in Section \ref{subsubsec:old_agn}. 

Sources E and F have moderately steep spectra. Their small angular sizes make them difficult to classify, so we refrain from doing so. 

Higher resolution observations at 1.4 GHz and other frequencies are required for precise spectral analysis, accurate classification, and detailed study of each newly identified radio source. 
\vspace{-1mm}

\subsection{The Warm Ionized Gas}
\label{subsec:warm_gas}

\subsubsection{Emission Line Maps}
\label{subsubsec:gmos_gas}

We present the H$\alpha$ flux, velocity, and velocity dispersion maps of the warm ($< 10^4$ K) ionized gas derived from the continuum-subtracted GMOS gas cube in Figure \ref{fig:gmos_ha}. The ionized gas surrounds the young stellar superclusters (YSCs), with the brightest knots of emission coincident with the young stars. The gas is marginally detected throughout the nuclei of the BCGs. The emission likely extends beyond the GMOS FOV (depicted as a rectangular box), but the emission near the edges lies within the noise of the data. It is unclear how much further out the emission extends and whether it envelops the full extent of the molecular gas, which we discuss later on in Section \ref{subsubsec:gmos_alma}.

The velocity map of the gas reveals that the ionized gas is significantly redshifted with respect to the stellar systemic velocity of the southern nucleus. The gas enshrouding the southern string of beads is redshifted up to $+800$ km s$^{-1}$ with respect to the systemic velocity, while the gas tracing the northern strings closely approaches the systemic velocity of the southern nucleus. The FWHM map shows that the gas is most disturbed near the southern nucleus. 

\subsubsection{Rate of Star Formation}
\label{subsubsec:gmos_sfr}

From the GMOS data, we measure a total H$\alpha$ flux of $1.9 \pm 0.03 \times 10^{-16} \text{ erg s}^{-1} \text{ cm}^{-2}$ across all spaxels. Although each ``bead" of star formation is likely embedded within dust supplied from the aging stars of the BCGs \citep{donahue_baryon_2022}, we are unable to reliably estimate the internal extinction for each spaxel via the Balmer decrement ($H\alpha/H\beta$). Although both $H\alpha$ and $H\beta$ emission lines are clearly detected, the Balmer decrement yields a median value of $\sim 2.2$, an unphysical value given the intrinsic line ratio of 2.86 appropriate for low-density gas with a temperature of $10^4$ K under the standard "case B" recombination scenario \citep{osterbrock_astrophysics_2006}. Such a flat H$\alpha$/H$\beta$ ratio is expected for environments where there is no dust attenuation or $n_e > 10^9$ cm$^{−3}$ \citep{adams_effect_1974}, neither of which are commonly expected in cool core clusters. The unphysical value may instead be due to difficulty in accurately subtracting the relatively weak stellar continuum, as described in Section \ref{subsec:gmosobs}. Without correcting for extinction, we use Equation (2) in \citep{kennicutt_global_1998} to obtain a total SFR of $0.60 \pm 0.009$ M$_\odot$ yr$^{-1}$, which we interpret as a lower limit. 

\citet{tremblay_30_2014} estimated an extinction-corrected SFR in the range of 5-10 M$_\odot$ yr$^{-1}$. They measured an SFR of $\sim 5 \pm 2 M_\odot$ yr$^{-1}$ from correcting the ALFOSC H$\alpha$ luminosity for extinction using the Balmer Decrement from SDSS DR10's reported H$\alpha$ and H$\beta$ fluxes\footnote{SDSS DR10 reported fluxes derived from the Princeton specBS pipeline, accessed at \url{http://das.sdss.org/SpecBS} \citep{adelman-mccarthy_sixth_2008}. An unphysical ($H\alpha/H\beta < 2.3$) Balmer decrement is obtained when using SDSS DR18 fluxes from the Portsmouth catalog, which is based on stellar kinematics as evaluated by \textsc{ppXF} \citep{thomas_stellar_2013}. This result agrees with the unphysical value we obtained from the GMOS datacube. Discrepancies between emission line fluxes reported by SpecBS and the Portsmouth catalog were similarly noted in \cite{lyu_high_2016}, although less drastic.  Without higher-resolution optical data to confirm which result is correct, we elect to adopt the physical value from SDSS DR10.} and an SFR of $9.55\ \pm 3.4 M_\odot$ yr$^{-1}$ when matching spectral energy distribution templates to extinction-corrected SDSS \textit{ugriz} magnitudes. We take the SFRs uncorrected for extinction as a lower limit on the SFR in the system. Without correcting for extinction, ALFOSC obtained a H$\alpha$ flux of $1.1 \times 10^{-15}$ ergs cm $^{-2}$ s$^{-1}$, which corresponds to an SFR of 3 M$_\odot$ yr$^{-1}$. We thus estimate the total SFR within the beaded star formation complex to be between $1-10$ M$_\odot$ yr$^{-1}$.

\begin{figure}
\centering \includegraphics[width=\linewidth]{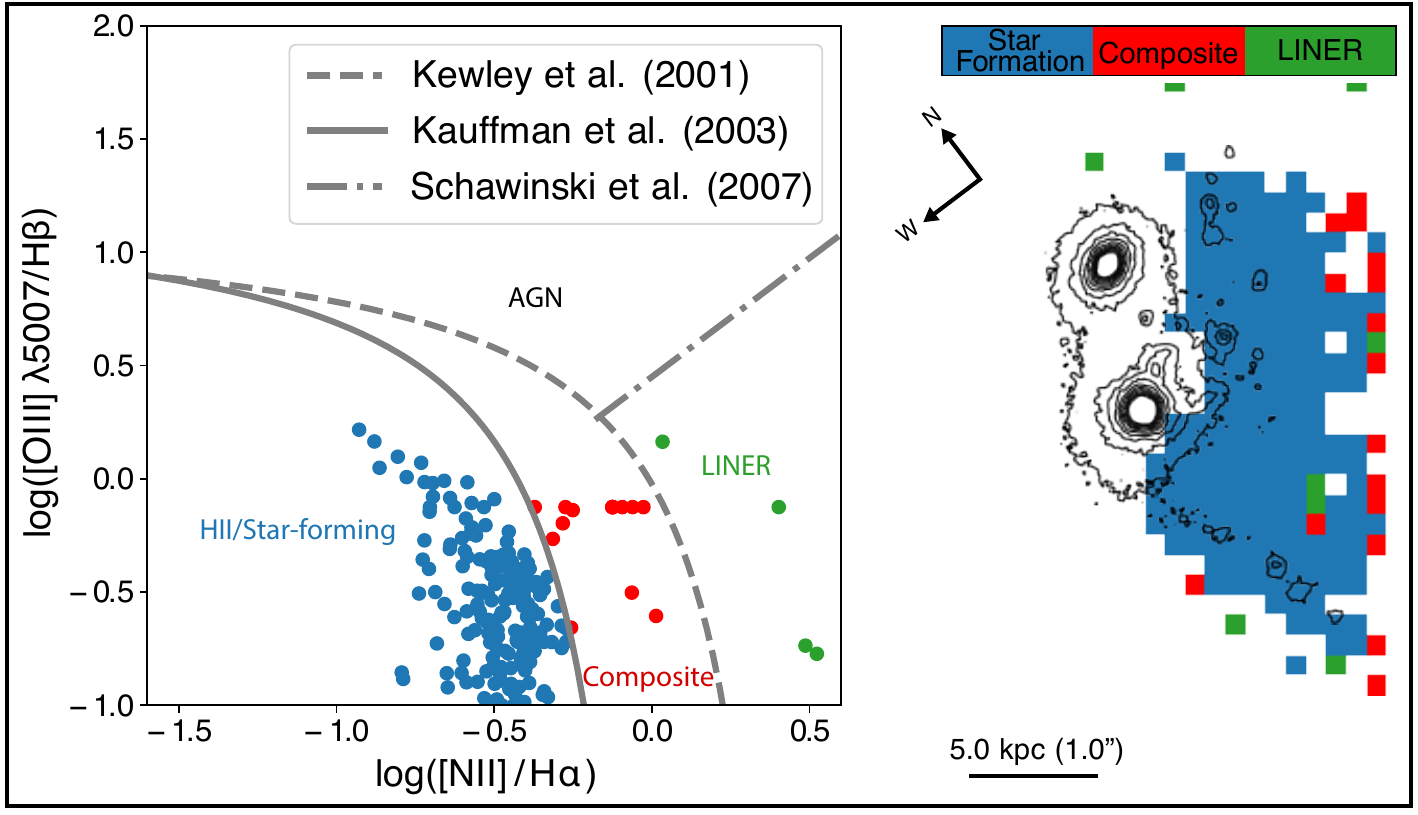}
\caption{Emission line diagnostic diagrams for spaxels with $S/N \geq 3$ in each emission line. \textit{Left:} Baldwin, Phillips \& Terlevich (BPT) \citep{baldwin_classification_1981} diagnostic plot. The spaxels are color-coded based on their location relative to boundaries between well-known empirical and theoretical classification schemes (see \citealt{kewley_theoretical_2001, kauffmann_host_2003, schawinski_suppression_2006}) shown in gray dashed and solid lines. The majority of the spaxels lie within the star-forming region. \textit{Right:} Sky distribution of the spaxels, color-coded according to their BPT diagram classification.}
\label{fig:gmos_bpt}
\end{figure}

\begin{figure*}
\centering \includegraphics[width=\linewidth]{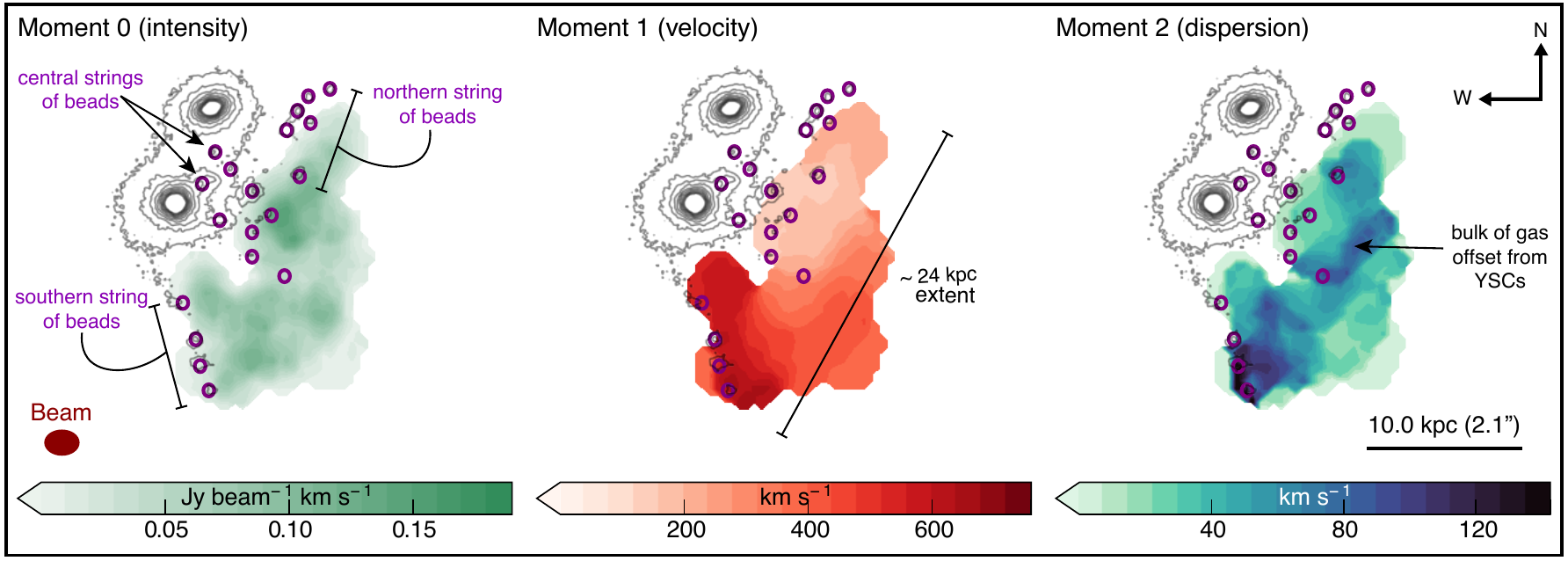}
\caption{
\textit{Left to Right:} ALMA integrated line intensity flux density (moment 0), intensity-weighted velocity (moment 1) and intensity-weighted velocity dispersion (moment 2) maps for SDSS J1531. The molecular gas extends $\sim 24$ kpc from North to South. There is a notable offset between the YSCs and the molecular gas distribution, which is discussed further in Section \ref{subsubsec:cold_shape}. Like the H$\alpha$ gas, the molecular gas is similarly redshifted up to $+600$ km s$^{-1}$ with respect to the southern BCG. The filled dark red circle represents the size of the ALMA beam. }
\label{fig:alma_mom}
\end{figure*}

\begin{figure*}
  \centering \includegraphics[width=\linewidth]{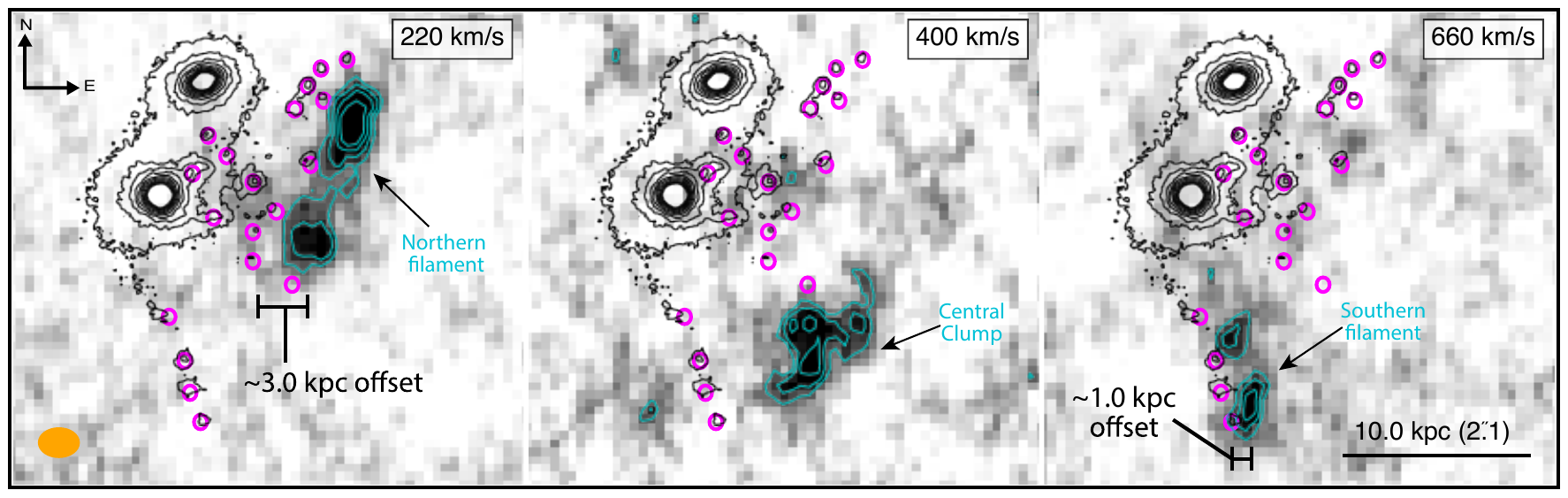}
\caption{ALMA CO (3-2) channel maps displaying the three distinct segments of molecular gas in velocity-space, and how the peak emission is offset by $1-3$ kpc from the YSCs. The left and right panels show emission morphologically similar to the YSCs at 220 and 660 km s$^{-1}$, respectively. The central panel at 400 km s$^{-1}$ shows emission comprising the central clump, mostly spatially distinct from the YSCs. The blue contours highlight $\geq 3 \sigma$ emission.}
\label{fig:alma_cube}
\end{figure*}

\subsubsection{Ionization Sources}
\label{subsubsec:gmos_ion}

 To identify potential sources of ionizing radiation in the star formation complex, we utilize the broad wavelength coverage of the optical spectra to compare key diagnostic emission line ratios. We created a spatially resolved Baldwin-Philips-Terlevich (BPT) diagram \citep{baldwin_classification_1981}, namely the  [\ion{N}{2}] - BPT diagram (\ion{O}{3} $\lambda 5007$ / H$\beta$ vs. \ion{N}{2} $\lambda 6583$ / H$\alpha$), shown in Figure \ref{fig:gmos_bpt} left). The classic [\ion{N}{2}] - BPT diagram is used to distinguish sources of ionizing radiation, namely between \ion{H}{2} regions/star formation, Seyfert, or a combination of Seyfert and star formation.  The solid line represents the demarcation line of pure star formation \citep{kauffmann_host_2003}, the dashed line represents the demarcation line of extreme star formation \citep{kewley_theoretical_2001}, and the dashed-dotted line represents the empirical division between LINER- and Seyfert like sources \citep{schawinski_suppression_2006}.  We have color-coded the data points based on the regions in which they sit. 
 
 Most of the spaxels lie inside the star formation region. Few spaxels lie near the outer edges, where the signal is weakest, in the "composite" and "LINER" regions. This suggests that the warm nebula is primarily ionized by the YSCs, with some potential contribution from AGN activity. We do not over-interpret the BPT classifications since cool core BCGs likely have several ionization sources \citep{ferland_collisional_2009,mcdonald_optical_2012}.

\subsection{The Cold Molecular Gas}
\label{subsec:cold_gas}
\subsubsection{Morphology: Clumps Offset from the Beads}
\label{subsubsec:cold_shape}

The morphology of the cold ($<10^2$ K) molecular gas is shown in Figure \ref{fig:alma_mom}, which displays the masked CO(3-2) integrated intensity map (moment 0), the intensity-weighted mean velocity field (moment 1), and the intensity-weighted velocity dispersion (moment 2). The integrated intensity map pieces clumps of molecular gas together to reveal a cloud-like structure spanning $\sim 24$ kpc across the sky, with three bright emission peaks to the north, southwest, and southeast. Surrounding the bulk of the emission are small pieces belonging to a smoother gas distribution below the $> 3.5 \sigma$ threshold mask. The ALMA observations do not resolve the 19 individual clumps detected in the NUV, hindering the opportunity for clump-by-clump analysis. Instead, the emission appears concentrated in three filaments and clumps that are spatially interconnected yet distinct in velocity-space. These clumps are shown in Figure \ref{fig:alma_spectrum} and further discussed in Section \ref{subsubsec:cold_vel}.

The molecular gas is peculiarly spatially offset with respect to the central BCGs and the young stellar superclusters. To quantify the offset, we compare the morphology of the gas in Figure \ref{fig:alma_cube} to that of the YSCs in the NUV (\textit{HST} F390w). The emission in the 220 km s$^{-1}$ channel resembles that of the northern YSCs. Although the molecular gas overlaps with the majority of the YSCs in the northern filament, the most northern string of beads is offset from the peak of the molecular gas by  $\sim$ 0\farcs6 ($\sim$ 2.8 kpc). Furthermore,  the bulk of the emission remains shifted to the east. Unlike the northern filament, the central clump is almost completely decoupled from the YSCs, and resembles none of the YSCs morphologically. The emission in the 660 km s$^{-1}$ channel matches the morphology of the southern string of beads. The beads and the molecular gas are separated by $\sim$ $0\farcs3$ ($\sim$ 1.4 kpc), with the molecular emission lying to the southeast.  In the mostly decoupled southern corner, the cold gas appears to fit into the southwestern bay-like feature like two pieces of a puzzle. 

It is important to note that the ALMA observations reveal faint, smooth emission across most of the YSCs, near the southern nuclei of the BCG, and in between the BCGs. However, this emission lies below the threshold applied to all CO(3-2) maps presented in this paper (see Section \ref{subsec:almaobs}). The detection of some faint molecular emission in the inner extent of the merging BCGs and across most of the YSCs suggests that if we observed the system at greater depths with ALMA, we might detect CO(3-2) completely enveloping the YSCs and certain areas of the BCGs. However, the bulk of the emission would most likely still lie to the east of the merging system.

Regardless, the coincidence between the morphology of the YSCs and that of the molecular gas suggests that they are associated. However, the question remains: why is the molecular gas spatially offset from the young stars? One possibility is that the molecular gas that originally enveloped the YSCs collapsed to form stars, which then ionized the surrounding gas. However, this does not explain why the collapse was localized to one side of the gas. We discuss potential scenarios to explain the gas' relation to the YSCs in Section \ref{subsec:cool_ff} and describe its relation to the warm ionized gas in Section \ref{subsubsec:gmos_alma}.

\subsubsection{Velocity Structure: Redshifted Gas Flows}
\label{subsubsec:cold_vel}
The first and second moment maps in Figure \ref{fig:alma_mom} reveal complex velocity structure spanning the cold molecular gas' extent. The projected line-of-sight velocities across the structure are significantly redshifted by $\sim +700$ km s$^{-1}$ with respect to the stellar velocity of the southern BCG. The velocity of the gas peaks near the southern string of beads, with a corresponding velocity dispersion peak of $\sim 150$ km s$^{-1}$.  The gas with the largest velocity dispersion traverses a diagonal path from the end of the southern string of beads to just beneath the northern string of beads.

\begin{figure}[H]
\centering \includegraphics[width=\linewidth]{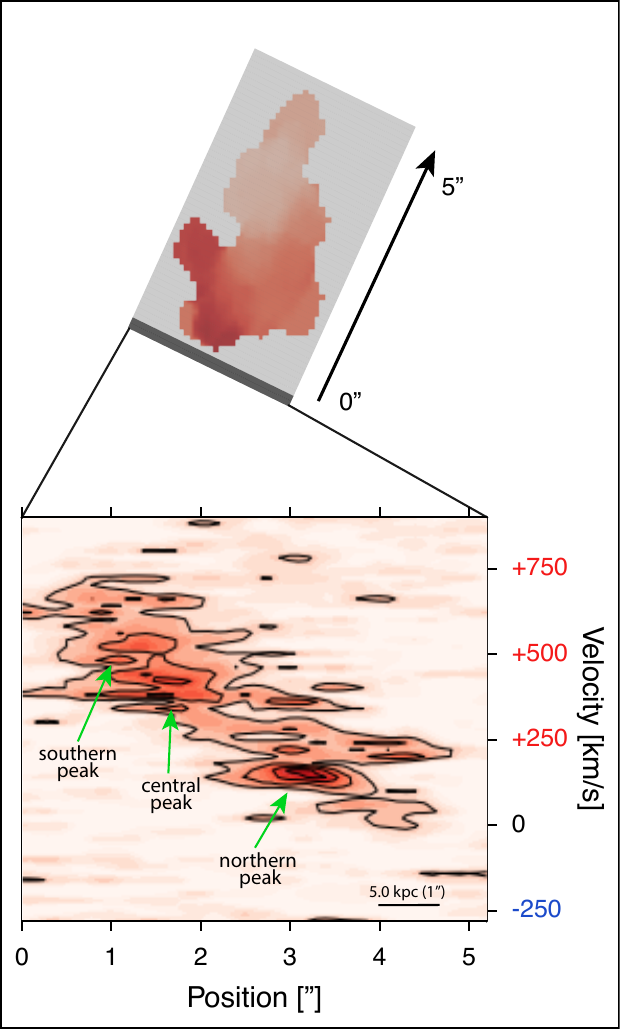}
\caption{Position-Velocity ($PV$) diagram created from the CO(3-2) data. The top image displays the moment one velocity map with the \textit{PV} extraction apertures overlaid in grey. The lower image shows the \textit{PV} diagram extracted from the grey rectangular region. The molecular gas exhibits a smooth/continuous velocity distribution.}
\label{fig:alma_pv}
\end{figure}

To analyze the gas flows in greater detail, we define a rectangular region that spans the total extent of the molecular gas \ and average the emission in each velocity channel over the width of the region to produce a position-velocity (PV) diagram, shown in Figure \ref{fig:alma_pv}. The PV diagram reveals a smooth decrease in velocity from the southern to the northern nuclei, with distinct velocity components present across the structure. There is one primary peak in the northern filament (Position 2 - 5\arcsec), one in the central clump  (Position 1 - 2\arcsec), and another in the southern filament (Position 0.5 - 1.5\arcsec). 

\begin{figure*}
\centering
\includegraphics[width=\linewidth]{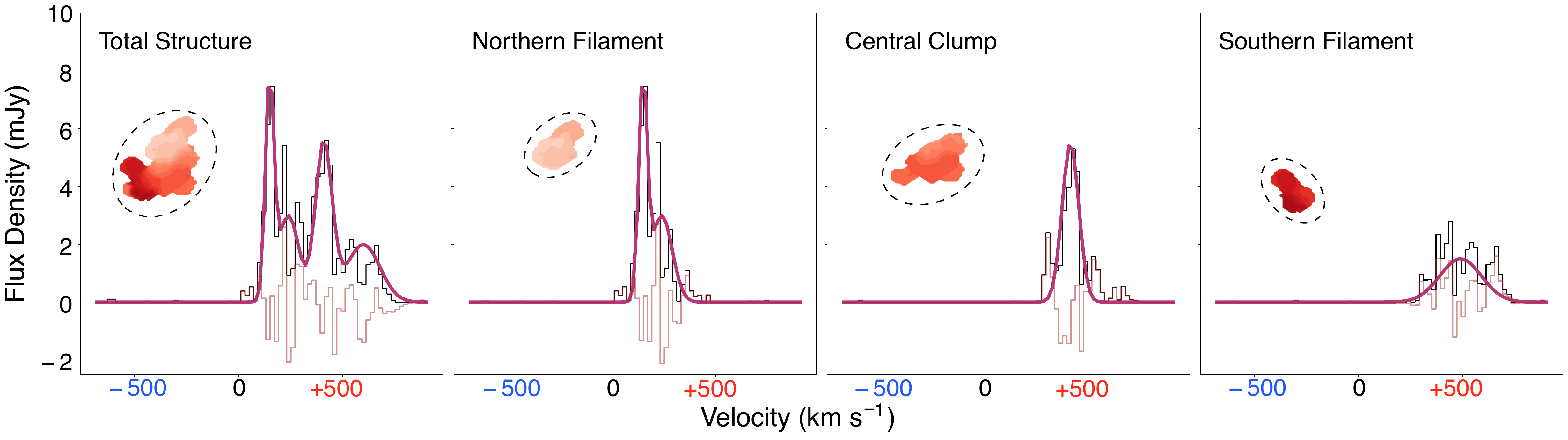}
\caption{$\geq 3\sigma$ ALMA CO(3-2) spectra extracted from different components of the molecular gas. From left to right, the panels display multi-Gaussian fits to the CO(3-2) spectra extracted from the region encompassing the entire extent of the molecular gas, the northern filament, the central clump, and the southern filament. The black curves represent the raw spectra extracted from each region, while the purple curves depict the fits to each component of the molecular gas within the velocity extent defined for each filament/clump. The residuals are shown in light pink. The multi-Gaussian fit of the total structure's spectrum estimates a cold gas mass of approximately $10^{10} M_\odot$, with the majority concentrated in the northern filament and a roughly even distribution between the central clump and southern filament.}
\label{fig:alma_spectrum}
\end{figure*}

The velocity of the gas in the northern filament, central clump, and southern filament ranges mostly from $+60 - 260$ km s$^{-1}$, $+280 - 460$ km s$^{-1}$, and $+480 - 680$ km s$^{-1}$, respectively. We present the intensity-weighted velocity map of each structure defined by the aforementioned velocity ranges in Figure \ref{fig:alma_spectrum}. The smooth velocity gradient across the total structure could indicate positive or negative radial velocities. This suggests that the molecular gas clumps in this system experienced radial velocity changes at different epochs. Specifically, the gas in the northern filament is closely associated with the stellar systemic velocity, while the gas in the southern filament appears more consistent with the average 

\noindent velocity for seven other galaxies found within $\sim 600$ kpc of the cluster core \citep{bayliss_geminigmos_2011}. This may indicate the presence of molecular gas that is mostly influenced by the bulk motions of the intracluster gas (near the southern filament), slowly entering the gravitational potential well of the BCGs (near the northern filament). We provide a more detailed interpretation of the gas motions in Section \ref{subsubsec:agn_feedback}.

\subsubsection{Mass Distribution}
\label{subsubsec:cold_mass}

Assuming a CO(3-2)/CO(1-0) line ratio of $\sim0.8$ \citep{edge_detection_2001},
we estimate the mass of the molecular H$_2$ nebula by following
the relation reviewed by \citet{bolatto_co--h2_2013}:
\begin{align}
 M_\mathrm{mol} & = \left(\frac{1.05 \times 10^4}{0.8}\right) ~ \left(\frac{X_{\mathrm{CO}}}{X_{\mathrm{CO,~MW}}} \right)                                                                    \\
                & \times \left(\frac{1}{1+z}\right) \left(\frac{S_{\mathrm{CO}}\Delta v}{\mathrm{Jy~km~s}^{-1}}\right)   \left(\frac{D_\mathrm{L}}{\mathrm{Mpc}}\right)^2 M_\odot, \nonumber
 \label{eqn:molmass}
\end{align}
where $S_{\mathrm{CO}}\Delta v$ is the integrated CO(3-2) intensity,
$z$ is the galaxy redshift ($z=0.335$), and $D_L$ is its luminosity distance (1767 Mpc in our adopted cosmology). The Galactic CO-to-H$_2$ conversion factor $X_{\mathrm{CO}}$ dominates the uncertainty in this relation
\citep[e.g.,][]{bolatto_co--h2_2013}. For SDSS 1531, we adopt the average value for the disk of the Milky Way, $X_{\mathrm{CO}} = X_{\mathrm{CO,~MW}} = 2 \times 10^{20}$ cm\ $\left(\mathrm{K~km~s}^{-1}\right)^{-1}$, which has a $\sim$30\% uncertainty. The true value of the conversion factor largely depends on the gas metallicity and whether the CO emission is optically thick. It is unclear whether the $X_{\mathrm{CO}}$ factor measured in the Milky Way and nearby spiral galaxies should match that of BCGs, as it often varies wildly in ultraluminous infrared galaxies \citep{bolatto_co--h2_2013}. However, \citet{vantyghem_13co_2017} recently reported one of the first detections of $^{13}$CO in a BCG (RX J0821+0752) and found a $X_{\mathrm{CO}}$ factor only a factor of 2 lower than that of the Milky Way. To account for our limited information about the system's metal abundance and to facilitate direct comparison with the molecular gas masses derived in other studies of BCGs \citep[e.g.,][]{tremblay_galaxy-scale_2018, russell_alma_2016,olivares_ubiquitous_2019,north_wisdom_2021}, we adopt $X_{\mathrm{CO}} = X_{\mathrm{CO,~MW}}$ as our most reasonable choice. A caveat to this selection is that we may overestimate the total molecular mass by a factor of a few. This should be interpreted as the overriding uncertainty on all mass estimates quoted below.

\begin{figure*}
  \centering \includegraphics[width=\linewidth]{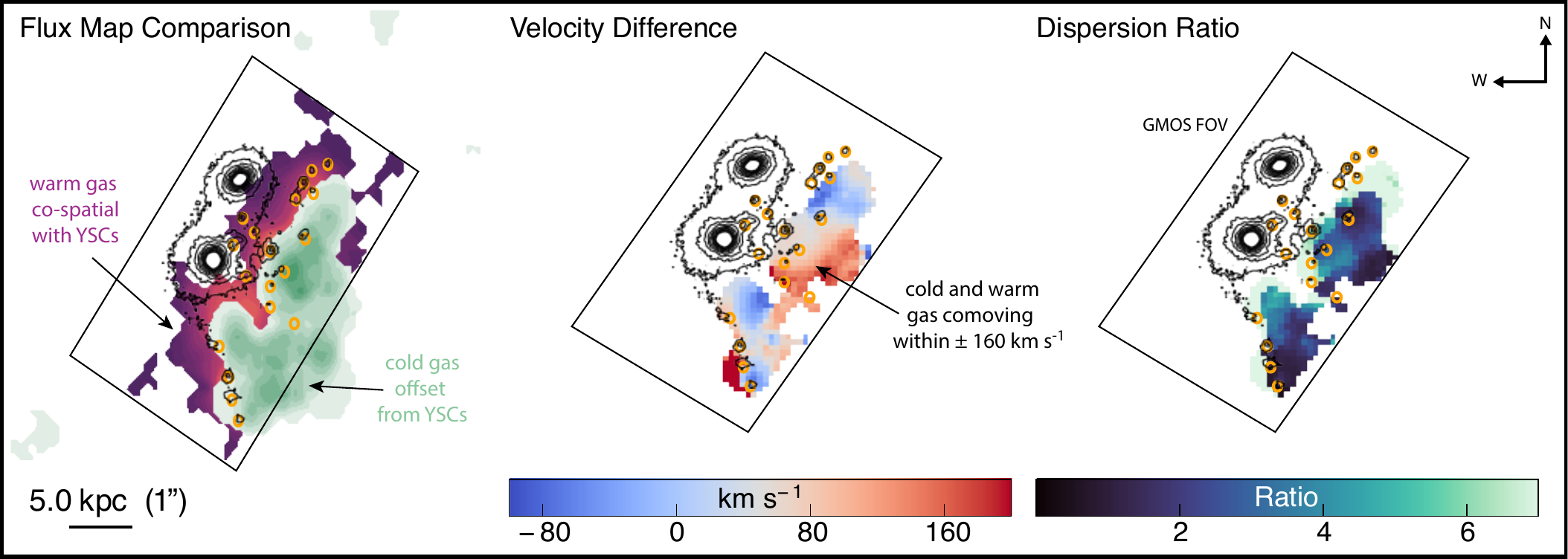}
\caption{\textit{Left to right}: Maps of the H$\alpha$ and CO(3-2) flux, velocity difference, and velocity dispersion ratio, respectively, corrected for differences in spatial resolution (see Section \ref{subsec:gmosobs}). Whether the H$\alpha$ emission extends beyond the molecular gas is uncertain due to the limited field covered by GMOS (black rectangular box). The edges of the velocity difference and dispersion ratio maps should be disregarded as they are artifacts of the subtraction/division. Though the ionized and molecular gas are not fully co-spatial, they are largely comoving.}
\label{fig:comp_haco}
\end{figure*}

To estimate $S_{\mathrm{CO}}\Delta v$, we fit a four-component Gaussian line to the CO (3-2) spectrum extracted from an elliptical aperture containing all $\geq 3\sigma$ emission in the calibrated cube binned to 20 km s$^{-1}$ channels (the spectrum is shown in the left panel of Figure \ref{fig:alma_spectrum}). The fit, from -690 km s$^{-1}$ to 900 km s$^{-1}$, yielded an emission integral of $\sim 1.7 \pm 0.3$ Jy km s$^{-1}$. Noting the aforementioned caveats, these results give a $H_2$ gas mass of $(3.7 \pm 0.7) \times 10^{10}\text{ M}_\odot$. With natural weighting, we obtain integrals ranging from $0.88 - 2.5$ Jy km s$^{-1}$ depending on the different binning selected, yielding mass estimates that range from $1.9 - 6.5 \times 10^{10}$ M$_\odot$.\footnote{With Briggs weighting, we obtain integrals ranging from $0.17 - 4.8$ Jy km s$^{-1}$ depending on the different binning selected, yielding mass estimates that range from $3.7 - 100 \times 10^{9}$ M$_\odot$. This discrepancy with the natural weighting is likely due to the Briggs weighting scheme resolving out emission. We deem the natural weighting more appropriate for our data.}

The remaining panels in Figure \ref{fig:alma_spectrum} show the spectra of the gas across the northern filament, central clump, and southern filament regions. The fit to each structure's spectra indicates that most of the gas mass is concentrated in the northern filament ($\sim 45 \%$), and distributed relatively evenly throughout the central clump and southern filament ($\sim 32.5\%$ each).

\subsubsection{Undetected Continuum}
\label{subsubsec:cold_cont}

ALMA detected no continuum within the vicinity of the cluster's BCGs in any of the three line-free spectral windows placed in Band 6 (see Section \ref{fig:alma_cube}). From the standard deviation of the low-resolution data, we place $2\sigma$ upper limits of $4.87 \times10^{-4}$ mJy beam$^{-1}$ on the continuum. If detected, the continuum would likely have originated from thermal emission from dust and/or synchrotron and hot dust emission from the central AGN. The non-detection is, therefore, consistent with the picture of an AGN that is currently inactive.

To place an upper limit on the mass of dust present, we modeled the emission $S_\nu$ as a modified blackbody using the following equation:
\begin{align}
S_\nu = \frac{\kappa_\nu M_d B(\nu, T_d)}{D^2}
\end{align}

\noindent where $M_d$ is the dust mass, $B(\nu, T_d)$ is the Planck function, which depends on frequency $\nu$ and dust temperature $T_d$, and $D$ is the distance to the galaxy. $\kappa_\nu$ is the dust absorption coefficient, described by a power law with dust emissivity index $\beta$ such that $\kappa_\nu \propto \nu^\beta$. Here, we utilized an empirical $\kappa_\nu$, where $\kappa_{500 \mu\mathrm{m}} = 0.051$ m$^2$ kg$^{−1}$, and $\beta = 1.8$ \citep{clark_empirical_2016}. Assuming a dust temperature of 25 K \citep[e.g.,][]{davis_wisdom_2017}, we obtain a maximum dust mass of $3.7 \times 10^8$ M$_\odot$ (and thus a molecular gas-to-dust ratio of $\gtrsim 100$).

\subsubsection{Comparison with the Warm Ionized Gas}
\label{subsubsec:gmos_alma}
Given that young stars form in molecular clouds, one would expect the stellar superclusters to be deeply embedded within the cold gas from which they formed and that the cold gas is co-spatial with the warm ionized gas, as previous analyses of cool core clusters have observed \citep[e.g.,][]{olivares_ubiquitous_2019, tremblay_galaxy-scale_2018, tremblay_cold_2016, vantyghem_molecular_2016}. A direct comparison between the integrated ALMA CO(3-2) intensity map and GMOS H$\alpha$ observations (Figure \ref{fig:comp_haco}, left) reveals that the molecular and ionized gas have similar physical extents, with the ionized gas extending slightly further than the molecular gas within the GMOS field of view. However, it remains unclear whether the ionized gas fully enshrouds the entire molecular structure with the current observations. It is, nonetheless, evident that the molecular gas lies $\sim$3 kpc to the southwest of the ionized gas (Figure \ref{fig:comp_haco}, left), with the offset being too large to be attributed to uncertainty in either ALMA or Gemini North's respective astrometric reference frame.

\begin{figure*}
\centering \includegraphics[width=\linewidth]{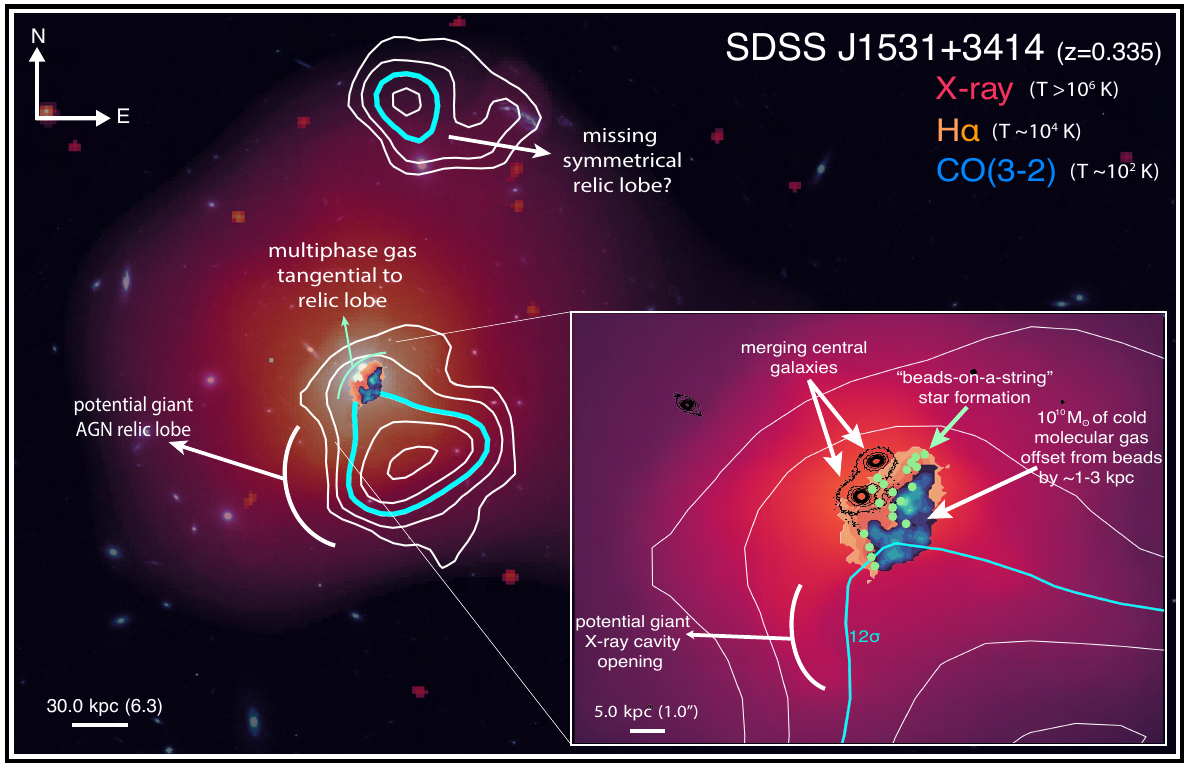}
\caption{A large-scale and small-scale (inset plot) view of the multiphase gas in SDSS 1531. Purple denotes the \textit{Chandra} X-ray surface brightness map tracing the hot intracluster gas, orange represents the warm ionized gas traced by Gemini/GMOS-N IFU H$\alpha$ flux, and blue shows the cold molecular gas traced by the ALMA CO$(3-2)$ line. LOFAR contours of Sources C and D are outlined in white, with the $12\sigma$ contour, which best outlines the extent of each potential relic lobe, highlighted in cyan. With the caveat that projection effects complicate interpretation, the $12\sigma$ radio contour perfectly fills the putative X-ray cavity opening, and the warm and cold gas phases are oriented perpendicular to the north of the cavity opening. The warm ionized gas fully envelops the YSCs, which are mostly offset from the molecular gas by $\sim 1-3$ kpc.}
\label{fig:comp_hagc}
\end{figure*}

The aforementioned studies also found the two gas components tend to be co-moving. To explore this possibility, we plot the difference in velocity and the velocity dispersion ratio between the overlapping CO(3-2) and H$\alpha$ gas in Figure \ref{fig:comp_haco} (center, right). Using the \textsc{REPROJECT} module from the \textsc{ASTROPY} package, the ALMA maps were resampled onto GMOS pixel grids.  The velocity difference map shows that the ionized and molecular gas exhibit similar overall velocity structure within $\pm 160$ km s$^{-1}$, which is within the GMOS data's $\sim 160$ km s$^{-1}$ velocity resolution. This indicates that the cold and warm ionized gas phases are likely co-moving, consistent with prior studies. 

The velocity dispersion ratio map shows that the H$\alpha$ velocity dispersion is mostly consistent with the CO(3-2) or slightly broader, consistent with multi-wavelength studies of cool core clusters \citep{olivares_ubiquitous_2019}. \citet{tremblay_galaxy-scale_2018} suggested that lines of sight are more likely to intersect warm gas than cold clouds as an explanation for the broader velocity distribution of the warm ionized gas compared to the cold gas.

\subsubsection{Gas Depletion Timescale}
\label{subsubsec:gmos_alma_gas_time}

To estimate the amount of time needed to deplete the entire $\sim 10^{10} M_\odot$ reservoir of molecular gas, we adopt a star formation rate of $1-10$ M$_\odot$ yr$^{-1}$. We calculate the depletion time as $t_{\text{dep}} = \frac{M_{\text{gas}}}{SFR}$ and recover timescales ranging from $\sim 1 - 10$ Gyr, which is approximately within a Hubble time ($t_H$ = 9.728 Gyr). The majority of cool core BCGs typically have gas depletion timescales of $\sim 1$ Gyr \citep{odea_infrared_2008}, except starburst BCGs, which can have depletion timescales as short as $\leq 30$ Myr \citep{mcdonald_state_2014}. Both ends of the depletion timescale range for SDSS 1531 lie above $\sim 1$ Gyr, implying that the star formation efficiency in this system is lower than in other BCGs. This is likely because only the BCG-facing border of the molecular gas has collapsed to form stars, which we discuss further in Section \ref{subsubsec:assemble}. 

\vspace{4mm}

\section{Discussion}
\label{sec:beads_origin}

The new X-ray, optical, and radio data presented in this paper reveal that the merging central ellipticals and the associated beaded strings of star formation in SDSS 1531 are situated within a rapidly cooling, highly magnetic ICM. The two most striking results of this study are 1) the remarkable spatial alignment between the diffuse radio emission from Source C and the concave X-ray surface brightness discontinuity, and 2) the $\sim 10$ billion solar mass molecular nebula's mostly offset spatial location from the young stellar superclusters. We propose that the alignment between the X-ray discontinuity and the radio source indicates the presence of a large X-ray cavity, likely blown during an older epoch of AGN feedback. We posit that the molecular gas originated from low-entropy gas entrained by the massive cavity and is slowly encircling and settling into the gravitational potential well of the BCGs. A combination of ram pressure, tidal interactions, and ionization by star formation likely contribute to the projected separation between the young stars and the molecular gas. A combination of ram pressure, tidal interactions, and ionization due to star formation contribute to the separation between young stars and molecular gas. In the following section, we provide a detailed description of this scenario based on the results presented in the previous section, which are summarized in Figure \ref{fig:comp_hagc}. We also evaluate alternative mechanisms that could have contributed to the observed molecular gas supply, such as its potential capture from previous encounters with gas-rich galaxies and/or an origin from the central ellipticals. A summary of our expectations for each scenario and whether or not the current observations support them is reported in Table \ref{tab:evidence}.

\subsection{ICM Condensation, AGN Feedback, and Star Formation in a Dynamic Cluster Environment}
\label{subsec:cool_ff}

\subsubsection{A Cluster-Scale Merger?} 
\label{subsubsec:cluster_dynamics}

\begin{figure}
  \centering \includegraphics[width=\linewidth]{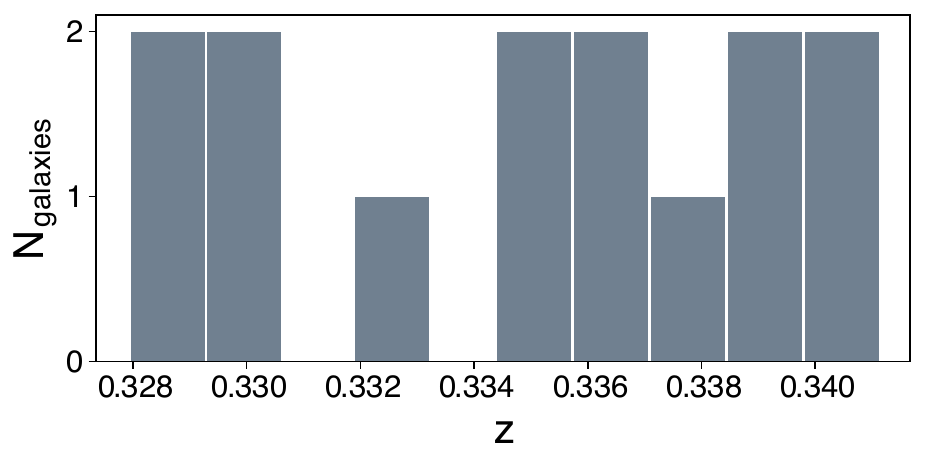}
 \caption{The redshift distribution of galaxies in the SDSS 1531 field, based on spectroscopic and photometric data from 14 galaxies}. The distribution shows hints of two distinct peaks, suggesting a possible bimodal nature of the galaxy population, which may indicate a subcluster merger.
\label{fig:hst_galz}
\end{figure}

Although the two central galaxies in SDSS 1531 are definitively engaged in a major merger, determining whether this event is part of a larger-scale subcluster merger demands more nuance. \citet{sharon_mass_2014} find that the mass distribution of the cluster is well described by two cluster-scale halos. Moreover, each central elliptical is of roughly equal stellar mass, and there is an apparent bifurcation in the optical line redshift distribution for fourteen of the nearby cluster members for which there is existing optical spectroscopy/photometry (see Figure \ref{fig:hst_galz}). Although the current redshift sample is too limited to draw any definitive conclusions, the presence of diffuse radio sources throughout the cluster signals broader systemic dynamical activity. Such cluster-scale synchotron emission is almost exclusively found within systems dynamically disturbed by mergers.

If the merging ellipticals in SDSS 1531 result from a cluster-scale merger, it is likely in the last stages of it. The large-scale X-ray gas distribution is relatively smooth and relaxed, with no secondary galaxy concentrations. The exact nature of the merger – whether it involved two clusters of equal size or a massive cluster and a smaller subcluster, or whether it was a head-on or off-center collision – is not discernible from the available data. In the context of the impact on the cool core, however, the literature provides varied insights. \citet{chadayammuri_fountains_2021}, \citet{valdarnini_study_2021}, and \citet{rasia_cool_2015} find that head-on major mergers significantly disrupt the structure of cool cores, at least for a few Gyr. On the other hand, \citet{zuhone_stirring_2010}, \citet{poole_impact_2008}, and \citet{hahn_rhapsody-g_2017} find that although cool cores are mostly disrupted by head-on major cluster mergers, low angular momentum mergers do not always destabilize the core. Among these varied findings, the consensus seems to be that while mergers can heat cluster cores, the parameters of a merger, such as mass ratios and impact parameters, likely determine the outcome. Currently, SDSS 1531 aligns with the portrait painted by the latter studies, for despite the dynamic activity within the cluster, it has maintained its cool core. Furthermore, the central 30 kpc of SDSS 1531's hot atmosphere still exhibits low entropy ($S \leq 30$ keV cm$^2$), thermally unstable ($t_{cool}/t_{ff} \sim 20$) gas with short cooling times ($t_{cool} < 1$ Gyr), characteristics often found in relaxed galaxy clusters with large molecular gas reservoirs, star formation, and nebular emission in their cores \citep{cavagnolo_entropy_2008}. 

\subsubsection{Thermally Unstable Cooling}
\label{subsubsec:cooling_dynamics}

The dynamic environment within SDSS 1531 offers several viable routes to condensation from the hot atmosphere. Defining the "classical" cooling rate as
\[\dot{M}_{cool} = \frac{M_{gas}(r < r_{cool})}{t_{cool}},\]
where $r_{cool} \sim 40 \text{ kpc}$ is the radius within which the cooling time is less than 3 Gyr \citep{mcdonald_revisiting_2018}, we obtain $\dot{M}_{\rm cool} \sim 185$ M$_\odot$ yr$^{-1}$. Assuming a consistent cooling rate over 3 Gyr, the hot atmosphere could easily supply up to $\sim 10^{12}$ M$\odot$ of cool material, with only 100 Myr needed to fill the entire $10^{10} \msun$ gas reservoir. 

If the multiphase gas observed directly cooled out of the hot atmosphere, the ``circumgalactic precipitation" model \citep{voit_global_2017} posits that the growth of small, local thermal instabilities can lead to unstable cooling when $t_{cool}/t_{ff} < 10$. For SDSS 1531, however, $t_{cool}/t_{ff} > 20$ within the cluster core. 

\begin{figure}
  \centering \includegraphics[width=\linewidth]{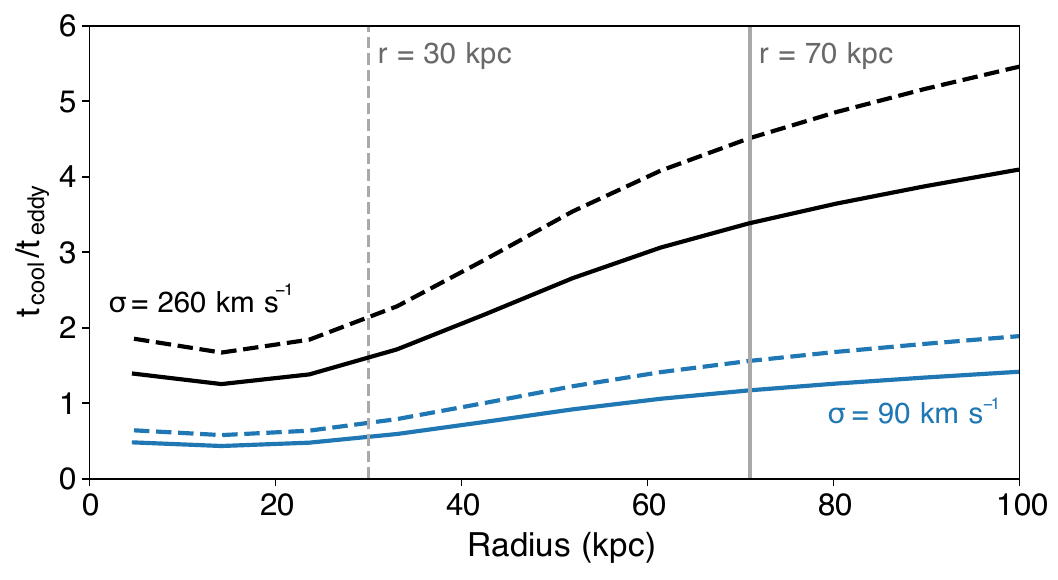}
 \caption{Profiles of the cooling time to eddy turnover time ratio ($t_{\rm cool}/t_{\rm eddy}$) in the ICM. The dashed and solid lines represent injection length scales $L = 30$ and 70 kpc, respectively, corresponding to the radial extent of the CO nebula and the radio cavity. The dark blue and black lines correspond to velocity dispersions $\sigma = 90$ and 260 km s$^{-1}$, respectively.} 
\label{fig:chandra_tcool_teddy}
\end{figure}

On the other hand, the Chaotic Cold Accretion (CCA) model \citep{gaspari_shaken_2018} proposes that the ICM condenses through turbulence driven by AGN outflows, mergers, and on smaller scales, supernovae and stellar winds. In this model, \cite{gaspari_shaken_2018} suggest that the ratio of $t_{\rm cool}$ to the eddy turnover time-scale better indicates thermal instability. The eddy turnover time $t_{\rm eddy}$ is defined as 

\begin{equation}
    t_{\rm eddy} = 2 \pi \frac{r^{2/3} L^{2/3}}{\sigma_{v, L}}
\end{equation}

\noindent where $L$ is the injection length scale of the turbulence and $\sigma_{v, L}$ is the velocity dispersion of the turbulence at the injection scale in the ICM. Given that neither of these parameters is directly observable for SDSS 1531, we use the velocity dispersion of the H$\alpha$ and CO nebulae as a proxy for the velocity dispersion of the ICM. The velocity dispersion of the H$\alpha$ and CO clouds range from $\sim 50 - 200$ km s$^{-1}$. Accounting for the conversion from line of sight to a three-dimensional velocity dispersion with a factor of $\sqrt{3}$, we adopt $\sigma_{v, L} = 80 - 300$ km s$^{-1}$. We infer the injection length scale $L$  from the extent of the CO emission and the length of the X-ray/radio cavity, ranging from $L = 30 - 70$ kpc. Figure \ref{fig:chandra_tcool_teddy} shows that the $t_{\rm cool}/t_{\rm eddy}$ profile approaches unity throughout the region where CO is observed, supporting the idea that the ICM is thermally unstable and rapidly cooling out to at least 30 kpc.

If the thermal instability originated from the hot atmosphere, possibly due to turbulence from the merger, we would expect the cooled gas to be relatively dust-free \citep{donahue_calcium_1993}. This is because dust grains sputter rapidly ($\sim 1$ Myr) in the ICM and can only form when the gas is shielded from UV and X-ray irradiation \citep{draine_destruction_1979}. The GMOS observations found minimal extinction across the extent of the warm ionized component of the multiphase gas, and the ALMA observations placed an upper limit of $\sim 10^6 M_\odot$ on the dust mass present. These observations, however, do not completely rule out the presence of dust within the cooled gas, in which case would favor an origin from within BCGs rather than directly condensing out of the hot atmosphere. 

In Section \ref{subsubsec:agn_feedback}, we argue that the observed turbulence-induced thermal instability was most likely engendered by uplift from the massive X-ray/radio cavity, a viable route to condensation in all three precipitation, CCA, and stimulated feedback models.


\subsubsection{An Older Epoch of AGN Activity?}
\label{subsubsec:old_agn}

Figure \ref{fig:comp_hagc} shows clear spatial alignment between potential X-ray cavity opening and the low frequency radio emission, hinting at AGN activity within the cluster. A similar large concave surface brightness discontinuity is also observed in the Ophiuchus cluster \citep{giacintucci_discovery_2020}, and on smaller scales in the Perseus cluster, Abell 1795, Abell 2390, and Centaurus cluster, where they were all interpreted as the inner walls of cavities resulting from AGN activity \citep{walker_exploring_2014, sanders_detecting_2016}. The concave X-ray surface brightness discontinuity likely does not fully wrap around the radio lobe due to its low surface-brightness contrast, thus requiring deeper X-ray observations for successful detection.

The steep radio emission, strongly detected at lower frequencies but absent at 1.4 GHz, likely originates from aged, relic plasma from a powerful AGN outburst in an older stage of the cluster's history. The velocity dispersion of the warm ionized gas is most disturbed near the southern BCG's nucleus. Moreover, a radio emission bridge connects the lobe to the BCG, suggesting the AGN outburst originated there. 

If Source C began its life as a buoyant bubble injected near the southern BCG and rose buoyantly to its current location in the plane of the sky at the terminal velocity $v_t$, we can estimate its age as
\begin{align}
    t_{cav} = R/v_t = R \cdot \sqrt{S C/2 g V},
\end{align}

\noindent where R is the projected distance from the southern BCG to the cavity, $V$ is the volume of the bubble, $S$ is the cross-sectional area of the bubble, $C=0.75$ is the drag coefficient \citep{churazov_evolution_2001}  and $g$ is the gravitational acceleration. We model the potential cavity as an ellipsoidal volume with an axis perpendicular to the plane of the sky and equal to the projected major axes ($r_{\rm maj}= 43.6$ kpc) and minor axes ($r_{\rm min}= 39.6$ kpc). 

Following \citet{birzan_systematic_2004}, we calculated the gravitational acceleration using the stellar velocity dispersion of the southern BCG under the approximation that the galaxy is an isothermal sphere, as $g \approx 2\sigma^2/R$. SDSS reports two values for the stellar velocity dispersion of the BCG: 1) $\sigma = 444 \pm 58 \text{ km s}^{-1}$ from the SDSS pipeline's \textsc{SpecObj} table, and 2) $\sigma = 356 \pm 34 \text{ km s}^{-1}$ from the Portsmouth catalog. We adopt the value reported from the SDSS pipeline as the stellar velocity dispersion and use the $\sim 100\text{ km s}^{-1}$  difference between the two measurements as a proxy for the systematic uncertainty in the velocity dispersion of the host galaxy. It would take at least $ \geq 150$ Myr for the cavity's hotspot to have traveled to its current distance of $\sim 67$ kpc moving at the terminal velocity ($v_t \sim 750 \kms$).

\setlength{\tabcolsep}{15pt}
\begin{deluxetable*}{ccccccccc}
\tabletypesize{\footnotesize}
\tablewidth{\linewidth}
\tablecaption{\textsc{Radio Properties Assuming Equipartition and/or pressure equilibrium}\label{tab:radio_sources}} 
\tablehead{
\colhead{Source} &
\colhead{$L_{\rm rad}$} &
\colhead{$\phi$} &
\colhead{$k$} &
\colhead{$P_{\rm min}$ } &
\colhead{$E_{\rm min}$ } &
\colhead{$B_{\rm min}$ } &
\colhead{$t_{e}$ }\\
 &
(\rm erg s$^{-1}$) &
&
&
(\rm dyn cm$^{-2}$) &
\colhead{(\rm erg)} &
\colhead{ (\rm $\mu$G)} &
\colhead{(\rm Myr)} 
}
\startdata
C & 1.5 $\times 10^{41}$ & 1 & 1 & 7.7$\times 10^{-13}$ & 1.2$\times 10^{58}$ & 2.9 & 54  \\
\nodata & \nodata & 0.1 & 5499 & 2.7$\times 10^{-10}$ & 4.2 $\times 10^{59}$& 54 & 4.7  \\
\hline
D & 7.1 $\times 10^{40}$ & 1 & 1 & 1.2$\times 10^{-12}$ & 2.8$\times 10^{57}$& 3.5 & 56  \\
\nodata & \nodata & 0.1 & 2699 & 2.7$\times 10^{-10}$ & 6.5$\times 10^{58}$& 54 & 5  \\
\enddata
\tablecomments{Summary of the radio properties derived for Sources C and D assuming equipartition only, and assuming equipartition and pressure equilibrium between the external gas pressure and the magnetized gas. }
\vspace{-6mm}
\end{deluxetable*}

We can also obtain a rough age of the cavity by assuming that synchrotron radiation and inverse-Compton losses were the only mechanisms by which the relativistic electrons lost energy. However, this method is less reliable because we only have two low-resolution spectral data points.  Assuming the equipartition of energy between relativistic particles and the magnetic field \citep{burbidge_estimates_1959}, we calculate the minimum magnetic field, pressure, and particle energy of the lobe using the following relations in \citet{odea_astrophysical_1987}:

\[\begin{aligned}
L_{\text{rad}} &= 1.2 \times 10^{27} \, D_L^2 S \nu_{\rm br}^{-\alpha} (1 + z)^{-(1 + \alpha)} (\nu_u^{1 + \alpha} - \nu_l^{1 + \alpha})(1 + \alpha)^{-1}  , \\
P_{\text{min}} &= \left( 2\pi \right)^{-3/7} (7/12) [ C_{12} L_{\text{radio}} (1+k) \phi V^{-1} ]^{4/7} ,  \\
B_{\text{min}} &= [ 2\pi (1 + k) C_{12} L_{\text{radio}} V^{-1} \phi^{-1} ]^{2/7} ,\\
E_{\text{min}} &= V \phi \left( 2\pi \right)^{-1} \left( L_{\text{radio}} (1 + k) C_{12} \right)^{4/7},\\
\end{aligned}\]

\noindent where $L_{\rm rad}$ is the radio luminosity in erg s$^{−1}$, $D_L$ is the luminosity distance in Mpc, $z$ is the source redshift, $S$ is the total flux density in Jy, $\nu$ is the frequency at which $S$ is measured in Hz, $\nu_u$ and $\nu_l$ are the upper and lower frequency cutoffs in Hz, respectively, $P_{\rm min} $ is the minimum pressure in dynes cm$^{−2}$, $k$ is the ratio of energy density in non-radiating particles to that in synchrotron-emitting particles, $V$ is the source volume, $C_{12}$ is a constant depending on the spectral index and frequency cutoffs, $\phi$ is the volume filling factor, $B_{\rm min}$ is the magnetic field at minimum pressure in $G$, and $E_{\rm min}$ is the particle energy (electrons and protons) at minimum pressure in ergs. All equipartition-derived values are given in Table \ref{tab:radio_sources}. 

The total radio luminosity was estimated assuming the radio spectrum extends from $\nu_l = 10$ MHz to $\nu_u=100$ GHz with a spectral index of $\alpha = −1.7 \pm 0.5$, and $S = 0.025$ Jy at $\nu = 144$ MHz. Over the cavity's $3.2 \times 10^5$ kpc$^3$ volume, we assume the particle energy is equally divided between the electrons and protons ($k=1$) and that the lobes are filled with radio plasma, which is justified by the fact that cavities must be mostly empty of thermal gas or they would not be evident in X-ray images \citep{mcnamara_chandra_2000, fabian_chandra_2000, blanton_chandra_2001}. We then estimate the lifetime of electrons in the radio component undergoing both synchrotron radiative and inverse-Compton losses on cosmic microwave background photos using the following relation from \citet{van_der_laan_aspects_1969}:

\begin{align}
t_e &= 2.6 \times 10^{4} \frac{B^{1/2}}{B^2 + B_R^2} \frac{1}{[(1 + z) \nu_b]^{1/2}} \, \rm{ yr}
\end{align}

\noindent where $B$ is the equipartition magnetic field $B_{\rm min}$, and the break frequency, $\nu_{\rm br}$, is assumed to be 144 MHz, the lowest available observing frequency. We estimated the magnetic field equivalent to the radiation, which is assumed to be primarily CMB photons, as $B_r \approx 4 \times 10^{-6} (1+z)^2$ G. With $B_{\rm min} = 2.9\mu$G, we derive an electron lifetime of $\sim 50$ Myr, which is about 3 times below our initial estimate.

The above calculation, however, yields a minimum pressure of $p \sim 7\times 10^{-13}$ dyne cm$^{-2}$, which is three orders of magnitude lower than the external gas pressure derived from the X-ray observations ($\sim 2 \times 10^{-10}$ dyne cm$^{-2}$). This scenario is clearly unphysical, for the bubble would collapse under the pressure of the external medium. Furthermore, the location of cool X-ray and multiphase gas near the rim surrounding the radio lobe argues strongly against supersonic motion near the lobe boundaries \citep{nulsen_interaction_2002}. To maintain equipartition and achieve local pressure equilibrium between the magnetized gas in the radio lobe and unmagnetized gas in the outer rim of the X-ray cavity, we must assume $1 + k \geq 5500$ and reduce the filling factor to $\phi \leq 10\%$. This, however, generates a much stronger magnetic field of $B_{\rm eq} \sim 50 \mu$G, and therefore an unrealistically short plasma age of $\sim 5$ Myr. This either means that equipartition does not hold in the relic lobe, or that there is additional pressure support, potentially from the lobe entraining hot thermal gas. This additional pressure support has been proposed to compensate for the significant pressure differences observed in other sources, such as the intermediate FRI/FRII sources in MS0735.6+7421 \citep{biava_constraining_2021} and Hydra A  \citep{croston_particle_2014}. 

There are additional puzzles with interpreting Source C as a relic AGN lobe. Firstly, AGN jets typically come in symmetric pairs, producing a pair of radio lobes in the ICM on two sides of the AGN. Assuming Source C represents one relic lobe, its counterpart is missing. Facing a similar dilemma in the Ophiuchus cluster, which also hosts a massive relic AGN lobe, \citet{giacintucci_discovery_2020} speculated that the counterpart may have propagated into a less-dense ICM on the other side of the cluster and completely faded away. 

While this may also be the case for SDSS 1531, we also explore the possibility that Source D is the missing counterpart given its similar morphology to Source C, and similarly steep spectral index ($\alpha \sim -1.3$). Adopting 150 Myr as the minimum age for both relic lobes, Source D would need to have been displaced at a minimum speed of $\sim 780 \kms$ to reach the projected $\sim 120$ kpc distance away from the southern BCG. This value is w below the measured velocity dispersion of $998^{+120}_{-194} \kms$ for 11 galaxies in the cluster, and also below the sound speed $c_s \sim 1100 \kms$ at 4.7 keV, an upper limit on the motions of gas within the ICM. Merging clusters, which SDSS 1531 likely is, can generate large-scale ($\sim 100$ kpc), long-lived ($\sim 1$ Gyr) motions of the ICM, called sloshing (for a review, see \citet{markevitch_shocks_2007}). Numerical simulations of cluster mergers predict gas velocities up to $\geq 1000 \kms$ \citep[e.g.,][]{roettiger_when_1993, roettiger_observational_1996, roettiger_anatomy_1998, ricker_off-axis_2001}. Therefore, it is plausible that turbulent gas motions in the ICM displaced the lobe.

If Source D is not the missing counterpart, and the lobe has not faded away in a less dense portion of the ICM, we also explore whether Source C could be a Wide-Angle-Tail (WAT) radio source whose tails are unresolved by the LOFAR beam. WATs are powerful, bent radio sources thought to be produced via the interaction of the outflowing radio jets with the magnetized ICM (for review, see \citet{odea_wide-angle-tail_2023}. As a result, WATs are preferentially found in merging clusters, which we believe to be the case for SDSS 1531. Although WATs are rarely found in cool core clusters because mergers strong enough to produce WATs are expected to disrupt the cool core \citep{ritchie_hydrodynamic_2002, zuhone_stirring_2010}, the few WATs observed in cool cores may not have had their cores disrupted yet due to a potential time delay between the merger and disruption. The main detractor from the WAT explanation lies in the morphology of Source C, which significantly deviates from typical WAT characteristics. WATs usually have tails of radio emission extending from the BCG, but Source C has an ``avocado" shape, with its major axis perpendicular to the expected tail orientation. 

If Source C is not a relic from a past AGN outburst, it could be a radio mini-halo. The $\geq 3\sigma$ emission fully traces the X-ray surface brightness distribution within the central 60 kpc of the cool core. Such large-scale emission is a hallmark of radio mini-halos, which are typically confined to the cores of relaxed clusters ( $\sim 50 - 500$ kpc) and characterized by higher emissivity \citep[e.g.,][]{giacintucci_new_2013, giacintucci_occurrence_2017}. While ram pressure might have altered the mini-halo's symmetrical profile to extend more to the southeast, this scenario is unlikely due to the striking concurrence between the X-ray cavity and the bulk of the radio emission. Furthermore, the image may artificially inflate the size of the source due to smearing from LOFAR's larger synthesized beam. In the event a mini-halo is present, its emission is likely intertwined with that from the AGN relic, but the low-resolution observation makes it difficult to distinguish them clearly, as observed in Abell 2029 \citep{govoni_search_2009}.

\subsubsection{Cooling Stimulated and Mitigated by AGN Feedback}
\label{subsubsec:agn_feedback}

As reviewed in the introduction, recent multi-wavelength observations have found evidence for multiphase gas draped around the rims of X-ray cavities in projection, supporting the idea that the gas condenses in the wake of the rising bubbles. Introduced by \citet{mcnamara_mechanism_2016}, the ``stimulated feedback" model proposes that X-ray bubbles blown by AGN outbursts lift cold, low entropy gas away from the location where the heating rate balances the cooling rate. If the overdensity does not return to its original position within its cooling time, the gas becomes thermally unstable, leading to condensation of small clouds that eventually rain back onto the BCG. 

Despite the current inactivity of the southern BCG's AGN, the southern end of the multiphase gas still retains a tangential connection to the cavity opening in projection, linking it to its potential origin from the previous epoch of AGN activity. During the bubble's inflation to its current projected location, it would have displaced ($M_{\rm disp} = \mu (m_p + m_e) n_e V \sim 10^{11} \msun$) of hot gas. The molecular gas has a lower mass of $\sim 10^{10} \msun$, implying a hot-to-cold accumulated mass ratio of 0.1 and indicating that the bulk of the cold reservoir could have been accumulated over the most recent cycle of jet activity.

To determine whether it is energetically possible for the cavity to have displaced the gas and mitigate the rapid ICM cooling, we can calculate the work done by the buoyantly rising cavity. \citet{birzan_lofar_2020} examined the low frequency radio emission in 19 nearby ($z<0.3$), and 6 higher redshift cool core clusters ($z>0.3$). When detected, they found that the low-frequency radio-emitting plasma rarely extended beyond the X-ray cavity edges, suggesting limited evidence for CR electrons leaking. Thus, the $\geq 12\sigma$ radio contour of Source C likely provides a better approximation for the actual extent of the cavity compared to the opening uncovered by the shallow X-ray observations. 

To estimate the bolometric X-ray cooling luminosity, $L_{cool}$, which describes the total luminosity within $r_{cool} = 40$ kpc, we use $L_{cool} = 4 \pi d_L^2 \cdot f_{SB}$, where $f_{SB}$ is the flux from the surface brightness profile, and $d_L$ is the luminosity distance. This gives an X-ray luminosity of $2.18 \pm 0.14 \times 10^{43} \text{ erg s$^{-1}$}$. Using the hybrid X-ray-radio method put forward in \citet{timmerman_measuring_2022}, which uses the volume measurement derived from radio observations and pressure measurement derived from X-ray observations, we can derive the cavity power $P_{\rm cav}$ as:

\[P_{cav} = \frac{E_{cav}}{t_{cav}} = \frac{4pV}{t_{cav}},\]

\noindent where $E_{\rm cav}$ is the enthalpy of the rising bubble and $t_{\rm cav}$ is the age of the cavity ($\sim 150$ Myr). 

For a surrounding pressure of $\sim 2.7 \times 10^{-10}$ dyne cm$^{-2}$, we obtain a cavity enthalpy of $\sim 1 \times 10^{61}$ erg, which is the same order of magnitude as the energetic outbursts observed in the few clusters with ``supercavities," namely Hydra A \citep{nulsen_cluster-scale_2005, wise_x-ray_2007}, MS 0735+7421 \citep{gitti_cosmological_2007}, Hercules A \citep{nulsen_powerful_2005}, and the more recent Ophiuchus cluster \citep{giacintucci_discovery_2020}. The mean cavity power is $2.2 \pm 0.5 \times 10^{45}$ erg s$^{-1}$, yielding $P_{\rm cav}/L_{\rm cool} \sim 86$. This indicates that AGN feedback alone would be sufficient to extinguish the cooling flow and raises questions regarding this energetic outburst's impact on the cool core, which we discuss further in Section \ref{subsubsec:context}.

Since the radio/X-ray cavity is likely a relic of past, rather than ongoing AGN activity, the cavity is likely no longer uplifting the large reservoir of molecular gas, as one would expect in its youth. Instead, the uplifted gas is likely now raining back down onto the BCGs. The gas, however, exhibits smooth radial velocity gradients, which could suggest either an inflow towards or outflow away from the gravitational potential of the BCGs.

Establishing the direction of gas motion is challenging without directly observing gas clouds in absorption. Absorption lines have been detected against the sub-mm nuclear continuum emission in a few systems such as NGC 5044, A2597, and Hydra A \citep{david_molecular_2014, tremblay_cold_2016, rose_constraining_2019, rose_does_2023}. In NGC 5044 and A2597 \citep{david_molecular_2014, tremblay_cold_2016}, the apparent motion of the gas relative to the AGN indicates inflowing clouds that could serve as fuel for the central AGN. In contrast, the motion of the gas in Hydra A \citep{rose_constraining_2019} suggests that it is on a stable, low ellipticity orbit around the central galaxy. For SDSS 1531, however, ALMA detected no mm continuum emission, likely because the gas has yet to be accreted by the AGN.

If the projected separation between the young stars and molecular gas is caused by ram pressure stripping (discussed in-depth in Section \ref{subsubsec:assemble}), it can provide insight into the direction of gas motion. Section 4 of \citet{li_effects_2018} investigated this using high-resolution hydrodynamical simulations ($\sim 244$ pc for the smallest cell) to examine the effect of ICM ram pressure on the cold clouds in the centers of cool-core clusters for different AGN feedback models. In their execution of the precipitation model, most cold gas condenses out of the ICM due to local thermal instabilities, causing the gas only to move inward. They find that when there is a detectable separation between the young stars and cold gas, the location of the young stars is always closer to the cluster center than the cold gas. In SDSS 1531, the young stars lead the molecular gas relative to the merging galaxies, and the gas is entirely redshifted, consistent with the scenario where the gas is falling toward the center from between the cluster center and the observer. This observation, coupled with the location of the molecular gas by the cavity rim, supports the idea that we are observing the gas at a stage where it has decoupled from the AGN-blown bubble. 

According to precipitation models, when the uplifted gas decouples from the ICM, it follows a drag-limited ballistic orbit, achieving speeds of up to $\sim 300-1000$ km s$^{-1}$. However, observations of condensed clouds in cluster cores often show that clouds drift with subvirial velocities, in agreement with the CCA model. In the CCA model, the condensed gas displays bulk velocities up to a few $100$ km s$^{-1}$. To test whether the bulk motions of the molecular gas follow a semi-ballistic trajectory, we follow \citet{lim_radially_2008} and assume that the gravitational potential can be modeled by a Hernquist profile \citep{hernquist_1990apj356359h_1990}. Thus, a freely falling gas cloud should accelerate to a velocity $v(r)$ of 

\begin{align}
\frac{v(r)^2}{2} - \frac{GM}{(r + a)} = \frac{v(r_0)^2}{2} - \frac{GM}{(r_0 + a)}
 \label{eqn:ff_vel}
\end{align}

\noindent with respect to the ICM. Following \cite{vantyghem_molecular_2016}, we modify the velocities in the above equation, such that $v(r)$ is $v(r) - v_{ICM}$, where $v_{ICM}$ is the velocity offset between the BCGs and the cooling gas, and $v(r_0)$ is the initial velocity of the cloud, which is the same as that of the ICM: $v(r_0) = v_{ICM}$. The inclination angle of the cloud's trajectory and $v_{ICM}$ are free parameters in this model. 

The remaining parameters in Equation \ref{eqn:ff_vel} are $r_0$, the initial radius from which the cloud originally formed, $r$, the current distance of the cloud from the center of the gravitational potential, $a$, the scale length related to the half-mass radius $r_{1/2}$ ($a = r_{1/2}/(2 + \sqrt{2}$) and $M$, the total gravitating mass of the BCGs.  The BCGs' total mass interior to 30 kpc is $\sim 10^{13} M_\odot$ \citep{sharon_mass_2014}. The K-band luminosity from the SDSS observations provides a rough estimate of $3.8 \times 10^{11}$ M$_\odot$ for the stellar mass of the BCG \citep{maraston_modelling_2009}. Assuming the half mass radius $r_{1/2}$ is approximately half of the $\sim 30$ kpc Petrosian radius of the BCsG, the scale length is $a \sim 4.4$ kpc. The amplitude of the velocity profile is primarily controlled by the total gravitating mass, which is degenerate with the inclination angle. Therefore, our results are not strongly affected by the adopted total mass value, since the inclination angle is uncertain and can be adjusted to compensate.

\begin{figure}
  \centering \includegraphics[width=\linewidth]{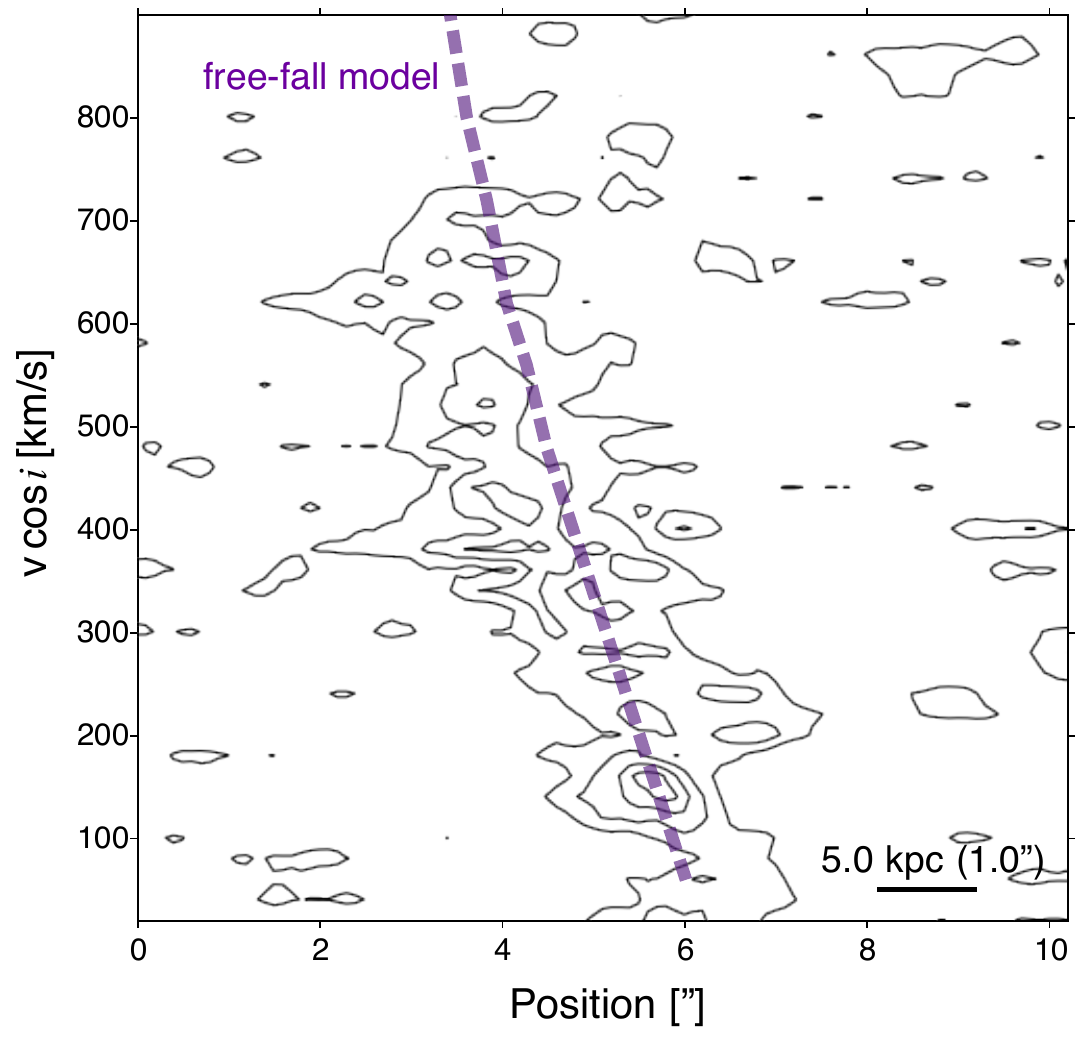}
 \caption{A free-fall model depicting the motion of a parcel of molecular gas released from a height of $\sim 18$ kpc along the gravitational potential of the SDSS 1531 BCG. The velocity profile of the parcel is represented by a dashed purple line, which is overlaid on the contours of the \textit{PV} diagram presented in Figure \ref{fig:alma_pv}. The model successfully describes the observed position-velocity distribution of the molecular gas.}
\label{fig:disc_freefall}
\end{figure}

In Figure \ref{fig:disc_freefall}, we present the velocity trajectory (purple) of a freely falling clump of gas dropped from an initial height of 16 kpc on top of the ALMA \textit{PV} diagram of the molecular gas. The trajectory assumes an inclination angle of 72$^\circ$ and $v_{ICM}$ of  $+600$ km s$^{-1}$. This simple gravitational free-fall model broadly reproduces the observed velocity gradient and velocities spanned by the molecular gas as it flows from south to north. Though the bulk motions of the cloud support gravitational infall, it is important to note that the molecular gas is not akin to a monolithic slab. Given the low volume filling factor of CO, the structure is more like a “mist” of smaller individual clumps and filaments seen in projection \citep[e.g.,][]{jaffe_infrared_2001, jaffe_hii_2005, wilman_integral_2006, emonts_co1-0_2013, mccourt_characteristic_2018, tremblay_galaxy-scale_2018}. The free-fall model fails to account for the fact that the individual clumps are likely not smoothly connected in velocity. Furthermore, the surface density of giant molecular clouds is such that the clouds reach terminal velocity and accelerate freely within the ICM \citep{combes_molecular_2018}. It is possible that some of the individual clumps are in approximate free-fall, while others are animated by the bulk velocities of the surrounding ICM velocity, as evidenced by the redshifted velocities characterizing the molecular filament.

\subsubsection{Assembling the Beads: A Cooling Wake, Ram Pressure and Tidal-Induced Star Formation}
\label{subsubsec:assemble}

The observed "beads-on-a-string" star formation complex in SDSS 1531 is likely a product of the dynamic cluster environment. Although the young stellar superclusters are separated from the molecular gas by $\sim 1-3$ kpc, the similarity in morphology between the beaded star formation and the edges of the cold molecular gas suggests that the gas played a critical role in fueling the star formation. Similar offsets between star formation and cold gas have been found in the Perseus Cluster (NGC 1275), Abell 1795, and in simulations of AGN feedback \citep{canning_filamentary_2014, tamhane_radio_2023, li_cooling_2015}.  Below, we discuss the potential contributions of a cooling wake, ram pressure, and tidal interactions to the observed separation between the star formation, the BCGs, and the molecular gas.

\paragraph{A Cooling Wake} The young stars and the cold molecular gas may be offset from the BCGs due to gas cooling in the wake of the central galaxies as they move through the cluster atmosphere. \citet{fabian_chandra_2001} proposed this scenario for Abell 1795, a cool core cluster where the BCG is offset by $+150$ km s$^{-1}$ from the cluster mean velocity, and by $+374$ km s$^{-1}$ within the central 270 kpc \citep{oegerle_dynamics_1994}. A1795 features a $\sim 50$ kpc trail of multiphase gas that extends to the southeast of the BCG, with the gas motions reflecting that of the cluster instead of the BCG. In SDSS 1531, the central ellipticals are offset by $\sim +100-400$ km s$^{-1}$ from the average cluster velocity, and the motions of the multiphase gas are similarly offset from the BCGs. Without a direct measurement of the ICM velocity, we cannot confirm whether the BCGs are in motion with respect to the cluster. If they are, the molecular gas was likely deposited above the BCGs and progressively slowed by dynamical friction and/or ram pressure as it approached the BCGs' gravitational potential well \citep{combes_molecular_2018}. This would explain why the gas is redshifted with respect to the southern nucleus and why the gas is not close enough to be accreted and fuel an AGN response \citep[e.g.,][]{tremblay_cold_2016, tremblay_galaxy-scale_2018}.

\paragraph{Ram Pressure} As discussed in Section \ref{subsubsec:cluster_dynamics}, the presence of a major merger between the central galaxies, a bifurcated cluster redshift distribution, and the presence of diverse sources of diffuse radio emission within the cluster collectively suggest a turbulent ICM environment in SDSS 1531. A significant pointer towards the effects of the turbulent ICM motions is the morphology of Source B, reminiscent of a radio ``jellyfish" galaxy. When such galaxies move through the dense ICM, they experience ram pressure forces strong enough to directly strip gas out of the disk and leave behind a wake of material trailing the galaxy. Complementing this, the relic AGN lobe extends southeastward from the southern BCG, mirroring the orientation of the molecular gas and the young stellar superclusters. Their parallel displacement suggests that ram pressure also plays a significant role in creating the observed projected offset between the YSCs and the cold molecular gas.

Since ram pressure primarily acts upon gas rather than stars, in order of critical density, it should slow down the ionized gas more efficiently than the molecular gas. In SDSS 1531, it is possible that the stars initially formed in the infalling molecular clouds and subsequently decoupled from the gas. As a result, the stars are now moving in the gravitational potential of the central galaxies without significant resistance, while the remaining molecular gas is still acted upon by ram pressure from the ICM.

To assess ram pressure's impact on the multiphase gas, we look to the few spatially resolved, high-resolution observations of molecular gas in jellyfish galaxies \citep[e.g.,][]{jachym_alma_2019, moretti_gasp_2020, cramer_alma_2020, zabel_alma_2019, zabel_alfocs_2020}. Each study shows that ram pressure from the ICM strips interstellar gas from infalling galaxies while the stars in the galaxy remain unaffected. Although the majority of these studies have found that the bulk of H$\alpha$-emitting gas is co-spatial in projection with CO, Figure 3 in \citet{moretti_gasp_2020} shows an offset on the order of $\sim 1$ kpc between H$\alpha$ and CO, similar to what is observed in SDSS 1531.

Jellyfish galaxies experience great ram pressure forces as they fall through the ICM to the center of the cluster potential well. The central galaxies in SDSS 1531, however, presumably sit at the center of the cluster potential well, absent a more detailed redshift survey to prove otherwise (e.g., Abell 1795; \citealt{oegerle_dynamics_1994}). Although the galaxies are not falling into the center at high $>1000$ km s$^{-1}$ speeds, the high gas density $\rho_{\text{ICM}}$ in the center of the cluster increases the strength of the ram pressure forces. 

To estimate the total ICM mass stripped $M_{\text{strip}}$ within a stripping radius $R_{\text{strip}}$, we use the analytically determined relations from \citet{gunn_feeding_1979} and \citet{domainko_enrichment_2006}, modified to account for the fact that the gravitational potential in BCGs is dominated by the total dark matter and gas mass and that elliptical galaxies do not have a stellar disk:
\begin{align}
M_{\text{strip}} = (x + 1)e^{-x} \cdot M_{\text{gas}},
\label{eqn:rps_mass}
\end{align}
where
\begin{align}
\label{eqn:rps_x}
\end{align}

 In the above equations, $M_{\text{total}}$ is the total mass of the BCGs, $M_{\text{gas}}$ is the ICM mass, $R_0$ is the Petrosian radius of the BCGs, and $v_{\text{gal}}$ is the relative velocity between the BCGs and the ICM.  We adopt a value of 30 kpc for the radius, based on the Petrosian radius \citep{tremblay_30_2014}. This radius encloses the BCGs, the beaded string of star formation, and the molecular gas. From our X-ray observations, we obtain $M_{HE\text{, total}}=3.5 \times 10^{12} M_\odot$ and $M_{\text{gas}}=5.9 \times 10^{10} M_\odot$ within $\sim 30$ kpc, using the total mass and gas density profiles outlined in Section \ref{subsec:chandraobs}. We also set $v_{\text{gal}} = 300$ km s$^{-1}$, the relative velocity between the merging galaxies. Within a radius of 30 kpc, $\rho_{\text{ICM}} \sim M_{\text{gas}}/V_{\text{30 kpc}}$ $ \sim 5.4 \times 10^{-26} g/\text{cm}^3$.  Using the simplifying equations and assumptions above, we find that ram pressure can strip up to $5.5 \times 10^{10} M_\odot$ of gas within $\sim 15$ kpc. This is comparable to the derived $\sim 10^{10}$ molecular gas mass. 

To estimate the time needed for ram pressure to separate the YSCs from the molecular gas, we consider a simple model. In this model, we assume that the infalling molecular gas is subject to ram pressure and that the YSCs formed at $t=0$. At $t=0$, the YSCs and gas have an initial relative velocity of $v_0=0$ km s$^{-1}$. From this time onwards, we assume that only ram pressure causes the separation between the stars and gas, with no additional acceleration or deceleration due to other factors such as turbulence and that both components experience roughly equivalent gravitational acceleration. We can estimate the gas' acceleration $a_{\text{mol}}$ due to ram pressure as:

\begin{equation}
a_{\text{mol}} = F_{\text{ram}}/m_{\text{mol}},
\end{equation}

\noindent where $F_{\text{ram}} = P_{\text{ram}} * A_{\text{mol}}$, where A is the cross-sectional area of the molecular gas, and $m_{\text{mol}}$ is the mass of the molecular gas. Modeling the projected extent of the molecular gas an ellipse with major and minor axes of $\sim 11.1$ kpc and $\sim 6.8$ kpc, respectively, we obtain a cross-sectional area of  $235$ kpc$^2$, and an acceleration of $1.8 \times 10^{-14}$ km s$^{-2}$. Under the simplifying assumption that the molecular gas is only moving away from the young stars, we can estimate the time $t_{\text{sep}}$ needed for the gas to travel a projected distance $d=1-3$ kpc as:

\begin{equation}
t_{\text{sep}} = \sqrt{\frac{2d}{a_{\text{mol}}}}.
\end{equation}

We obtain a timescale of $\sim 60-100$ Myr for the YSCs to separate from the molecular gas, which is consistent with the rough $<300$ Myr age of the YSCs \citep{tremblay_30_2014}. 

\paragraph{Tidal Interactions} The beaded star formation may have been stimulated by tidal forces resulting from the major merger between the BCGs. When galaxies merge, tidal forces can pull and distort the stars and gas within them, moving stars from the disk to the spheroid component \citep{toomre_mergers_1977, kaviraj_tidal_2012}. Tidal forces can also compress and shock gas into rapidly forming stars, resulting in a tidal-induced "starburst" \citep{wang_offnuclear_2004}. Gas-rich (wet) mergers typically host such starbursts due to the abundant fuel available for star formation  \citep[e.g.,][]{lin_redshift_2008, perez_chemical_2011, athanassoula_forming_2016}. In contrast, gas-poor (dry) mergers have less fuel, making starbursts less common in these systems \citep[e.g.,][]{bell_dry_2006, naab_properties_2006, lin_redshift_2008}.

Although gas-rich mergers between elliptical galaxies have been observed in some systems \citep[e.g.,][]{kaviraj_tidal_2012, george_structural_2017}, the massive ellipticals in SDSS 1531 are more likely to be gas-poor given the lack of significant molecular or ionized gas detected near their nuclei. However, tidal interactions can still impact the large reservoir of molecular gas that cooled from the ICM, providing the gas needed to stimulate star formation as expected in a wet merger. Since tidal forces scale with distance as $r^3$, they are strongest in the region between the merging galaxies and decrease in strength outwards. These forces may have compressed the gas along the western border of the gas, stimulating a burst of star formation. Since tidal interactions impact both stars and gas, they could also explain why some of the strings of star formation appear tightly wound around the nuclei of the two central galaxies in projection. Tidal forces may have been too weak to compress the shielded gas further east, accounting for the absence of star formation in that region. Further investigations with hydrodynamical simulations would help further understand the role of tidal interactions in SDSS 1531.

\subsubsection{SDSS 1531 In Context}
\label{subsubsec:context}

\paragraph{Another Beads-on-a-String System} To our knowledge, SDSS 1531 is one of only two known cool core clusters that hosts a "beads on a string" star formation complex near its center. The second, the SPARCS 104922.6+564032.5 BCG (hereafter SPARCS 1049), resides among several merging cluster members and features diffuse emission in a tidal-tail-like structure adorned with $\sim 66$ kpc long beads-on-a-string star formation \citep{webb_extreme_2015}. The star formation is similarly offset from the central galaxy, though to a much larger extent of $\sim 25$ kpc. 

While initial investigations linked the distinct morphology of SPARCS 1049 to starbursts induced by gas-rich mergers, and a gas-rich BCG \citep{webb_extreme_2015}, more recent observations \textit{Chandra} \citep{hlavacek-larrondo_evidence_2020} and the IRAM interferometer NOEMA \citep{castignani_environmental_2020}
have uncovered rapid ICM cooling and a significant $\sim 10^{11} M_\odot$ reservoir of cold molecular gas, offset from the center, likening its characteristics to SDSS 1531. SPARCS 1049's $\sim 860 M_{\odot} yr^{-1}$ star formation rate matches the rapid ICM cooling rate, making it one of the few known ``true" classical cooling flow systems \citep{hlavacek-larrondo_evidence_2020}. Observations also suggest that the AGN is currently inactive \citep{trudeau_multiwavelength_2019}. 

SDSS 1531, with a $\sim 1-10 M_\odot$ SFR, has likely diverged from SPARCS 1049's current evolutionary path due to an earlier epoch of powerful AGN activity suppressing its cooling rate. Although the redshift distribution of 27 cluster members in SPARCS 1049 provides no evidence for significant subcluster structure \citep{webb_extreme_2015}, the large-scale star formation and merger activity within its core suggests a dynamic cluster environment similar to that of SDSS 1531. In both clusters, the ongoing mergers and residual cluster motions likely stimulated the `beads on a string' star formation morphology. As a result, SDSS 1531 could represent a future evolutionary stage for SPARCS 1049, after AGN feedback severely limits the observed runaway cooling and star formation. 

\paragraph{Another Fossil AGN Outburst detected in a Cool Core system}

Although AGN feedback is a common occurrence in cool core clusters, the potential relic AGN lobes found in SDSS 1531 and Ophiuchus \citep{giacintucci_discovery_2020} stand out due to the massive amount of energy ($\sim 10^{61}$ ergs) required to excavate each lone supercavity. Neither cavity is fully resolved by the X-ray data, and both clusters show limited signs of young AGN activity. Both are also expected to have undergone cluster-scale merger activity, with the merger in Ophiuchus strong enough to shift the peak of the cooling gas off from the BCG. In theory, such large-scale mergers and powerful AGN outbursts should heat and displace a large amount of the ICM in the cooling cluster cores, contributing significantly to their eventual disruption. In MACS J1931.8−2634, which also hosts a powerful AGN outburst and cluster-scale merger, \cite{ehlert_extreme_2011} argued that the feedback from both processes has significantly disrupted the cooling core, evidenced by the cluster's flattening entropy profile, and the metalliicity profile's consistently flat slope, which suggest that large masses of metal-rich gas were stripped from the center of the cluster and dispersed to the surrounding regions. In Ophiuchus, the metallicity and entropy profiles are still centrally peaked, though \cite{werner_deep_2016} argued that these profiles were truncated to smaller radii by the activity within the core. In SDSS 1531, the entropy profile appears typical for a cool core cluster. However, deeper X-ray observations are needed to determine the extent to which AGN feedback and the merger have potentially disrupted the cool core. Nevertheless, the description of AGN feedback as ``gentle" \citep{mcnamara_mechanical_2012} still holds for all three systems. Though feedback may have disrupted their cool cores, they do not appear to be transitioning to a non-cool core state, and if they are, the journey is far less chaotic than it could be. 


\paragraph{Kinematics of the Molecular Gas} Molecular gas observed in cool core BCGs typically exhibits extended, filamentary morphologies with complex, chaotic velocity structures (e.g., 13 out of 15 clusters studied in \citealt{olivares_ubiquitous_2019}). Moreover, the majority of the molecular gas mass is typically concentrated in extended filaments (e.g., only 2 out of 12 BCGs studied by \citealt{russell_driving_2019} have $\leq 10\%$ of their mass located in filaments).  In contrast, SDSS 1531's molecular gas is mostly distributed in compact clumps. Additionally, the entire molecular gas structure exhibits a linear velocity gradient. RXC J2014.8-2430 similarly features a clumpy spatial distribution of molecular gas but does not exhibit a similar velocity structure.

Like SDSS 1531, the 2A 0335+096 galaxy cluster harbors two merging central galaxies with velocity offsets of $\sim 300$ km s$^{-1}$ \citep{vantyghem_massive_2021}. Although the spatial distribution of the molecular gas in 2A 0335+096 is more filamentary than clumpy, it exhibits a similar linear velocity gradient that can be replicated by a simple free-fall model. However, the gas is situated in a filament between the central galaxies and spans a significantly smaller spatial extent than the gas in SDSS 1531.

Other galaxy clusters that have been observed to possess molecular gas distributions and velocity structures similar to SDSS 1531 include MACS 1931 BCG \citep{fogarty_dust_2019}, Hydra-A \citep{olivares_ubiquitous_2019}, and Abell 262 \citep{north_wisdom_2021}. The molecular gas in Hydra-A is thought to have a profile consistent with a rotating molecular disk, given the linear gas gradient, double-peaked spectra, and a central broader component of the velocity dispersion. The molecular gas in A262 is believed to either have an outflow or cooling filament of gas stimulated by CCA. In contrast, the molecular gas in SDSS 1531 does not have a velocity profile and structure consistent with a disk. 

\subsection{Alternative Origins}
\label{subsec:alt_origin}

\subsubsection{Molecular gas captured from previous encounters with gas-rich galaxies?}
\label{subsubsec:merger_origin}

A commonly proposed alternative for the origin of multiphase gas in cluster cores is that the gas was accreted from interactions with gas-rich spiral or dwarf galaxies. This theory is consistent with the high luminosity, dusty, and small spatial extent of many molecular filaments observed in cool core clusters. It could also provide a unified picture of AGN and radio galaxies, wherein the material responsible for activating the supermassive black hole is thought to be driven by a stochastic merger process \citep{sparks_imaging_1989}.

We cannot completely rule out this option for SDSS 1531, as BCGs are also predicted to grow through mergers \citep{ostriker_cannibalism_1977}, and there are at least two gas-rich spiral galaxies within the central 50 kpc. One of the aforementioned spiral galaxies has a redshift of $z=0.329$ and shines brightly in the \textit{HST} NUV filter, indicating the presence of young stars. We detect no molecular gas on the galaxy with the current ALMA observations, suggesting that it could have been stripped or that our observations aren't deep enough to resolve the molecular gas content. The merger hypothesis is also viable because captured gas would be expected to rotate around the central galaxies, which we presume to be the case here. 

The mass and velocity distribution of the observed molecular gas are the most significant challenges to the proposed scenario. Although the velocity distribution of the molecular gas shows signs of distinct velocity peaks, as expected in a major merger event \citep[e.g.,][]{gao_molecular_2001, greve_interferometric_2005, schulz_interstellar_2007}, the overall structure is still remarkably coherent, and cannot be easily accounted for by the contributions of multiple galaxies on different orbits. Furthermore, assuming the eight galaxies within the central 100 kpc of the cluster core that shine brightly in the \textit{HST} NUV filter contain significant young stellar populations and thus high gas fractions, we can estimate an average of $\sim 10^8$ M$_\odot$ of $H_2$ (e.g., ESO 137-001, \citealt{jachym_alma_2019}) was stripped and flowed toward the BCGs. This results in $\sim 10^{9} M_\odot$ gas reservoir, an order of magnitude below the estimated gas mass of $\sim 10^{10}$ M$_\odot$. 

\subsubsection{Molecular gas native to the BCGs? }
\label{subsubsec:native_origin}

Lastly, we consider the possibility that the molecular gas is native to the merging central galaxies. In this scenario, the cold gas was expelled during the ongoing major merger, and the beaded star formation results from the gas dynamically responding to the gravitational torques and shear induced by the strong tidal field created by the merger. Of all the scenarios proposed, we find this to be the least likely. The molecular gas does not appear significantly disturbed morphologically, as one would expect if from the result of a major merger (e.g., SP423 \citealt{mcdonald_detailed_2019}). Moreover, major mergers involving elliptical galaxies have been found to trigger minimal or no star formation due to the small amounts of gas present \citep{lin_redshift_2008, cattaneo_downsizing_2008}. Therefore, it is unlikely that the gas from the merger could account for the $\sim 10^{10}$ M$_\odot$ of $H_2$ observed, particularly considering minimal amounts of molecular and ionized gas were detected on the nuclei themselves.

However, the presence of YSCs between the merging BCGs offers some merit to this scenario, albeit on a smaller scale. If the merger enhances gas compression and stimulates star formation \citep{wang_hiohrecombination_2020}, we would expect to find young stars between the merger participants, which is clearly observed. The bulk of the molecular gas lies east of the BCGs, so the little gas native to the BCGs could be responsible for the YSCs in between them. It would naturally follow that the YSCs and the ongoing merger entirely ionized the molecular gas in this region.

\setlength{\tabcolsep}{0pt}
\begin{deluxetable}{lc}[ht]
\tabletypesize{\footnotesize}
\tablewidth{2 \linewidth} 
\tablecaption{\textsc{Observational Evidence }\label{tab:evidence}}
  \tablehead{
    \colhead{Evidence} &
    \colhead{Y/N/Inc} 
}
  \startdata
    \textbf{Origin I}: ICM Condensation, AGN Feedback \& Star Formation & \\
    $t_{\text{cool}} < 1 \text{Gyr}$ & Y \\
    $t_{\text{cool}}/t_{ff} < 30$ & Y \\
    $S \leq 30 \text{keV cm}^2$& Y \\
    Presence of X-ray cavity filled by radio AGN lobe & Y \\
    Bipolar X-ray cavities and radio AGN lobes & Inc \\
    Multiphase gas coincident with X-ray/radio cavity & Y \\
    Uplifted X-ray gas massive enough to supply molecular & Y \\
    gas reservoir &  \\
    Dusty Molecular/Ionized Gas & Inc \\
    Molecular gas flows consistent with ballistic motion & Y \\
    Molecular gas flows consistent with subvirial velocities & N \\
    Optical line ratios consistent with ionization by SF & Y\\
    Central galaxies in motion with respect to the cluster & Inc\\
    Strong ram pressure forces given high $\rho_{ICM}$ and & Y\\
    significant velocity offset between central galaxies & \\ 
    Offset between H$\alpha$ and CO within the ram pressure & Y \\
    stripping radius &  \\
    Morphology of beaded star formation consistent with & Y \\
    strong tidal interactions & \\
    \hline
    \textbf{Origin II}: Molecular Gas Captured from Previous Interactions& \\
    Dusty Molecular/Ionized Gas & Inc \\
    Presence of nearby star-forming galaxies & Y \\
    Presence of enough nearby star-forming galaxies to & Inc\\
    supply observed molecular gas mass & \\
    Multiphase gas exhibits disturbed spatial morphology & N\\
    Multiphase gas exhibits disturbed kinematics & N\\
    \hline
    \textbf{Origin III}: Molecular Gas Native to the BCGs & \\
    Morphology of beaded star formation consistent with strong & Y \\
    tidal interactions & \\
    Multiphase gas exhibits disturbed spatial morphology & N\\
    Multiphase gas exhibits disturbed kinematics & N\\
    YSCs located between merging elliptical galaxies & Y\\
  \enddata
 \tablecomments{ A  summary of expectations for each scenario explored in Section \ref{sec:beads_origin} to explain the origin of the cold molecular gas and whether the current observations support these expectations (Y - Yes), find them unlikely (N - No), or are inconclusive (Inconclusive - Inc).}
\vspace*{-10mm}
\end{deluxetable}
 
 \section{Conclusions, Summary \& Future Work}

Initially observed by the \textit{Subaru} ground-based telescope \citep{hennawi_new_2008}, SDSS 1531 originally appeared to contain one large and disturbed central BCG, with the faint blue excess attributed to an artifact of gravitational lensing. However, subsequent observations from \textit{HST}, \textit{SDSS}, \textit{ALFOSC}, and more revealed \textit{two} elliptical galaxies engaged in a major merger, and that the blue excess was not an artifact of gravitational lensing, but a remarkable chain of stellar superclusters-- a kpc-scale manifestation of the Jeans instability \citep{tremblay_30_2014}. Less than a decade later, with new observational data from \textit{Chandra}, LOFAR, GMOS, and ALMA, this paper presents four main results: 
\begin{enumerate}[leftmargin=*,itemsep=0pt]
    \item \textit{SDSS 1531 is a cool core cluster--} SDSS 1531 satisfies all the criteria for a cool core cluster, including a cuspy emission measure profile ($\alpha = 1.4$) with a clear central overdensity, $t_{cool} < t_H$)within the inner 100 kpc, and $t_{ff}/t_{cool} \sim 20$  within the central 10 kpc. The gas within the $r_{\text{cool}} = 100$ kpc cools at a rate of $\dot{M} \sim 185 M_\odot$ yr$^{-1}$, which corresponds to a cumulative gas mass of $10^{12} M_\odot$ within 3 Gyr if cooling were uninterrupted.
    
    \item \textit{There is compelling evidence for an old, extremely powerful AGN outburst.} The \textit{Chandra} X-ray observations reveal a concave surface brightness discontinuity near the edge of the cool core. LOFAR low-frequency observations fill this discontinuity, in alignment with the picture of a radio AGN lobe pushing aside the X-ray gas to create an X-ray cavity. The source's steep spectrum suggests that the emission stems from aged plasma from an AGN outburst that occurred during an earlier epoch in the cluster's history. The energy required to excavate the cavity is $\sim 10^{61}$ ergs, making it one of the most powerful AGN outbursts observed. If the missing symmetrical lobe has not faded into a less dense portion of the ICM, we propose that the nearby radio source D, which bears a similar morphology and spectral index to Source C, could be the missing lobe. 
    
\vspace*{-3mm}
    \item \textit{Comoving cold and warm gas are tangentially connected to the cavity opening-- } In projection, a bright cloud-like structure of cold molecular and warm ionized gas lies to the north of the X-ray cavity opening, suggesting that the origin of the gas is tied to the older AGN outburst. The multiphase gas likely cooled in the wake of the buoyantly rising cavity and is now infalling back to the BCGs. The warm gas envelops the young stellar superclusters and is redshifted up to $\sim 800$ km s$^{-1}$. The massive $\sim 10^{10} M_\odot$ reservoir of cold gas is mostly comoving with the warm gas, with the central regions differing in velocity by up to $\sim 160$ km s$^{-1}$. The cold gas lies mostly to the southeast of the young stellar superclusters, offset by $\sim 1-3$ kpc.

    \item \textit{The "beads-on-a-string" star formation complex is likely a product of the dynamic cluster environment--} The beaded star formation and the edges of the cold molecular gas suggest that the gas played a critical role in powering the observed star formation. A cooling wake from the central galaxies moving through the intracluster medium and/or strong ram pressure forces may have caused the observed offset between the young stellar superclusters and the molecular gas. Tidal interactions resulting from the major merger between the central ellipticals may have stimulated the beaded star formation via gas compression and contributed to the observed morphology of the stars and gas.
\end{enumerate}

Further constraining the potential origins of the star-forming gas will require follow-up observations across multiple wavelengths in addition to numerical simulations.  A forthcoming spectroscopic survey of the cluster with the Multiple Mirrors Telescope will be crucial for identifying whether a sub-cluster merger is mitigating ICM cooling in SDSS 1531, confirming our estimate of galaxies with high gas fractions within the cluster core that could have been stripped to power the observed cold gas reservoir, and constraining the velocity offset between the central galaxies and the average cluster member. 

To confirm the presence of an extremely powerful, old AGN outburst in the cluster and investigate its link to the multiphase gas, we would need deeper, more sensitive radio observations that sample several lower and higher frequencies. Additional molecular line observations could probe the extent of ram pressure's impact on the spatial offset between the molecular and ionized gas, as we would expect to see spatial offsets between different CO line maps in order of critical density. The molecular line data could also allow the creation of line ratio maps to explore potential excitation mechanisms within the molecular filament, which we predict to be most excited near the western border. Deeper \textit{Chandra} observations will be critical for confirming the presence or absence of X-ray cavities, while future X-ray micro-calorimetric observations could establish if large-scale motions of the surrounding ICM dominate the motion of the cold and warm gas. This scenario could also be investigated using numerical simulations of cluster mergers that include the effects of radiative cooling and star-forming gas to determine if merger-induced gas motions can produce the observed separations between the cooled gas and newly formed stars.

\clearpage
\acknowledgments 

We thank the referee for thoughtful and constructive feedback that greatly improved the manuscript. The scientific results reported in this paper are based on observations from numerous ground- and space-based observatories, namely the \textit{Chandra} X-ray Observatory, LOFAR, VLA, Gemini North, and ALMA. LOFAR, the Low Frequency Array designed and constructed by ASTRON (Netherlands Institute for Radio Astronomy), has facilities in several countries, that are owned by various parties (each with their own funding sources), and that are collectively operated by the International LOFAR Telescope (ILT) foundation under a joint scientific policy. ALMA (data from ADS/JAO.ALMA\#2015.1.01426.S) is a partnership of ESO (representing its member states), NSF (USA), and NINS (Japan), together with NRC (Canada) and NSC and ASIAA (Taiwan), in cooperation with the Republic of Chile. The Joint ALMA Observatory is operated by ESO, AUI/NRAO, and NAOJ. VLA data was accessed from the National Radio Astronomy Observatory (NRAO), a facility of the National Science Foundation, operated under cooperative agreement by Associated Universities, Inc. The international Gemini Observatory, a program of NSF’s NOIRLab, is managed by the Association of Universities for Research in Astronomy (AURA) under a cooperative agreement with the National Science Foundation on behalf of the Gemini Observatory partnership: the National Science Foundation (United States), National Research Council (Canada), Agencia Nacional de Investigaci\'{o}n y Desarrollo (Chile), Ministerio de Ciencia, Tecnolog\'{i}a e Innovaci\'{o}n (Argentina), Minist\'{e}rio da Ci\^{e}ncia, Tecnologia, Inova\c{c}\~{o}es e Comunica\c{c}\~{o}es (Brazil), and Korea Astronomy and Space Science Institute (Republic of Korea). The Gemini Observatory is located atop Maunakea in Hawaii. We are deeply grateful to work with observations taken from the mauna, and acknowledge the mauna's cultural and historical significance to the indigenous Kānaka Maoli community.  The observations made with the NASA/ESA \textit{Hubble Space Telescope}, associated with Program HST-GO-13003, were obtained from the data archive at the Space Telescope Science Institute (STScI). STScI is operated by the Association of Universities for Research in Astronomy, Inc. under NASA contract NAS 5-26555.  We also present data obtained in part with ALFOSC, which is provided by the Instituto de Astrofisica de Andalucia (IAA) under a joint agreement with the University of Copenhagen and the Nordic Optical Telescope (NOT). NOT is owned in collaboration by the University of Turku and Aarhus University, and operated jointly by Aarhus University, the University of Turku and the University of Oslo, representing Denmark, Finland, and Norway, the University of Iceland and Stockholm University at the Observatorio del Roque de los Muchachos, La Palma, Spain, of the Instituto de Astrofisica de Canarias.

OO acknowledges support from the National Science Foundation Graduate Research Fellowship and thanks Professors Karin Öberg, Douglas Finkbeiner and Lars Hernquist for helpful comments. GRT acknowledges support for program HST-GO-16173.001-A provided through a grant from the STScI under NASA contract NAS5-26555. SB and CO acknowledge support from the Natural Sciences and Engineering Research Council (NSERC) of Canada.


\textit{Facilities: ALMA, CXO, Gemini-N, HST, LOFAR, Nordic Optical, VLA}

\textit{Software:} \textsc{Astropy} \citet{robitaille_astropy_2013, collaboration_astropy_2018, collaboration_astropy_2022}, \textsc{CASA} \citep{mcmullin_casa_2007}, \textsc{CIAO} \citep{fruscione_ciao_2006}, \textsc{IPython} \citep{perez_ipython_2007}, \textsc{Matplotlib} \citep{caswell_matplotlibmatplotlib_2022}, \textsc{Numpy} \citep{van_der_walt_numpy_2011}, \textsc{PySpecKit} \citep{ginsburg_pyspeckit_2011}, \textsc{scipy} \citep{virtanen_scipy_2020}

\clearpage

\bibliography{beads_bib}

\end{document}